\documentclass[aip,jcp,amsmath,amssymb,amsbsy,preprint]{revtex4-2}

\usepackage{graphicx}
\usepackage{dcolumn}
\usepackage{bm}

\usepackage[version=3]{mhchem} 
\usepackage[utf8]{inputenc}
\usepackage[T1]{fontenc}
\usepackage{mathptmx}
\usepackage{etoolbox}

\usepackage{comment}
\usepackage{cases}
\usepackage{enumerate}
\usepackage{algorithm}
\usepackage{algorithmicx}
\usepackage{algpseudocode}
\usepackage{paralist}
\usepackage{threeparttable}
\usepackage{threeparttablex}
\usepackage{booktabs}
\usepackage{makecell}
\usepackage{multirow}
\usepackage{color}
\usepackage{bbm}

\newcommand{\ii}{\mathbbm{i}}

\newcommand{\tr}{\mathrm{Tr}}
\newcommand{\bsigma}{\boldsymbol{\sigma}}
\newcommand{\balpha}{\boldsymbol{\alpha}}

\algnewcommand{\Inputs}[1]{
  \State \textbf{Inputs:}
  \Statex \hspace*{\algorithmicindent}\parbox[t]{.8\linewidth}{\raggedright #1}
}
\algnewcommand{\Initialize}[1]{
  \State \textbf{Initialize:}
  \Statex \hspace*{\algorithmicindent}\parbox[t]{.8\linewidth}{\raggedright #1}
}
\algnewcommand{\Outputs}[1]{
  \State \textbf{Outputs:}
  \Statex \hspace*{\algorithmicindent}\parbox[t]{.8\linewidth}{\raggedright #1}
}
\algnewcommand\algorithmicswitch{\textbf{switch}}
\algnewcommand\algorithmiccase{\textbf{case}}
\algdef{SE}[SWITCH]{Switch}{EndSwitch}[1]{\algorithmicswitch\ #1\ \algorithmicdo}{\algorithmicend\ \algorithmicswitch}%
\algdef{SE}[CASE]{Case}{EndCase}[1]{\algorithmiccase\ #1}{\algorithmicend\ \algorithmiccase}%
\algtext*{EndSwitch}%
\algtext*{EndCase}%

\newfloat{Algorithm}{htbp}{alg}
\floatname{Algorithm}{Algorithm}

\makeatletter
\def\@email#1#2{%
 \endgroup
 \patchcmd{\titleblock@produce}
  {\frontmatter@RRAPformat}
  {\frontmatter@RRAPformat{\produce@RRAP{*#1\href{mailto:#2}{#2}}}\frontmatter@RRAPformat}
  {}{}
}%
\makeatother

\UseRawInputEncoding
\begin{document}


\title{Unified construction of relativistic Hamiltonians}

\author{Wenjian Liu}\email{liuwj@sdu.edu.cn}
\affiliation{Qingdao Institute for Theoretical and Computational Sciences,
Institute of Frontier and Interdisciplinary Science, Shandong University, Qingdao, Shandong 266237, P. R. China}

\begin{abstract}

It is shown that four-component (4C), quasi-four-component (Q4C), and exact two-component (X2C)
relativistic Hartree-Fock (HF) equations can be implemented in an unified manner,
by making use of the atomic nature of the small components of molecular 4-spinors.
A model density matrix approximation can first be invoked for the small-component charge/current density functions,
which gives rise to a static, pre-molecular mean field (pmf) to be combined with the one-electron term.
As a result, only the nonrelativistic-like two-electron term of the 4C/Q4C/X2C Fock matrix needs
to be updated during the iterations. A `one-center small-component'  approximation can then be
invoked in the evaluation of relativistic integrals. That is, all atom-centered small-component
basis functions are regarded as extremely localized nearby the position of the atom to which they belong,
such that they have vanishing overlaps with all small- or large-component functions centered at other nuclei.
Under these approximations, the 4C, Q4C, and X2C mean-field and many-electron Hamiltonians share precisely the same structure
and accuracy. Beyond these is the effective quantum electrodynamics Hamiltonian that can be constructed in the same way.
Such approximations lead to errors that are orders of magnitude smaller than other sources of errors
(e.g., truncation errors in the one- and many-particle bases as well as uncertainties of experimental measurements)
and are hence safe to use for whatever purposes. The quaternion forms of the 4C, Q4C, and X2C equations are
also presented in the most general way, based on which the corresponding Kramers-restricted open-shell (RKOHF)
variants are formulated for `high-spin' open-shell systems.
\end{abstract}

\maketitle
\onecolumngrid

\section{Introduction}
The last two decades have witnessed fast progresses in relativistic quantum chemistry, as evidenced by
both theoretical developments\cite{Schwerdtfegerbook1,Grantbook,Dyallbook,Baryszbook,Kaldorbook,Reiherbook,Liubook,PhysRep,IJQCrelH,IJQCeQED,X2C2016,Essential2020,RelChina2020,NakaiRev2021,LiuWires} and efficient implementations in modern computational software\cite{ADFrev2001,UTChem-2003,NTChem-IJQC2015,RAQET-JCC2018,BAGEL-WIREs2018,
Quantum-package2019,Dalton-JCTC2020,NWChem-JCP2020,PSI4-JCP2020,CFOUR2020,Molpro-JCP2020,OpenMolcas-JCP2020,Turbomole-JCP2020,Orca-JCP2020,QChem-5-JCP2021,Hyperion-JCTC2022,
BERTHA-JCTC2020,PySCF-JCP2020,Dirac-JCP2020,Chronus-WIREs2020,ReSpect-JCP2020,BDFrev2020}.
Much of the success is due to the advent of the so-called exact through a one-step block-diagonalization of 
the matrix Dirac equation, similar to the normalized elimination of the small component (NESC)\cite{NESC}.
It is even simpler than approximate two-component (A2C) approaches\cite{CPD,ZORA1,DKH21986,DKH21989} resulting from
order-by-order expansions of the operator Dirac equation and has hence become the main workhorse of relativistic quantum chemistry.
Noticeably, the various implementations\cite{SaueX2C,Q4C,Q4CX2C,mmf-X2C2009,X2CAMF2018,X2CAMF2022,2ePCE2022} of
the same X2C equation\cite{X2C2005,X2C2009} have adopted seemingly different assumptions and even different terminologies. It is therefore timely to establish
an unified formulation to standardize the implementations (see Sec. \ref{SecX2C}), such that
different codes can produce precisely the same results, just like the nonrelativistic ones.
Not only so, the four-component (4C)\cite{swirles1935relativistic}
and quasi-four-component (Q4C)\cite{Q4C} approaches also remain to be unified into the same framework.
To this end, the Dirac-Hartree-Fock (DHF) equation represented in a restricted kinetically balanced (RKB) basis\cite{RKB}
is first recapitulated in Sec. \ref{SecDEQ}, to establish the notations.
For completeness, the quaternion form of DHF is formulated in the most general way, based on which
a four-component Kramers-restricted open-shell HF (KROHF) approach is proposed
for the first time to describe `high-spin' open-shell systems with both double group and time-reversal symmetries (see Appendix \ref{AppendixA}).
The projected four-component (P4C) equation, arising from a special discretization\cite{RosenAtomicP,wood1986relativistic,LiuPhD,LiuBDF1996,BDF1,LCA4S} of the
operator DHF equation,
is then presented in Sec. \ref{SecQ4C}, which leads naturally to Q4C\cite{LiuMP,X2CBook2017} by
invoking a model density matrix (MDM) approximation for the small-component charge/current density functions.
This gives rise to a static, pre-molecular mean field (pmf) that can be added to the one-electron term.
As a result, only the nonrelativistic-like two-electron term of Q4C needs to be updated during the iterative cycles for self-consistency.
That is, Q4C is four-component in structure but two-component in computation, thereby justifying the name\cite{Q4C}.
It turns out that the same MDM approximation can be applied to DHF and X2C as well, so as to
recast them to the same form as Q4C. The `extended atomic mean-field' (eamf) variant of X2C
proposed recently by Knecht and coworkers\cite{2ePCE2022} can also be reproduced this way.
Moreover, the effective, HF-like one-body potential\cite{eQED,PhysRep} describing the leading quantum electrodynamics (QED) effect
can also be included from the outset, particularly when it is fitted into a model spectral form\cite{ShabaevModelSE,ShabaevModelSEcode2018}.
After having discussed in depth these mean-field relativistic approaches, the corresponding many-body relativistic Hamiltonians
will be presented in Sec. \ref{SecMB}. The presentation is closed with concluding remarks in Sec. \ref{SecConclusion}.

Plain and boldface letters are used to denote operators and matrices, respectively. The Einstein summation convention over repeated indices
is always employed.

\section{DHF}\label{SecDEQ}
The Dirac operator-based relativistic mean-field description of a system of $N$ electrons and $\tilde{N}$ positrons
gives rise to the following energy expression\cite{DyallPositron,Essential2020}
\begin{align}
E_{\mathrm{ep}}&=\sum_k n_k h_k^k + \frac{1}{2}\sum_{k,l}n_k n_l\bar{g}^{kl}_{kl}, \quad k,l \in \mathrm{PES, NES}, \label{E1ep}\\
h_p^q&=h_{pq}=\langle\psi_p|h|\psi_q\rangle,\label{hmat}\\
g_{pr}^{qs}&= g_{pq,rs}=(\psi_p\psi_q|V(1,2)|\psi_r\psi_s),\quad \bar{g}_{pr}^{qs}=g_{pr}^{qs}-g_{pr}^{sq},
\end{align}
where $\{n_k\}$ are the occupation numbers of the 4-spinors $\{\psi_k\}$:
$+1$ for the $N$ occupied positive-energy states (PES), $-1$ for the $\tilde{N}$ occupied negative-energy states (NES), and 0
for the remaining unoccupied PESs and NESs.
The one-body operator $h$ is composed of the Dirac operator $D$ and nuclear attraction $V$,
\begin{align}
h&=D+V,\label{hop}\\
D&=c\balpha  \cdot\boldsymbol{p}+(\beta-1)c^2, \\
\balpha  &=\begin{pmatrix}0&\bsigma \\
\bsigma &0\end{pmatrix},\quad \beta=\begin{pmatrix}1&0\\
0&-1\end{pmatrix}, \quad\boldsymbol{p}=-\ii \hbar\boldsymbol{\partial},\\
\sigma_x&=\begin{pmatrix}0&1 \\  1&0 \end{pmatrix},\quad
\sigma_y =\begin{pmatrix}0&-\ii\\  \ii&0 \end{pmatrix},\quad
\sigma_z =\begin{pmatrix}1&0 \\  0&-1\end{pmatrix},\\
V(\boldsymbol{r})&=-\sum_{A}\frac{Z_A}{|\boldsymbol{r}-\boldsymbol{R}_A|}.
\end{align}
The leading term in the
electron-electron interaction $V(1,2)$ is the Coulomb (C) interaction,
\begin{equation}
V^C(1,2)=\frac{1}{r_{12}}=g_{12},\label{Cop}
\end{equation}
which describes the electrostatic, charge-charge interaction between two electrons. On top of this, the Gaunt (G) interaction,
\begin{align}
V^G(1,2)&=-\balpha_1\cdot\balpha_2 g_{12},\label{Gop}
\end{align}
can further be included to account for the magnetic, current-current interaction between two electrons.
The combined CG interaction, $V^C(1,2)+V^G(1,2)$, covers all two-electron spin-orbit, spin-spin, and orbit-orbit couplings\cite{MC-DPT1} and is therefore
accurate enough for most chemical problems. Nevertheless, the gauge term $V^g(1,2)$ in the Breit (B) interaction $V^B(1,2)$,
\begin{align}
V^B(1,2)&=V^G(1,2)+V^{g}(1,2),\label{Bop}\\
V^{g}(1,2)&=\frac{1}{2} g_{12}\balpha_1\cdot\balpha_2-\frac{1}{2}g_{12}^3 (\balpha_1\cdot\boldsymbol{r}_{12}) (\balpha_2\cdot \boldsymbol{r}_{12}),\quad \boldsymbol{r}_{12}=\boldsymbol{r}_1-\boldsymbol{r}_2\label{Bgop1}\\
&=-\frac{1}{2}V^G(1,2)
 -\frac{1}{2}\balpha_1\cdot \mathbf{b}_{12} \cdot\balpha_2,\label{Bgop2}\\
  \mathbf{b}_{12}&=\frac{\boldsymbol{r}_{12}\otimes\boldsymbol{r}_{12}}{r_{12}^3},\quad b_{12}^{ij}=\frac{(\boldsymbol{r}_{12})_i(\boldsymbol{r}_{12})_j}{r_{12}^3},\quad i,j\in x,y,z,
\end{align}
is known to be important for deep core electrons\cite{PyykkoChemRev2012,LiXSDCB-MCSCF} and should hence be
further included in accurate descriptions of properties sampling the wave function in the vicinity of a nucleus.
Overall, the electron-electron interaction can be written as
\begin{align}
V(1,2)&=g_{12}+\mathrm{c_g}g_{12}\balpha_1\cdot\balpha_2+\mathrm{c_b}\balpha_1\cdot\mathbf{b}_{12}\cdot\balpha_2,\label{V12full}
\end{align}
which reduces to the C, CG, and CB interactions  by setting $(\mathrm{c_g},\mathrm{c_b})$ to $(0,0)$, $(-1,0)$, and $(-1/2, -1/2)$, respectively.
To avoid over notation, we will refer to the second term in Eq. \eqref{Bgop2} simply as the gauge term.

To minimize the energy $E_{\mathrm{ep}}$ \eqref{E1ep} subject to the orthonormal conditions $\langle\psi_p|\psi_q\rangle=\delta_{pq}$, we  introduce the following canonical
Lagrangian
\begin{align}
L=E_{\mathrm{ep}} -\sum_{k}n_k [\langle \psi_k|\psi_k\rangle-1]\epsilon_{k}, \quad k \in \mathrm{PES, NES}.\label{LagE}
\end{align}
The condition $\frac{\delta L}{\delta \psi_i^\dag}=0$ then gives rise to
\begin{align}
F n_i|\psi_i\rangle &= \epsilon_{i}n_i|\psi_i\rangle,\quad i\in \mathrm{PES, NES},\label{fpositron2}
\end{align}
where the Fock operator $F$ reads\cite{LiuMP}
\begin{align}
F&= h + \sum_k n_k\bar{g}_{\cdot k}^{\cdot k},\quad k \in \mathrm{PES, NES}\label{fpositron}\\
&=\begin{pmatrix}h^{LL}&h^{LS}\\ h^{SL}&h^{SS}\end{pmatrix}+\begin{pmatrix}G^{LL}&G^{LS}\\ G^{SL}&G^{SS}\end{pmatrix}=\begin{pmatrix}F^{LL}&F^{LS}\\ F^{SL}&F^{SS}\end{pmatrix},\label{FockOperator}\\
h^{LL}&=V,\quad h^{LS}=c\bsigma\cdot\boldsymbol{p},\quad h^{SL}=c\bsigma\cdot\boldsymbol{p},\quad h^{SS}=V-2c^2,\\
G^{LL}&=J^{C,LL}+J^{C,SS}-K^{C,LL}-\mathrm{c_g}K^{G,SS}-\mathrm{c_b}K^{g,SS}=G^{LL\dag},\label{GLLop}\\
G^{LS}&=-K^{C,LS}+\mathrm{c_g}(J^{G,LS}+J^{G,SL}-K^{G,SL})+\mathrm{c_b}(J^{g,LS}+J^{g,SL}-K^{g,SL})=G^{SL\dag},\label{GLSop}\\
G^{SL}&=-K^{C,SL}+\mathrm{c_g}(J^{G,LS}+J^{G,SL}-K^{G,LS})+\mathrm{c_b}(J^{g,LS}+J^{g,SL}-K^{g,LS})=G^{LS\dag},\label{GSLop}\\
G^{SS}&=J^{C,LL}+J^{C,SS}-K^{C,SS}-\mathrm{c_g}K^{G,LL}-\mathrm{c_b}K^{g,LL}=G^{SS\dag}.\label{GSSop}
\end{align}
As it stands, Eq. \eqref{fpositron2} determines only the occupied PESs and NESs but which can be extended to the unoccupied ones, viz.,
\begin{align}
F |\psi_p\rangle &= \epsilon_{p}|\psi_p\rangle,\quad p\in \mathrm{PES, NES}.\label{DHFeq}
\end{align}
The validity of the canonical Lagrangian \eqref{LagE} stems from the fact that the Fock operator $F$ \eqref{fpositron}
is Hermitian, such that $\langle\psi_p|\psi_q\rangle=0$ holds automatically for $p\ne q$.
The energetically lowest $N$ PESs and highest $\tilde{N}$ NESs are to be occupied in each iteration when solving Eq. \eqref{DHFeq} iteratively.
That is, the non-Aufbau, interior roots of Eq. \eqref{DHFeq} are to be sought in a self-consistent manner.
Note in passing that Eq. \eqref{E1ep} and hence Eq. \eqref{DHFeq} are approximations to
the mean-field QED theory\cite{PhysRep,Essential2020} of electrons and positrons by neglecting the vacuum polarization and electron self-energy.
Further setting the occupation numbers to zero for all NESs, Eq. \eqref{E1ep}/\eqref{DHFeq} reduces to the usual Dirac-Hartee-Fock (DHF) energy/stationarity
condition for $N$ electrons alone. In this case, the expressions \eqref{E1ep} and \eqref{DHFeq} can also be derived\cite{swirles1935relativistic}
by starting with the empty Dirac picture (where all NESs are assumed to be unoccupied),
instead of the filled Dirac picture underlying the QED formulation\cite{PhysRep,Essential2020}.
Yet, it should be kept in mind that the agreement between the empty and filled Dirac pictures holds only for one-electron properties, but not for
any two-body property (including electron correlation)\cite{PCCPNES,eQED}.

In practice, the molecular 4-spinors (M4S) $\{\psi_p\}$ have to be expanded in a suitably chosen basis. Among the various prescriptions\cite{IKB},
the so-called restricted kinetic balance (RKB)\cite{RKB} is most recommended due to its simplicity, viz.,
\begin{align}
|\psi_p\rangle&=\sum_{\mu=1}^{n}|\xi_{\mu}\rangle C_{\mu p}, \quad p\in[1,4n],\\
|\xi_{\mu}\rangle&=\begin{pmatrix}|\chi^L_\mu\rangle&0\\ 0&|\chi_{\mu}^S\rangle\end{pmatrix},\label{RKBbasis}\\
|\chi^L_\mu\rangle&=\begin{pmatrix}|g_{\mu}\rangle & 0 \\ 0&|g_{\mu}\rangle\end{pmatrix}=\sigma_0 |g_\mu\rangle,\quad \sigma_0=\mathbf{I}_2,\label{RKBlarge}\\
|\chi^S_\mu\rangle&=\frac{1}{2c}\Pi|\chi^L_\mu\rangle=\frac{1}{2c}\Pi|g_\mu\rangle=-|\chi^S_\mu\rangle^\dag,\quad \Pi=\bsigma\cdot\boldsymbol{p},\label{RKBsmall}\\
C_{\mu p}&=\begin{pmatrix}C_{\mu p}^L\\ C_{\mu p}^S\end{pmatrix},\quad C^X_{\mu p}=\begin{pmatrix}C^{X\alpha}_{\mu p}\\ C^{X\beta}_{\mu p}\end{pmatrix},
\quad X\in L, S,\label{Coeff}
\end{align}
where $\{g_{\mu}\}_{\mu=1}^{n}$ are prechosen $n$ scalar functions. Note that each of the four columns of $|\xi_{\mu}\rangle$ \eqref{RKBbasis}
corresponds to an independent four-component (4C) basis function, such that there are in total $4n$ basis functions and hence $4n$ M4Ss $\{|\psi_p\rangle\}$, half of which are PESs and the other half of which are NESs. In terms of such a basis expansion, Eq. \eqref{DHFeq}
is converted to an algebraic equation,
\begin{align}
\mathbf{F}\mathbf{C}&=\mathbf{M}\mathbf{C}\mathbf{E},\label{GHF}\\
F_{\mu\nu}&=h_{\mu\nu}+G_{\mu\nu}[\mathbf{D}],\quad M_{\mu\nu}=\langle \xi_{\mu}|\xi_{\nu}\rangle,\label{Fmat0}\\
h_{\mu\nu}&=\langle \xi_{\mu}|h|\xi_{\nu}\rangle,\label{h1e} \\
G_{\mu\nu}[\mathbf{D}]&=(J^C_{\mu\nu}[\mathbf{D}]-K^C_{\mu\nu}[\mathbf{D}])+\mathrm{c_g}(J^G_{\mu\nu}[\mathbf{D}]-K^G_{\mu\nu}[\mathbf{D}])
+\mathrm{c_b}(J^g_{\mu\nu}[\mathbf{D}]-K^g_{\mu\nu}[\mathbf{D}])\\
&=[(\Omega_{\mu\nu}|g_{12}|\rho^{4c})-(\Omega_{\mu\lambda}|g_{12}|\mathbf{D}_{\lambda\kappa}\Omega_{\kappa\nu})]\nonumber\\
&+\mathrm{c_g}[(\boldsymbol{\Xi}_{\mu\nu}|g_{12}\cdot|\boldsymbol{j}^{4c})-(\boldsymbol{\Xi}_{\mu\lambda}|g_{12}\cdot| \mathbf{D}_{\lambda\kappa}\boldsymbol{\Xi}_{\kappa\nu})]\nonumber\\
&+\mathrm{c_b}[(\boldsymbol{\Xi}_{\mu\nu}|\cdot\mathbf{b}_{12}\cdot|\boldsymbol{j}^{4c})
-(\boldsymbol{\Xi}_{\mu\lambda}|\cdot\mathbf{b}_{12}\cdot|\mathbf{D}_{\lambda\kappa}\boldsymbol{\Xi}_{\kappa\nu})],\label{Gmat0}\\
\mathbf{D}&=\mathbf{C}_+\mathbf{n}\mathbf{C}_+^\dag=\begin{pmatrix}\mathbf{C}_+^L\\
\mathbf{C}_+^S\end{pmatrix}\mathbf{n}\begin{pmatrix}\mathbf{C}_+^{L\dag}&\mathbf{C}_+^{S\dag}\end{pmatrix}\nonumber\\
&=\begin{pmatrix}\mathbf{D}^{LL}&
\mathbf{D}^{LS}\\ \mathbf{D}^{SL}&\mathbf{D}^{SS}\end{pmatrix},\quad
\mathbf{D}^{XY}=\mathbf{C}^{X}_+\mathbf{n}C^{Y\dag}_+,\quad X, Y= L, S,\label{AO4cDmat}
\end{align}
where $\mathbf{n}$ is the diagonal occupation number matrix,
whereas the overlap charge and current distribution functions are defined as\cite{ReSpectJCP2020}
\begin{align}
\Omega_{\mu\nu}&=\xi_{\mu}^\dag\xi_{\nu}
=\begin{pmatrix}\Omega_{\mu\nu}^{LL}&0\\
0&\Omega_{\mu\nu}^{SS}\end{pmatrix}=
\begin{pmatrix}\chi_{\mu}^{L\dag}\chi_{\nu}^L&0\\
0&\chi_{\mu}^{S\dag}\chi_{\nu}^S\end{pmatrix}\label{Omegadef}\\
&=\frac{1}{4c^2}\begin{pmatrix}4c^2\sigma_0 g_{\mu}g_{\nu}&0\\
0&\sigma_0g_{\mu}^ig_{\nu}^i+\ii\epsilon_{ijk}\sigma_i g_{\mu}^jg_{\nu}^k\end{pmatrix},\quad i,j,k\in x,y,z,\label{4COmegadef}\\
\Xi_{\mu\nu}^i&=\xi_{\mu}^\dag\alpha_i \xi_{\nu}
=\begin{pmatrix}0&\Xi_{\mu\nu}^{i,LS}\\
\Xi_{\mu\nu}^{i,SL}&0\end{pmatrix}
=\begin{pmatrix}0&\chi_{\mu}^{L\dag}\sigma_i\chi_{\nu}^S\\
\chi_{\mu}^{S\dag}\sigma_i\chi_{\nu}^L&0\end{pmatrix}
\label{Gammadef}\\
&=\frac{\ii}{2c}\begin{pmatrix}0&-\sigma_0 g_{\mu}g_{\nu}^i-\ii \epsilon_{ijk}g_{\mu}g_{\nu}^j\sigma_k\\
\sigma_0 g_{\mu}^ig_{\nu}-\ii\epsilon_{ijk}g_{\mu}^jg_{\nu}\sigma_k&0\end{pmatrix},\quad i,j,k\in x,y,z,\label{4CGammadef}
\end{align}
where $g_{\mu}^i=\partial_i g_{\mu}$ ($i\in x,y,z$). The number charge and current densities can then be calculated as
\begin{align}
\rho^{4C}&=n_i\psi_i^{\dag} \psi_i=n_i\psi_i^{L\dag} \psi_i^L+n_i\psi_i^{S\dag} \psi_i^S=\tr[\Omega^{LL}\mathbf{D}^{LL}]+\tr[\Omega^{SS}\mathbf{D}^{SS}]=
\tr[\Omega\mathbf{D}],\label{4Cdensity}\\
\boldsymbol{j}^{4C}&=n_i\psi_i^{\dag}\balpha\psi_i=n_i\psi_i^{L\dag}\bsigma\psi_i^S+n_i\psi_i^{S\dag}\bsigma\psi_i^L
=\tr[\boldsymbol{\Xi}^{LS}\mathbf{D}^{SL}]+\tr[\boldsymbol{\Xi}^{SL}\mathbf{D}^{LS}]
=\tr[\boldsymbol{\Xi}\mathbf{D}].\label{4Ccurrent}
\end{align}
As already indicated by Eqs. \eqref{Omegadef} and \eqref{Gammadef},
Eq. \eqref{GHF} can be recast into block form
\begin{align}
\begin{pmatrix}\mathbf{F}^{LL}&\mathbf{F}^{LS}\\ \mathbf{F}^{SL}&\mathbf{F}^{SS}\end{pmatrix}
\begin{pmatrix}\mathbf{C}^L_+&\mathbf{C}^L_-\\
\mathbf{C}^S_+&\mathbf{C}^S_-\end{pmatrix}&=\begin{pmatrix}\mathbf{M}^{LL}&\mathbf{0}\\ \mathbf{0}&\mathbf{M}^{SS}\end{pmatrix}
\begin{pmatrix}\mathbf{C}^L_+&\mathbf{C}^L_-\\
\mathbf{C}^S_+&\mathbf{C}^S_-\end{pmatrix}\begin{pmatrix}\mathbf{E}_+&\mathbf{0}\\ \mathbf{0}&\mathbf{E}_-\end{pmatrix},\label{DEQMat}\\
\begin{pmatrix}\mathbf{F}^{LL}&\mathbf{F}^{LS}\\ \mathbf{F}^{SL}&\mathbf{F}^{SS}\end{pmatrix}&=
\begin{pmatrix}\mathbf{h}^{LL}&\mathbf{h}^{LS}\\ \mathbf{h}^{SL}&\mathbf{h}^{SS}\end{pmatrix}+
\begin{pmatrix}\mathbf{G}^{LL}[\mathbf{D}]&\mathbf{G}^{LS}[\mathbf{D}]\\ \mathbf{G}^{SL}[\mathbf{D}]&\mathbf{G}^{SS}[\mathbf{D}]\end{pmatrix},\label{block-F}
\end{align}
where
\begin{align}
M^{LL}_{\mu\nu}&=\langle\chi^L_{\mu}|\chi^L_{\nu}\rangle=\sigma_0 S_{\mu\nu},\quad S_{\mu\nu}=\langle g_{\mu}|g_{\nu}\rangle,\\
M^{SS}_{\mu\nu}&=\langle\chi^S_{\mu}|\chi^S_{\nu}\rangle=\frac{1}{2c^2}\sigma_0 T_{\mu\nu},\quad T_{\mu\nu}=\frac{1}{2}\langle g_{\mu}|\boldsymbol{p}^2|g_{\nu}\rangle,\\
h^{LL}_{\mu\nu}&=\langle\chi_u^L|V|\chi_{\nu}^L\rangle=\sigma_0 V_{\mu\nu},\quad V_{\mu\nu}=\langle g_{\mu}|V|g_{\nu}\rangle,\\
h^{LS}_{\mu\nu}&=\langle\chi_u^L|c\Pi|\chi_{\nu}^S\rangle=\sigma_0 T_{\mu\nu}=h^{SL*}_{\nu\mu},\\
h^{SS}_{\mu\nu}&=\langle\chi_u^S|V-2c^2|\chi_{\nu}^S\rangle=\frac{1}{4c^2}W_{\mu\nu}-\sigma_0 T_{\mu\nu},\quad W_{\mu\nu}=\langle \Pi g_{\mu}|V|\Pi g_{\nu}\rangle,
\end{align}
\begin{align}
G^{LL}_{\mu\nu}[\mathbf{D}]&=(\mu^L\nu^L|g_{12}|i^Li^L)-(\mu^Li^L|g_{12}|i^L\nu^L)+(\mu^L\nu^L|g_{12}|i^Si^S)\nonumber\\
&-\mathrm{c_g}(\mu^L\bsigma i^S|g_{12}\cdot|i^S\bsigma \nu^L)
-\mathrm{c_b}(\mu^L\bsigma i^S|\cdot\mathbf{b}_{12}\cdot| i^S\bsigma \nu^L)\label{4CGLLorb}\\
&=(\Omega_{\mu\nu}^{LL}|g_{12}|\tr_2[\mathbf{D}^{LL}_{\lambda\kappa}\Omega^{LL}_{\kappa\lambda}])
-(\Omega_{\mu\lambda}^{LL}|g_{12}|\mathbf{D}^{LL}_{\lambda\kappa}\Omega^{LL}_{\kappa\nu})+(\Omega_{\mu\nu}^{LL}|g_{12}|\tr_2[\mathbf{D}^{SS}_{\lambda\kappa}\Omega^{SS}_{\kappa\lambda}])\nonumber\\
&-\mathrm{c_g}(\boldsymbol{\Xi}_{\mu\lambda}^{LS}|g_{12}\cdot|\mathbf{D}^{SS}_{\lambda\kappa}\boldsymbol{\Xi}_{\kappa\nu}^{SL})
-\mathrm{c_b}(\boldsymbol{\Xi}_{\mu\lambda}^{LS}|\cdot\mathbf{b}_{12}\cdot|\mathbf{D}^{SS}_{\lambda\kappa}\boldsymbol{\Xi}_{\kappa\nu}^{SL}),\label{4CGLL}\\
G^{LS}_{\mu\nu}[\mathbf{D}]&=-(\mu^L i^L|g_{12}|i^S\nu^S)\nonumber\\
&+\mathrm{c_g}[(\mu^L\bsigma \nu^S|g_{12}\cdot|i^L\bsigma  i^S)
-(\mu^L\bsigma i^S|g_{12}\cdot|i^L\bsigma \nu^S)+(\mu^L\bsigma \nu^S|g_{12}\cdot|i^S\bsigma  i^L)]\nonumber\\
&+\mathrm{c_b}[(\mu^L\bsigma \nu^S|\cdot\mathbf{b}_{12}\cdot|i^L\bsigma  i^S)
-(\mu^L\bsigma i^S|\cdot\mathbf{b}_{12}\cdot|i^L\bsigma \nu^S)
+(\mu^L\bsigma \nu^S|\cdot\mathbf{b}_{12}\cdot|i^S\bsigma  i^L)]\label{4CGLSorb}\\
&=-(\Omega^{LL}_{\mu\lambda}|g_{12}|\mathbf{D}^{LS}_{\lambda\kappa}\Omega^{SS}_{\kappa\nu})\nonumber\\
&+\mathrm{c_g}[(\boldsymbol{\Xi}_{\mu\nu}^{LS}|g_{12}\cdot|\tr_2[\mathbf{D}^{SL}_{\lambda\kappa}\boldsymbol{\Xi}^{LS}_{\kappa\lambda}])
-(\boldsymbol{\Xi}^{LS}_{\mu\lambda}|g_{12}\cdot| \mathbf{D}^{SL}_{\lambda\kappa}\boldsymbol{\Xi}^{LS}_{\kappa\nu})+(\boldsymbol{\Xi}^{LS}_{\mu\nu}|g_{12}\cdot|\tr_2[\mathbf{D}^{LS}_{\lambda\kappa}\boldsymbol{\Xi}^{SL}_{\kappa\lambda}])]\nonumber\\
&+\mathrm{c_b}[\boldsymbol{\Xi}_{\mu\nu}^{LS}|\cdot\mathbf{b}_{12}\cdot|\tr_2[\mathbf{D}^{SL}_{\lambda\kappa}\boldsymbol{\Xi}^{LS}_{\kappa\lambda}])
-(\boldsymbol{\Xi}^{LS}_{\mu\lambda}|\cdot\mathbf{b}_{12}\cdot| \mathbf{D}^{SL}_{\lambda\kappa}\boldsymbol{\Xi}^{LS}_{\kappa\nu})
+(\boldsymbol{\Xi}^{LS}_{\mu\nu}|\cdot\mathbf{b}_{12}\cdot|\tr_2[\mathbf{D}^{LS}_{\lambda\kappa}\boldsymbol{\Xi}^{SL}_{\kappa\lambda}])],\label{4CGLS}\\
G^{SL}_{\mu\nu}[\mathbf{D}]&=-(\mu^S i^S|g_{12}|i^L\nu^L)\nonumber\\
&+\mathrm{c_g}[(\mu^S\bsigma \nu^L|g_{12}\cdot|i^S\bsigma  i^L)
-(\mu^S\bsigma i^L|g_{12}\cdot|i^S\bsigma \nu^L)+(\mu^S\bsigma \nu^L|g_{12}\cdot|i^L\bsigma  i^S)]\nonumber\\
&+\mathrm{c_b}[(\mu^S\bsigma \nu^L|\cdot\mathbf{b}_{12}\cdot|i^S\bsigma  i^L)
-(\mu^S\bsigma i^L|\cdot\mathbf{b}_{12}\cdot|i^S\bsigma \nu^L)
+(\mu^S\bsigma \nu^L|\cdot\mathbf{b}_{12}\cdot|i^L\bsigma  i^S)]\label{4CGSLorb}\\
&=-(\Omega^{SS}_{\mu\lambda}|g_{12}|\mathbf{D}^{SL}_{\lambda\kappa}\Omega^{LL}_{\kappa\nu})\nonumber\\
&+\mathrm{c_g}[\boldsymbol{\Xi}_{\mu\nu}^{SL}|g_{12}\cdot|\tr_2[\mathbf{D}^{LS}_{\lambda\kappa}\boldsymbol{\Xi}^{SL}_{\kappa\lambda}])
-(\boldsymbol{\Xi}^{SL}_{\mu\lambda}|g_{12}\cdot| \mathbf{D}^{LS}_{\lambda\kappa}\boldsymbol{\Xi}^{SL}_{\kappa\nu})+(\boldsymbol{\Xi}^{SL}_{\mu\nu}|g_{12}\cdot|\tr_2[\mathbf{D}^{SL}_{\lambda\kappa}\boldsymbol{\Xi}^{LS}_{\kappa\lambda}])]\nonumber\\
&+\mathrm{c_b}[(\boldsymbol{\Xi}_{\mu\nu}^{SL}|\cdot\mathbf{b}_{12}\cdot|\tr_2[\mathbf{D}^{LS}_{\lambda\kappa}\boldsymbol{\Xi}^{SL}_{\kappa\lambda}])
-(\boldsymbol{\Xi}^{SL}_{\mu\lambda}|\cdot\mathbf{b}_{12}\cdot|\mathbf{D}^{LS}_{\lambda\kappa}\boldsymbol{\Xi}^{SL}_{\kappa\nu})
+(\boldsymbol{\Xi}^{SL}_{\mu\nu}|\cdot\mathbf{b}_{12}\cdot|\tr_2[\mathbf{D}^{SL}_{\lambda\kappa}\boldsymbol{\Xi}^{LS}_{\kappa\lambda}])],\label{4CGSL}\\
G^{SS}_{\mu\nu}[\mathbf{D}]&=(\mu^S\nu^S|g_{12}|i^Si^S)-(\mu^Si^S|g_{12}|i^S\nu^S)+(\mu^S\nu^S|g_{12}|i^Li^L)\nonumber\\
&-\mathrm{c_g}(\mu^S\bsigma i^L|g_{12}\cdot|i^L\bsigma \nu^S)
-\mathrm{c_b}(\mu^S\bsigma i^L|\cdot\mathbf{b}_{12}\cdot| i^L\bsigma \nu^S)\label{4CGSSorb}\\
&=(\Omega_{\mu\nu}^{SS}|g_{12}|\tr_2[\mathbf{D}^{SS}_{\lambda\kappa}\Omega^{SS}_{\kappa\lambda}])
-(\Omega_{\mu\lambda}^{SS}|g_{12}|\mathbf{D}^{SS}_{\lambda\kappa}\Omega^{SS}_{\kappa\nu})
+(\Omega_{\mu\nu}^{SS}|g_{12}|\tr_2[\mathbf{D}^{LL}_{\lambda\kappa}\Omega^{LL}_{\kappa\lambda}])\nonumber\\
&-\mathrm{c_g}(\boldsymbol{\Xi}_{\mu\lambda}^{SL}|g_{12}\cdot|\mathbf{D}^{LL}_{\lambda\kappa}\boldsymbol{\Xi}_{\kappa\nu}^{LS})
-\mathrm{c_b}(\boldsymbol{\Xi}_{\mu\lambda}^{SL}|\cdot\mathbf{b}_{12}\cdot|\mathbf{D}^{LL}_{\lambda\kappa}\boldsymbol{\Xi}_{\kappa\nu}^{LS}).\label{4CGSS}
\end{align}
In the above notation, $\mathbf{D}^{XY}$, $\Omega^{XY}$, and $\boldsymbol{\Xi}^{XY}$ ($X,Y=L,S$) are all 2-by-2 block matrices,
each block of which is a $n$-by-$n$ matrix.
Alternatively, every element $\mathbf{D}^{XY}_{\mu\nu}$, $\Omega^{XY}_{\mu\nu}$ or $\boldsymbol{\Xi}^{XY}_{\mu\nu}$ can be understood as a 2-by-2 matrix over
spin labels. Therefore, the symbol $\mathrm{Tr}_2$ means trace over the spin degrees of freedom, e.g.,
\begin{align}
\tr_2[\mathbf{D}^{XY}_{\lambda\kappa}\Omega^{YX}_{\kappa\lambda}]&=\tr_2[\Omega^{YX}_{\kappa\lambda}\mathbf{D}^{XY}_{\lambda\kappa}],\quad \lambda,\kappa\in[1,n] \\
&=D^{X\alpha Y\alpha}_{\lambda\kappa}\Omega^{Y\alpha X\alpha}_{\kappa\lambda}+D^{X\alpha Y\beta}_{\lambda\kappa}\Omega^{Y\beta X\alpha}_{\kappa\lambda}
+D^{X\beta Y\alpha}_{\lambda\kappa}\Omega^{Y\alpha X\beta}_{\kappa\lambda}+D^{X\beta Y\beta}_{\lambda\kappa}\Omega^{Y\beta X\beta}_{\kappa\lambda}.\label{Tr2def}
\end{align}
Note in particular that $\mathbf{D}^{XY}_{\lambda\kappa}$ and $\Omega^{YZ}_{\kappa\lambda}/\boldsymbol{\Xi}^{XY}_{\kappa\lambda}$
do not commute in block form (i.e., $\lambda,\kappa\in[1,n]$). Nevertheless, they do commute with each other when fully expanded
(i.e., $\lambda,\kappa\in[1,2n]$). Given the density matrix $\mathbf{D}$ \eqref{AO4cDmat},
the energy $E_{\mathrm{ep}}$ \eqref{E1ep} can be calculated as
\begin{align}
E_{\mathrm{ep}}&=\frac{1}{2}\tr (\mathbf{h}+\mathbf{F})\mathbf{D}.
\end{align}

Some remarks are in order:
\begin{enumerate}[(I)]
\item The RKB prescription \eqref{RKBsmall} for generating $2n$ small-component basis functions $\{\chi_{\mu}^S\}$ directly
from the $n$-scalar functions $\{g_{\mu}\}_{\mu=1}^n$ guarantees the correct nonrelativistic limit (nrl) of the PESs but the NESs
are still in error of $\mathcal{O}(c^0)$\cite{IKB}. As such, it does not provide full variational safety.
Depending very much on the construction of $\{g_{\mu}\}_{\mu=1}^n$, some bounds failures (or prolapse\cite{Prolapse}) of $O(c^{-4})$ may occur.
Nevertheless, such bounds failures will diminish when approaching to the basis set limit, at a rate that is
not much different from the nonrelativistic counterpart\cite{kutzelnigg2007completeness}. Another point that deserves to be mentioned is that, when $\{g_{\mu}\}_{\mu=1}^n$ are spherical Gaussians, the principal quantum number
must be set to the angular momentum $l$ plus one (i.e., $1s$, $2p$, $3d$, $4f$, $5g$, etc.). Otherwise,
terrible variational collapse would occur\cite{IKB}. Although this is usually the default option,
it is not mandatory in the nonrelativistic case.
\item\label{ERItruncation} It is the appearance of small-component basis functions that
renders the DHF calculation very expensive. As shown in Appendix \ref{AppendixA}, there are
in total 25 (325) real-valued scalar integrals to evaluate and process just for a single term of
the two-electron matrix element $G_{\mu\nu}[\mathbf{D}]$ under the Coulomb (Gaunt/Breit) interaction.
However, the situation is not really that bad.
Since the small component $\psi_p^S(\boldsymbol{r})$ ($=\sum_A\sum_{\mu_A}\chi_{\mu_A}^S(\boldsymbol{r}-\boldsymbol{R}_A)C_{\mu_A p}^S$)
of a M4S $\psi_p(\boldsymbol{r})$ is appreciable only nearby the positions of the nuclei\cite{Visscher1997SSSS},
the products $C_{\mu_A p}^{S\dag}\chi_{\mu_A}^{S\dag}(\boldsymbol{r}-\boldsymbol{R}_A)\chi_{\nu_B}^S(\boldsymbol{r}-\boldsymbol{R}_B)C_{\nu_B q}^S$
in $\psi_p^{S\dag}(\boldsymbol{r})\psi_q^S(\boldsymbol{r})$ are negligibly small
when $A$ and $B$ refer to different atoms in the molecule (NB: $\chi_{\mu_A}^S$ denotes a small-component function $\chi_{\mu}^S$ centered at
the position $\boldsymbol{R}_A$ of atom $A$).
An immediate deduction is that
the small-component distribution functions $\Omega^{SS}_{\mu\nu}$ (of $\mathcal{O}(c^{-2})$) and density matrix elements $\mathbf{D}^{SS}_{\nu\mu}$
can be confined to one-centered only, i.e.,
\begin{align}
\Omega^{SS}_{\mu_A\nu_B}(\boldsymbol{r})\approx\Omega^{SS}_{\mu_A\nu_B}(\boldsymbol{r})\delta_{AB},\quad
\mathbf{D}^{SS}\approx\sum^\oplus_A\mathbf{D}^{SS}_A,\quad \mathbf{D}^{SS}_A=\mathbf{C}_{A,+}^S\mathbf{n}_A\mathbf{C}_{A,+}^{S\dag},
\end{align}
thereby leading to
\begin{align}
\rho^{S}=\tr[\Omega^{SS}\mathbf{D}^{SS}]\approx\tilde{\rho}^S=\sum_A \rho^{S}_A\label{MolDenS}
\end{align}
for the molecular small-component density.
Such approximations imply the following relations\cite{Visscher2002SS}
\begin{align}
(L_AL_B|S_CS_D)&\approx (L_AL_B|S_CS_D)\delta_{CD},\nonumber\\
(S_AS_B|L_CL_D)&\approx (S_AS_B|L_CL_D)\delta_{AB},\nonumber\\
(S_AS_B|S_CS_D)&\approx(S_AS_B|S_CS_D)\delta_{AB}\delta_{CD}.\label{SSERI}
\end{align}
for the electron repulsion integrals (ERI).
The $(S_AS_A|S_BS_B)$ type of ERIs, which are of $\mathcal{O}(c^{-4})$,
can further be confined to one-centered only, i.e.,
\begin{align}
(S_AS_A|S_BS_B)\approx (S_AS_A|S_BS_B)\delta_{AB}. \label{SSSS1c}
\end{align}
The situation is even better when atomic 4-spinors
(A4S; from DHF calculations of averaged and nonpolarized atomic configurations)
are taken as the basis\cite{RosenAtomicP,wood1986relativistic,LiuPhD,LiuBDF1996,BDF1,LCA4S}, for their small
components do not overlap each other discernibly when they come from different atoms.
Similarly, the Gaunt/Breit integrals $(LS|SL)$, $(SL|LS)$, $(LS|LS)$, and $(SL|SL)$ can be confined to at most two-centered,
\begin{align}
(X_AY_B|Y_CX_D)&\approx (X_AY_B|Y_CX_D)\delta_{AB}\delta_{CD},\quad X, Y = L, S,\quad X\ne Y,\nonumber\\
(X_AY_B|X_CY_D)&\approx (X_AY_B|X_CY_D)\delta_{AB}\delta_{CD},\quad X, Y = L, S,\quad X\ne Y.\label{LSLS1c}
\end{align}
Eqs. \eqref{SSERI} to \eqref{LSLS1c} can be termed collectively `one-center small-component' (1CSC) approximation
(i.e., a small-component function is always regarded as extremely localized nearby the position of a nucleus,
such that it has vanishing overlaps with all functions centered at other nuclei).
Since such ERIs involve at most three centers, the use of local symmetries (planar, cylindrical, and spherical
symmetries for three-, two-, and one-center integrals) would be very beneficial.
Moreover, since only the spin-free part of the whole gauge term $V^g$ in Eq. \eqref{Bgop2}
contributes to the energy at $\mathcal{O}(c^{-2})$ (which is even simpler than that of $V^G$)\cite{MC-DPT1},
It can be anticipated that only the spin-free part of $V^g$ needs to be included in practical applications\cite{LiXSscalarBreitInt}.
The surving ERIs can further be evaluated efficiently with the RKB-based resolution of the identity (RKB-RI)\cite{RKB-RI-ERI}.
\item Only the $N_o$ ($=N+\tilde{N}\ll 4n$) occupied states need to be determined when solving Eq. \eqref{GHF}/\eqref{DEQMat} iteratively.
Such states appear as a narrow, interior portion of the whole orbital spectrum. Therefore, it is of great value
to invoke an algorithm that can directly access such interior roots. The recently proposed iterative vector interaction (iVI)\cite{iVI,iVI-TDDFT} is one
of such algorithms. It is not only very robust but also memory efficient, by working with a search space of fixed dimension
that is determined automatically
by the number of target states or a preset energy window. A speedup factor of $(4n)^3/[N_o(4n)^2]=4n/N_o\gg 1$
can be gained as compared with the full matrix diagonalization.
\end{enumerate}

The above presentation provides the basics for an efficient implementation of DHF. Additional
gain in efficiency stems from the full use of double point group and time reversal symmetries,
which can be achieved in two distinct ways, by constructing
Kramers-paired double group symmetry functions according to fermion irreducible representations (irrep)\cite{meyer1988construction,visscher1996construction,matveev2004efficient,Symm2009,peng2011symmetrized} or by
combining quaternion algebra (which incorporates time reversal symmetry\cite{RoschQuaternion1983,SaueMolPhys1997}) with corresponding
boson irreps for the real and imaginary parts of each component of a quaternion spinor\cite{SaueQuaternion1999,armbruster2017quaternionic}.
The former is suited for a $j$-adapted spinor basis, whereas the latter
is suited for a scalar basis. The quaternion form of the DHF equation \eqref{GHF} underlying the latter
was formulated\cite{SaueQuaternion1999} only for closed-shell systems under the Coulomb interaction.
As a matter of fact, it can be formulated more generally (see Appendix \ref{AppendixA})
via a quaternion unitary transformation\cite{Biquaternions} that can block-diagonalize any matrix
(e.g., open-shell Fock matrix with the Gaunt/Breit interaction) with identical diagonal blocks.
Based on such a general quaternion DHF, a four-component relativistic Kramers restricted open-shell DHF (KROHF) scheme for `high-spin' open-shell systems
can be formulated (see Appendix \ref{AppendixA}), in the same way as the two-component counterpart\cite{Nakai-KROHF}.

\section{Q4C}\label{SecQ4C}
The DHF equation \eqref{GHF}/\eqref{DEQMat} generates both PESs and NESs.
The latter are usually discarded at a correlated level under the no-pair approximation (NPA).
The question is how to avoid the molecular NESs from the outset if the NPA is
doomed to make. Actually, this can be done in two different ways\cite{Q4CX2C}: freeze or remove the NESs.
The former stays in the four-component framework but can be made operationally the same as a two-component theory,
whereas the latter works with a two-component, electron-only Hamiltonian. In essence,
the two paradigms stem from the same physical origin and can hence be made and have been made completely equivalent in terms of
simplicity, accuracy, and efficiency\cite{Q4CX2C}.

To realize the first paradigm, we first examine the $S/L$ ratio between the small and large components of a PES $\psi_i$,
which can be obtained from the second row of Eq. \eqref{fpositron2},
\begin{align}
\psi_i^S&=(\epsilon_i-F^{SS})^{-1}F^{SL}\psi_i^L\label{S2La}\\
&\triangleq(\epsilon_i-V^{SS}_{\mathrm{eff}}+2c^2)^{-1}(c\bsigma \cdot\boldsymbol{p}+V^{SL}_{\mathrm{eff}})\psi_i^L\label{S2Lb} \\
&\approx\frac{1}{2c}R_i\bsigma \cdot\boldsymbol{p}\psi_i^L,\label{ExactKB}\\
R_i(\boldsymbol{r})&=[1+\frac{1}{2c^2}(\epsilon_i-V^{SS}_{\mathrm{eff}}(\boldsymbol{r}))]^{-1}
 \stackrel{c\rightarrow\infty}\rightarrow 1,\label{Rop}
\end{align}
where the potential terms in $G^{SL}$ \eqref{GSLop} and $G^{SS}$ \eqref{GSSop} have been grouped into $V^{SL}_{\mathrm{eff}}$
and $V^{SS}_{\mathrm{eff}}$, respectively, when going from Eq. \eqref{S2La} to Eq. \eqref{S2Lb}. The former
is further neglected when going from  Eq. \eqref{S2Lb} to Eq. \eqref{ExactKB}, for it is of $\mathcal{O}(c^{-1})$.
The major effect of $\bsigma \cdot\boldsymbol{p}$ is to change the parity of the large component to that of the small component.
So the $S/L$ ratio is determined mainly by the $R_i(\boldsymbol{r})$ operator \eqref{Rop}, which is
extremely short ranged\cite{SaueChapter}: Each $R_i(\boldsymbol{r})$ becomes
a constant factor just slightly away from the position of a heavy atom (e.g., $0.05$ a.u.
(roughly the radii of $2s$ and $2p$ shells) in the case of Rn; cf. Fig. 1 in Ref. \citenum{Q4CX2C}). Imagine we have first solved
the (radial) Dirac equation for each isolated
(spherical and unpolarized) atom $A$ and thus obtained the
corresponding A4Ss $\{\varphi_{Ap}\}$. Then, the atoms are
brought together to synthesize the molecule. While both the large and small
components of $\varphi_{Ap}$ will change, the $S/L$ ratio will \emph{not}!\cite{Q4CX2C,LCA4S}.
The mathematical realization\cite{RosenAtomicP,wood1986relativistic,LiuPhD,LiuBDF1996,BDF1,LCA4S} of such a physical picture is to expand
the positive-energy M4Ss $\psi_i$ in terms only of the \emph{positive-energy} A4Ss $\{\varphi_{Ap}\}$, viz.,
\begin{align}
|\psi_i\rangle&=\sum_{A}^{N_A}\sum_{p\in A} |\varphi_{Ap}\rangle \bar{C}_{+,Ap, i}
=\sum_{A}^{N_A}\sum_{p\in A}\begin{pmatrix}|\varphi_{Ap}^L\rangle\\ |\varphi_{Ap}^S\rangle\end{pmatrix}\bar{C}_{+, Ap, i}\label{LCA4Sa}\\
&=\sum_{A}^{N_A}\sum_{p\in A}\sum_{\mu\in A}\begin{pmatrix}|\chi^L_{A\mu}\rangle a_{A\mu, Ap}\\ |\chi^S_{A\mu}\rangle b_{A\mu, Ap}\end{pmatrix}\bar{C}_{+, Ap, i}
=\sum_{A}^{N_A}\sum_{\mu\in A}\begin{pmatrix}|\chi^L_{A\mu}\rangle \bar{A}_{A\mu, i}\\ |\chi^S_{A\mu}\rangle \bar{B}_{A\mu, i}\end{pmatrix},\label{LCA4S}
\end{align}
which gives rise to the following projected four-component (P4C) approach
\begin{align}
\mathbf{F}_+^{\mathrm{P4C}}\bar{\mathbf{C}}_+&=\bar{\mathbf{M}}_+\bar{\mathbf{C}}_+\bar{\mathbf{E}}_+,\label{P4Ceq}\\
\mathbf{F}_+^{\mathrm{P4C}}&=\bar{\mathbf{h}}^{\mathrm{P4C}}+\bar{\mathbf{G}}^{\mathrm{P4C}}[\bar{\mathbf{P}}],\label{FP4C}\\
\bar{\mathbf{h}}^{\mathrm{P4C}}&=\sum_{X,Y=L,S}\langle[\varphi^X]|h^{XY}|[\varphi^Y]\rangle=\sum_{X,Y=L,S}\bar{\mathbf{h}}^{XY}\label{P4C1eA4S}\\
&=\mathbf{a}^\dag\mathbf{h}^{LL}\mathbf{a}+\mathbf{a}^\dag\mathbf{h}^{LS}\mathbf{b}+\mathbf{b}^\dag\mathbf{h}^{SL}\mathbf{a}
+\mathbf{b}^\dag\mathbf{h}^{SS}\mathbf{b},\label{P4C1e}\\
\bar{\mathbf{G}}^{\mathrm{P4C}}[\bar{\mathbf{P}}]&=\sum_{X,Y=L,S}\langle[\varphi^X]|G^{XY}|[\varphi^Y]\rangle=\sum_{X,Y=L,S}\bar{\mathbf{G}}^{XY}[\bar{\mathbf{D}}]\label{P4C2eA4S}\\
&=\mathbf{a}^\dag \mathbf{G}^{LL}[\bar{\mathbf{D}}]\mathbf{a}
+\mathbf{a}^\dag \mathbf{G}^{LS}[\bar{\mathbf{D}}]\mathbf{b}
+\mathbf{b}^\dag \mathbf{G}^{SL}[\bar{\mathbf{D}}] \mathbf{a} +
\mathbf{b}^\dag \mathbf{G}^{SS}[\bar{\mathbf{D}}]\mathbf{b},\label{P4C2e}\\
\mathbf{a}&=\sum_A^{\oplus} \mathbf{a}_A,\quad  \mathbf{b}=\sum_A^{\oplus} \mathbf{b}_A, \\
\bar{\mathbf{D}}&=\begin{pmatrix}\bar{\mathbf{D}}^{LL}&\bar{\mathbf{D}}^{LS}\\
\bar{\mathbf{D}}^{SL}&\bar{\mathbf{D}}^{SS}\end{pmatrix}=\begin{pmatrix}\mathbf{a}\bar{\mathbf{P}}\mathbf{a}^\dag&\mathbf{a}\bar{\mathbf{P}}\mathbf{b}^\dag\\
\mathbf{b}\bar{\mathbf{P}}\mathbf{a}^\dag&\mathbf{b}\bar{\mathbf{P}}\mathbf{b}^\dag\end{pmatrix},\quad
 \bar{\mathbf{P}}=\bar{\mathbf{C}}_+\mathbf{n}\bar{\mathbf{C}}_+^\dag,\label{P4Cden}\\
\bar{\mathbf{M}}_+&=\bar{\mathbf{M}}^{LL} +\bar{\mathbf{M}}^{SS}=\mathbf{a}^\dag\mathbf{M}^{LL}\mathbf{a}+\mathbf{b}^\dag\mathbf{M}^{SS}\mathbf{b}.\label{P4Cmetric}
\end{align}
The dimension of $\mathbf{F}_+^{\mathrm{P4C}}$ \eqref{FP4C} in the $j$-adapted spinor basis $\{\varphi_p\}_{p=1}^{2n}$ is $2n$ instead of $4n$. That is, molecular NESs are excluded completely.
What is neglected here is rotations between the PESs of an atom and the NESs of the other atoms
in the molecule, a kind of polarization of the atomic vacua. Being of $\mathcal{O}(c^{-4})$,
such an approximation introduces no discernible errors to molecular spectra
\cite{BDF1,Q4C,Q4CX2C}.
Yet, it can be envisaged that
P4C will break down when the interatomic distance
between two heavy atoms is very short (which may occur in highly charged molecular systems).
In this case, the A4Ss may be replaced with diatomic 4-spinors (DA4S) obtained by
diagonalizing
\begin{align}
\mathbf{F}^{\mathrm{Frag}}_{\mu\nu}=h^{\mathrm{Frag}}_{\mu\nu}+G^{\mathrm{Frag}}_{\mu\nu}[\sum_{A\in F}^\oplus \mathbf{D}_A],\quad \mu,\nu\in F,\label{FragF}
\end{align}
for every pair $F$ of atoms. Here, $\mathbf{D}_{A}$ is the four-component density matrix of atom $A$.
In essence, it is the interatomic interaction strength that
is taken here as a perturbation parameter to expand the projector for
the molecular no-pair relativistic Hamiltonian\cite{Q4CX2C,ProgChem}.

The structures of $\bar{\mathbf{h}}^{\mathrm{P4C}}$ \eqref{P4C1e}, $\bar{\mathbf{G}}^{\mathrm{P4C}}$ \eqref{P4C2e}, and $\bar{\mathbf{M}}_+$ \eqref{P4Cmetric}
are reminiscent\cite{Essential2020} of the NESC (normalized elimination of the small component) approach\cite{NESC}, which becomes
more transparent by introducing the following formal relations
\begin{align}
\mathbf{b}_A&=\mathbf{X}_A\mathbf{a}_A,\quad \mathbf{b}=\mathbf{X}\mathbf{a},\quad
\mathbf{X}=\sum_A^{\oplus} \mathbf{X}_A,\label{ATOMX}
\end{align}
such that
\begin{align}
\bar{\mathbf{h}}^{\mathrm{P4C}}&=\mathbf{a}^\dag \mathbf{L}_+^{\mathrm{NESC,1e}}\mathbf{a},\label{P4CNESC1e}\\
\bar{\mathbf{G}}^{\mathrm{P4C}}[\bar{\mathbf{P}}]&=\mathbf{a}^\dag \mathbf{L}_+^{\mathrm{NESC,2e}}[\bar{\mathbf{D}}]\mathbf{a},\label{P4CNESC2e}\\
\mathbf{L}_+^{\mathrm{NESC,1e}}&=\mathbf{V}+\mathbf{T}\mathbf{X}+\mathbf{X}^\dag\mathbf{T}
+\frac{1}{4c^2}\mathbf{X}^\dag\mathbf{W}\mathbf{X}-\mathbf{X}^\dag\mathbf{T}\mathbf{X},\label{NESC1e}\\
\mathbf{L}^{\mathrm{NESC,2e}}_+[\mathbf{Z}]&=\mathbf{G}^{LL}[\mathbf{Z}]+\mathbf{G}^{LS}[\mathbf{Z}]\mathbf{X}+\mathbf{X}^\dag\mathbf{G}^{SL}[\mathbf{Z}]
+\mathbf{X}^\dag\mathbf{G}^{SS}[\mathbf{Z}]\mathbf{X}\label{NESC2e}\\
&\triangleq\mathbf{G}^{\mathrm{NR}}[\mathbf{Z}^{LL}]+\mathbf{G}^{\mathrm{NESC}}[\mathbf{Z}],\label{NESC2eNR}\\
G^{\mathrm{NR}}_{\mu\nu}[\mathbf{Z}]&=(\Omega^{LL}_{\mu\nu}|g_{12}|\tr_2[\mathbf{Z}_{\lambda\kappa}\Omega^{LL}_{\kappa\lambda}])
-(\Omega^{LL}_{\mu\lambda}|g_{12}|\mathbf{Z}_{\lambda\kappa}\Omega^{LL}_{\kappa\nu}),\label{GNR}\\
\bar{\mathbf{M}}_+&=\mathbf{a}^\dag\tilde{\mathbf{M}}_+\mathbf{a},\quad
\tilde{\mathbf{M}}_+=\mathbf{M}^{LL}+\mathbf{X}^\dag\mathbf{M}^{SS}\mathbf{X}.\label{RelMetric}
\end{align}
It looks like that P4C is just NESC working with the A4S basis and meanwhile the atomic approximation (AtomX) \eqref{ATOMX}
to the molecular decoupling (transfer) matrix $\mathbf{X}$,
\begin{align}
\mathbf{C}_+^S=\mathbf{X}\mathbf{C}_+^L.\label{L2SX}
\end{align}
However, P4C and NESC were introduced in completely different ways. The former
is by construction a direct four-component approximation to the DHF equation \eqref{GHF}, whereas the latter is a
two-component theory for the large components of the PESs, viz.,
\begin{align}
\mathbf{L}^{\mathrm{NESC}}_+\mathbf{C}^L_+&=\tilde{\mathbf{M}}_+\mathbf{C}^L_+\mathbf{E}_+,\quad
\mathbf{L}^{\mathrm{NESC}}_+=\mathbf{L}^{\mathrm{NESC,1e}}_+ +\mathbf{L}^{\mathrm{NESC,2e}}_+.\label{NESCeigen}
\end{align}
The AtomX \eqref{ATOMX} is not mandatory but can be introduced\cite{LCA4S} to $\mathbf{L}^{\mathrm{NESC}}_+$ later on. On the practical side,
it is not mandatory for NESC to use a generally contracted basis, even with the AtomX
(e.g., $\mathbf{a}_A$ and $\mathbf{b}_A$ can be set to unit matrices for an uncontracted RKB basis).
On the other hand, P4C need not be limited to a RKB basis. Rather, it can also adopt, e.g., numerical A4Ss
by solving the atomic radial DHF equations with grids (cf. the second equality of Eq. \eqref{LCA4Sa} and Eqs. \eqref{P4C1eA4S}, and \eqref{P4C2eA4S}).
A more important distinction between P4C and NESC lies in that the eigenvectors of the P4C eigenequation \eqref{P4Ceq} give
the total density represented in the A4S basis $\{\varphi_{Ap}\}$, whereas those of the NESC eigenequation \eqref{NESCeigen} give only the large-component
density represented in the basis $\{\chi^L_{\mu}\}$.

At this stage, the computational efficiency is gained only
in the matrix diagonalization step (by a factor of 8 in the case of
full diagonalization or by a factor of 4 in the case of partial diagonalization by iVI\cite{iVI,iVI-TDDFT}), which is very little
for a moderate basis. The real gain in efficiency
can only be achieved by further invoking a `model small component approximation' (MSCA), which was proposed\cite{Q4C} originally in the context
of density functional theory (DFT). Therein,
the molecular small-component density $\rho^{S}$ is first approximated as the superposition of
the atomic ones (cf. Eq. \eqref{MolDenS}),
\begin{align}
\rho^{S}=\tr[\Omega^{SS}\bar{\mathbf{P}}]\approx\tilde{\rho}^S=\sum_A \tilde{\rho}^S_A=\tr[\Omega^{SS}\bar{\mathbf{P}}^0],\quad \Omega^{SS}_{pq}=\varphi_p^{S\dag}\varphi_q^S,
 \quad \bar{\mathbf{P}}^0=\sum_A^{\oplus}\mathbf{n}_A,\label{P0Mat}
\end{align}
such that the molecular density $\rho^{4c}$ can be calculated as
\begin{align}
\rho^{4c}&\approx\tilde{\rho}^{4c}=\rho^{L}+\tilde{\rho}^S, \quad
\rho^{L}=\tr[\Omega^{LL}\bar{\mathbf{P}}],\quad \Omega^{LL}_{pq}=\varphi_p^{L\dag}\varphi_q^L.\label{PfullMat}
\end{align}
The matrix elements of the local Kohn-Sham (KS) potential  $V_{\mathrm{eff}}[\rho^{4c}]$
are then approximated as
\begin{align}
\langle\varphi_p^L|V_{\mathrm{eff}}[\rho^{4c}]|\varphi_q^L\rangle+\langle\varphi_p^S|V_{\mathrm{eff}}[\rho^{4c}]|\varphi_q^S\rangle
&\approx\langle\varphi_p^L|V_{\mathrm{eff}}[\tilde{\rho}^{4c}]|\varphi_q^L\rangle+\langle\varphi_p^S|V_{\mathrm{eff}}[\rho^{4c}_{\mathrm{mod}}]|\varphi_q^S\rangle,\label{ModPot}\\
\rho^{4c}_{\mathrm{mod}}&=\sum_A\rho^{4c}_A=
\tr[(\Omega^{LL}+\Omega^{SS})\bar{\mathbf{P}}^0],\label{ModDen}
\end{align}
such that, under the LCA4S and MSCA, only the first iteration of the calculation is four-component but subsequent iterations are just two-component
(i.e., only $\rho^L$ need to be updated), thereby justifying the name quasi-four-component (Q4C)\cite{Q4C}.
Even the first iteration is much cheaper than a regular four-component iteration, for the sparsity associated with $\bar{\mathbf{P}}^0$ renders
the evaluation of the two-electron matrix elements only at a fractional cost of a regular evaluation of the two-electron matrix elements.
It has been shown\cite{Q4C,X2CBook2017} that Q4C-KS is indeed very accurate, e.g., with
errors being only a few milli-Hartrees for the energy levels of the innermost shells of E117$_2$
and completely negligible for molecular spectroscopic constants.
The MSCA\cite{Q4CX2C}, which can also be termed `model density approximation',
can trivially be generalized to a model density matrix (MDM) approximation to incorporate HF exchange.
That is, the molecular density matrix elements $\bar{\mathbf{D}}_{\lambda\kappa}$ (cf. Eq. \eqref{P4Cden}) in
all small-component-containing terms of $\mathbf{L}^{\mathrm{NESC,2e}}_+[\bar{\mathbf{D}}]$
(cf. Eq. \eqref{NESC2eNR} and $\mathbf{G}^{XY}[\bar{\mathbf{D}}]$ in Eqs. \eqref{4CGLL}, \eqref{4CGLS}, \eqref{4CGSL}, and \eqref{4CGSS}) are replaced with those of the model density matrix $\bar{\mathbf{D}}^0=\sum_A^\oplus\bar{\mathbf{D}}_A$. This amounts to replacing $\mathbf{G}^{\mathrm{NESC}}[\bar{\mathbf{D}}]$
in Eq. \eqref{NESC2eNR} with $\mathbf{G}^{\mathrm{NESC}}[\bar{\mathbf{D}}^0]$, thereby giving rise to
the Q4C-HF equation\cite{LiuMP,X2CBook2017} represented in the RKB basis employed to expand the A4Ss $\{\varphi_{Ap}\}$,
\begin{align}
{\mathbf F}_+^{\mathrm{Q4C}}\bar{\mathbf{C}}_+&=\bar{\mathbf{M}}_+\bar{\mathbf{C}}_+\bar{\mathbf{E}}_+,\\
{\mathbf F}_+^{\mathrm{Q4C}}&=\bar{\mathbf{h}}^{\mathrm{Q4C}}_{\mathrm{eff}}+\bar{\mathbf{G}}^{\mathrm{NR}}[\bar{\mathbf{P}}],\label{effQ4C}\\
\bar{\mathbf{h}}^{\mathrm{Q4C}}_{\mathrm{eff}}&=\mathbf{a}^\dag\mathbf{L}_+^{\mathrm{NESC,1e}}\mathbf{a}+\bar{\mathbf{G}}^{\mathrm{Q4C}}_{\mathrm{eff}},\quad
\bar{\mathbf{G}}^{\mathrm{Q4C}}_{\mathrm{eff}}=\mathbf{a}^\dag\mathbf{G}^{\mathrm{NESC}}[\bar{\mathbf{D}}^0]\mathbf{a},\label{effQ4C1e}\\
\bar{\mathbf{G}}^{\mathrm{NR}}[\bar{\mathbf{P}}]&=\mathbf{a}^\dag\mathbf{G}^{\mathrm{NR}}[\bar{\mathbf{D}}^{LL}]\mathbf{a},\quad \bar{\mathbf{D}}^{LL}=\mathbf{a}^\dag\bar{\mathbf{P}}\mathbf{a},\label{effQ4CGNR}
\end{align}
where $\bar{\mathbf{G}}^{\mathrm{Q4C}}_{\mathrm{eff}}$ can be
interpreted as a correction of two-electron picture-change errors (2ePCE)\cite{PCE},
although everything is done here within the four-component framework.
Its construction is very cheap due to the sparsity associated with $\bar{\mathbf{D}}^0$.

Usually only the occupied and low-lying virtual A4Ss $\{\varphi_p\}_{p=1}^{2n_B}$ are
needed to form the backbone of the basis, which is to be augmented with
some uncontracted flat functions $\{f_{\nu}\}_{\nu=2n_B+1}^{2n}$
for describing the deformation and polarization of the atoms when forming
the molecule. As $R_{\nu}(\vec{r})\approx 1$ in the valence region (cf. Fig. 1 in Ref. \citenum{Q4CX2C}),
such flat functions can simply be taken as
 \begin{eqnarray}
 f_{\nu}=\begin{pmatrix}\chi^L_{\nu}\\ \chi_{\nu}^S\end{pmatrix},\quad \chi_{\nu}^S=
 \frac{1}{2c}\Pi \chi^L_{\nu}\approx 0,\quad \nu\in[2n_B+1,2n].\label{RKB0}
 \end{eqnarray}
 Unlike Eq. \eqref{RKBbasis}, the large and small component basis functions are here combined together. Such
a `HF+P' type of spinor basis, single-zeta (or double-zeta) for core shells and multiple-zeta for valence shells,
is very effective\cite{Q4CX2C,Q4C}. The elements of $\mathbf{F}^{\mathrm{Q4C}}_+$
among the added flat functions $\{f_{\nu}\}_{\nu=2n_B+1}^{2n}$ can be treated nonrelativistically,
\begin{align}
F^{\mathrm{Q4C}}_{+,\mu\nu}&=\langle \chi^L_{\mu}|T|\chi^L_{\nu}\rangle+\langle \chi^L_{\mu}|V|\chi^L_{\nu}\rangle+G^{\mathrm{NR}}_{\mu\nu}[\bar{\mathbf{P}}],
\quad \mu,\nu\in[2n_B+1,2n],\label{AugBas}
\end{align}
while those between $\{\varphi_p\}_{p=1}^{2n_B}$ and $\{f_{\nu}\}_{\nu=2n_B+1}^{2n}$ can be treated in the same way as Eq. \eqref{effQ4C} or simply
approximated as
\begin{align}
F^{\mathrm{Q4C}}_{+,p\nu}&=\bar{h}^{\mathrm{Q4C}}_{\mathrm{eff},p\nu}+\bar{G}^{\mathrm{NR}}_{p\nu}[\bar{\mathbf{P}}]=[F^{\mathrm{Q4C}}_{+,\nu p}]^*,
\quad p\in[1,2n_B],\quad \nu\in[2n_B+1,2n],\label{Q4Ccross}\\
\bar{G}^{\mathrm{NR}}_{p\nu}[\bar{\mathbf{P}}]&=(\mathbf{a}^\dag \mathbf{G}^{\mathrm{NR}}[\bar{\mathbf{D}}^{LL}])_{p\nu},\label{barGNRpv}\\
\bar{h}^{\mathrm{Q4C}}_{\mathrm{eff},p\nu}&=\langle \varphi_p^L|T|\chi^L_{\nu}\rangle+\langle \varphi_p^L|V|\chi^L_{\nu}\rangle
+\bar{G}^{\mathrm{Q4C}}_{\mathrm{eff},p\nu}[\bar{\mathbf{D}}^0]\\
&=[\mathbf{a}^\dag (\mathbf{T}+\mathbf{V}+\mathbf{G}^{\mathrm{NESC}}[\bar{\mathbf{D}}^0])]_{p\nu},\\
G^{\mathrm{NESC}}_{\mu\nu}[\bar{\mathbf{D}}^0]&=G^{\mathrm{NESC,LL}}_{\mu\nu}[\bar{\mathbf{D}}^0]+G^{\mathrm{NESC},SL}_{\mu\nu}[\bar{\mathbf{D}}^0],\\
\bar{G}^{\mathrm{Q4C}}_{\mathrm{eff},p\nu}[\bar{\mathbf{D}}^0]&=\bar{G}^{\mathrm{Q4C},LL}_{\mathrm{eff},p\nu}[\bar{\mathbf{D}}^0]
+\bar{G}^{\mathrm{Q4C},SL}_{\mathrm{eff},p\nu}[\bar{\mathbf{D}}^0]
=(\mathbf{a}^\dag\mathbf{G}^{\mathrm{NESC}}[\bar{\mathbf{D}}^0])_{p\nu},\label{GNESCcross}\\
\bar{G}^{\mathrm{Q4C},LL}_{\mathrm{eff},p\nu}[\bar{\mathbf{D}}^0]&= (\Omega_{p\nu}^{LL}|g_{12}|\tr_2[\bar{P}^0_{rs}\Omega^{SS}_{sr}])-\mathrm{c_g} (\Xi_{pr}^{LS}|g_{12}\cdot|\bar{P}^0_{rs}\Xi_{s\nu}^{SL})\nonumber\\
&-\mathrm{c_b} (\Xi_{pr}^{LS}|\cdot\mathbf{b}_{12}\cdot|\bar{P}^0_{rs}\Xi_{s\nu}^{SL})\label{GNESCLLpv}\\
&=(\mathbf{a}^\dag \mathbf{G}^{\mathrm{NESC},LL}[\bar{\mathbf{D}}^0])_{p\nu},\label{Q4C-GSSLL}\\
\mathbf{G}^{\mathrm{NESC},LL}_{\mu\nu}[\bar{\mathbf{D}}^0]
&=(\Omega_{\mu\nu}^{LL}|g_{12}|\tr_2[\bar{D}^{0,SS}_{\lambda\kappa}\Omega^{SS}_{\kappa\lambda}])
-\mathrm{c_g} (\Xi_{\mu\lambda}^{LS}|g_{12}\cdot|\bar{D}^{0,SS}_{\lambda\kappa}\Xi_{\kappa\nu}^{SL})\nonumber\\
&-\mathrm{c_b} (\Xi_{\mu\lambda}^{LS}|\cdot\mathbf{b}_{12}\cdot|\bar{D}^{0,SS}_{\lambda\kappa}\Xi_{\kappa\nu}^{SL}),\\
\bar{G}^{\mathrm{Q4C},SL}_{\mathrm{eff},p\nu}[\bar{\mathbf{D}}^0]&=-(\Omega_{pr}^{SS}|g_{12}|\bar{P}^0_{rs}\Omega^{LL}_{s\nu})\nonumber\\
&+\mathrm{c_g} [(\Xi_{p\nu}^{SL}|g_{12}\cdot|\tr_2[\bar{P}^0_{rs}(\Xi_{sr}^{SL}+\Xi_{sr}^{LS})])
-(\Xi_{pr}^{SL}|g_{12}\cdot|\bar{P}^0_{rs}\Xi_{s\nu}^{SL})]\nonumber\\
&+\mathrm{c_b} [(\Xi_{p\nu}^{SL}|\cdot\mathbf{b}_{12}\cdot|\tr_2[\bar{P}^0_{rs}(\Xi_{sr}^{SL}+\Xi_{sr}^{LS})])
-(\Xi_{pr}^{SL}|\cdot\mathbf{b}_{12}\cdot|\bar{P}^0_{rs}\Xi_{s\nu}^{SL})]\label{GNESCSSpv}\\
&=(\mathbf{a}^\dag\mathbf{G}^{\mathrm{NESC},SL}[\bar{\mathbf{D}}^0])_{p\nu},\\
\mathbf{G}^{\mathrm{NESC},SL}_{\sigma\nu}[\bar{\mathbf{D}}^0]
&=X_{\mu\sigma}^*\{-(\Omega_{\mu\lambda}^{SS}|g_{12}|\bar{D}^{0,SL}_{\lambda\kappa}\Omega^{LL}_{\kappa\nu})\nonumber\\
&+\mathrm{c_g}[(\Xi_{\mu\nu}^{SL}|g_{12}\cdot|\tr_2[\bar{D}^{0,LS}_{\lambda\kappa}\Xi_{\kappa\lambda}^{SL}+
\bar{D}^{0,SL}_{\lambda\kappa}\Xi_{\kappa\lambda}^{LS}])
-(\Xi_{\mu\lambda}^{SL}|g_{12}\cdot|\bar{D}^{0,LS}_{\lambda\kappa}\Xi_{\kappa\nu}^{SL})]\nonumber\\
&+\mathrm{c_b}[(\Xi_{\mu\nu}^{SL}|\cdot\mathbf{b}_{12}\cdot|\tr_2[\bar{D}^{0,LS}_{\lambda\kappa}\Xi_{\kappa\lambda}^{SL}+
\bar{D}^{0,SL}_{\lambda\kappa}\Xi_{\kappa\lambda}^{LS}])
-(\Xi_{\mu\lambda}^{SL}|\cdot\mathbf{b}_{12}\cdot|\bar{D}^{0,LS}_{\lambda\kappa}\Xi_{\kappa\nu}^{SL})]\}\label{Q4C-GSSSS}.
\end{align}
The corresponding overlap matrix elements read
\begin{align}
\bar{M}_{+,\mu\nu}&=M^{LL}_{\mu\nu}+M^{SS}_{\mu\nu},\quad \mu,\nu\in[2n_B+1,2n],\\
\bar{M}_{+,p\nu}&=[\mathbf{a}^\dag(\mathbf{M}^{LL}+\mathbf{M}^{SS})]_{p\nu},\quad p\in[1,2n_B].
\end{align}
The `rule' for obtaining the elements $\bar{G}^{\mathrm{Q4C},XY}_{\mathrm{eff},p\nu}[\bar{\mathbf{D}}^0]$ is to set the $\chi^S_{\nu}$-containing terms
of $G^{XY}_{p\nu}$ (cf. Eqs. \eqref{4CGLLorb}, \eqref{4CGLSorb}, \eqref{4CGSLorb}, and \eqref{4CGSSorb}) to zero, thereby leaving only
$\bar{G}^{\mathrm{Q4C},LL}_{\mathrm{eff},p\nu}[\bar{\mathbf{D}}^0]$ \eqref{GNESCLLpv} and $\bar{G}^{\mathrm{Q4C},SL}_{\mathrm{eff},p\nu}[\bar{\mathbf{D}}^0]$
\eqref{GNESCSSpv} stemming from $G^{LL}_{p\nu}$ \eqref{4CGLLorb} and $G^{SL}_{p\nu}$ \eqref{4CGSLorb},
respectively. The use of the same expansion coefficients $\{\bar{\mathbf{C}}_{+,i}\}$ for the small and large components of the M4Ss $\{\psi_i\}$ is also made
here (cf. Eq. \eqref{LCA4Sa}). Eq. \eqref{RKB0} along with the expressions \eqref{AugBas} and \eqref{Q4Ccross} has been termed RKB0\cite{Q4CX2C},
which is very accurate for spectroscopic constants even of the heaviest molecular systems\cite{E111,E114,E113,BDF3,BDFECC}.

Additional remarks on Q4C should be made here:
\begin{enumerate}[(a)]
\item The above `HF+P' type of spinor basis, subject to the RKB and RKB0 conditions, is most compact.

\item Once the A4Ss are symmetrized according to both double point group and time reversal symmetries\cite{Symm2009},
the quaternion form of Q4C-HF can readily be obtained by the $\mbox{}^{\mathrm{Q}}\mathbf{U}$ transformation \eqref{QUmatC}
of  $\mathbf{F}^{\mathrm{Q4C}}_+$ (cf. Eqs. \eqref{effQ4C}, \eqref{AugBas}, and \eqref{Q4Ccross}). The Q4C variant of KROHF
can then be formulated (see Appendix  \ref{AppendixA}).

\item Since the large and small components of $\varphi_p$ share
the same coefficients, the A4S integrals $(\varphi_p\varphi_q|V(1,2)|\varphi_r\varphi_s)$ can be transformed
to the MO representation as a whole (instead of component-wise). Therefore, both the integral transformation and
correlation steps in Q4C are computationally the same as those in two-component approaches\cite{LiuMP,X2CBook2017}.
\item Unlike two-component approaches, Q4C does not suffer from 2ePCEs in both the mean-field and correlation steps.

\item The MDM approximation can also be applied to the DHF Fock matrix \eqref{DEQMat}, leading to
\begin{align}
\mathbf{F}&=\mathbf{h}_{\mathrm{eff}}^{\mathrm{4C}}+\mathbf{G}^{\mathrm{4C}}[\mathbf{D}],\label{pmf4C}\\
\mathbf{h}_{\mathrm{eff}}^{\mathrm{4C}}&=\begin{pmatrix}\mathbf{h}^{LL}+\mathbf{G}^{\prime LL}[\mathbf{D}^0] & \mathbf{h}^{LS}+\mathbf{G}^{LS}[\mathbf{D}^0]\\ \mathbf{h}^{SL}+\mathbf{G}^{SL}[\mathbf{D}^0]&\mathbf{h}^{SS}+\mathbf{G}^{SS}[\mathbf{D}^0] \end{pmatrix},\quad \mathbf{D}^0=\sum_A^\oplus \mathbf{D}_A,\\
\mathbf{G}^{\prime LL}[\mathbf{D}^0]&=\mathbf{G}^{LL}[\mathbf{D}^0]-\mathbf{G}^{\mathrm{NR}}[\mathbf{D}^{0,LL}],\\
\mathbf{G}^{\mathrm{4C}}[\mathbf{D}]&=\begin{pmatrix}\mathbf{G}^{\mathrm{NR}}[\mathbf{D}^{LL}] & \mathbf{0}\\ \mathbf{0}&\mathbf{0} \end{pmatrix},
\end{align}
where only $\mathbf{G}^{\mathrm{NR}}[\mathbf{D}^{LL}]$ needs to be updated in each iteration.
\end{enumerate}

\section{X2C}\label{SecX2C}
By definition, a two-component relativistic theory is a Schr\"odinger-like equation that describes only electrons relativistically.
In this sense, NESC\cite{NESC} is not yet a genuine two-component relativistic theory, since the
eigenequation \eqref{NESCeigen}, along with the relativistic metric \eqref{RelMetric}, determines only the large components of the M4Ss, whereas
the small components have to be constructed explicitly via relation \eqref{L2SX}.
That is, NESC is still within the Dirac picture, \emph{unlike} P4C and Q4C. The real significance of NESC
lies in the initiative of matrix formulations of two-component relativistic theories.
That is, the starting point is the matrix DHF equation \eqref{DEQMat} instead of the operator DHF equation \eqref{DHFeq}.
The former can be block-diagonalized exactly, so as to achieve exact decoupling of the PESs from the NESs,
ending up with an ``exact two-component'' (X2C)\cite{X2CName} relativistic theory for electrons.
This is different from approximate two-component (A2C) theories\cite{CPD,ZORA1,DKH21986,DKH21989} that
are correct only to a finite order in relativity. The block-diagonalization of Eq. \eqref{DEQMat}
can be done in one step\cite{X2C2005,X2C2009,X2C2007kutz,SaueX2C}, two steps\cite{BSS1,BSS2,Jensen2005}, and multiple steps\cite{ReiherDKH-I,ReiherDKH-II,DKH-Peng2009}.
It has been shown\cite{LiuMP} that
the three types of matrix formulations share the same decoupling condition and differ only in the renormalization.
There exist even closed mappings in between\cite{LiuMP}. Since
the initio free-particle transformation invoked in the two-step and multiple-step formulations
is only necessary to ensure variational stability and regularization of finite-order expansions\cite{DKH21986,DKH21989} but not needed
for an exact decoupling, and becomes very clumsy in the presence of magnetic interactions,
it is clear that it is the one-step formulation\cite{X2C2005,X2C2009,X2C2007kutz,SaueX2C} that should be advocated.
It deserves to be mentioned that the same one-step matrix X2C equation\cite{X2C2005,X2C2009}
can also be obtained by the (formal) Foldy-Wouthuysen (FW) transformation\cite{FW1950} of the operator DHF equation \eqref{DHFeq},
followed by making use of the RKB-RI to convert the operators therein
to matrix forms\cite{LiuMP}. That is, the matrix and operator (more precisely, operator-like) formulations of the one-step X2C are actually identical.
This applies also to the two-step\cite{BSS1,BSS2,Jensen2005} and multiple-step\cite{ReiherDKH-I,ReiherDKH-II,DKH-Peng2009}
formulations: instead of the usual operator formulations followed by the use of the RKB-RI, the same matrix
equations therein can also be obtained\cite{X2CSOC1} by starting with the partitioned matrix DHF equation \eqref{DEQMat}.
In essence, the use of the RKB-RI in the \emph{unexpanded} one-step, two-step, and multiple-step operator formulations
is merely a formal step and does not introduce any error even with a finite RKB basis. This is different
from A2C approaches, where the use of the RKB-RI for their matrix representations is indeed an approximation \cite{Essential2020}.

The one-step block-diagonalization of the matrix DHF equation \eqref{DEQMat} starts with
the formal relation \eqref{L2SX} between the small- and large-component coefficients of the PESs, and
a similar relation
\begin{eqnarray}
\mathbf{C}^L_-= \tilde{\mathbf{X}}\mathbf{C}^S_-\label{Xdef}
\end{eqnarray}
between the small (upper)- and large (lower)-component coefficients of the NESs.
In terms of such relations, the following transformation matrix $\mathbf{U}$ can be constructed\cite{LiuMP}
\begin{align}
\mathbf{U}&=\begin{pmatrix}\mathbf{U}^{LL}&\mathbf{U}^{LS}\\
\mathbf{U}^{SL}&\mathbf{U}^{SS}\end{pmatrix}=\boldsymbol{\Omega}_{D}\boldsymbol{\Omega}_{N}=\begin{pmatrix}\mathbf{R}_+&\tilde{\mathbf{X}}\mathbf{R}_-\\
\mathbf{X}\mathbf{R}_+&\mathbf{R}_-\end{pmatrix},\label{Utrans2}\\
\boldsymbol{\Omega}_{D}&=\begin{pmatrix}
\mathbf{I} & \tilde{\mathbf{X}} \\
\mathbf{X} & \mathbf{I} \end{pmatrix},\quad
\boldsymbol{\Omega}_{N}=\begin{pmatrix}
\mathbf{R}_{+} & \mathbf{0} \\
\mathbf{0} & \mathbf{R}_{-}\end{pmatrix},
\end{align}
where $\boldsymbol{\Omega}_{D}$ does the decoupling, whereas $\boldsymbol{\Omega}_{N}$
establishes the renormalization. Note that the matrix $\mathbf{U}$ defined here is just
the matrix representation\cite{X2C2009} of the FW transformation\cite{FW1950} in a finite RKB basis.
The decoupling matrix $\mathbf{X}$ (cf. Eq. \eqref{L2SX}) is to be determined by
the condition
\begin{align}
(\mathbf{U}^\dag \mathbf{F} \mathbf{U})^{SL}=\sum_{X,Y=L,S}(\mathbf{U}^\dag)^{SX}\mathbf{F}^{XY}\mathbf{U}^{YL}=\mathbf{0}.
\end{align}
More specifically\cite{X2C2005},
\begin{align}
\mathbf{F}^{SL}+\mathbf{F}^{SS}\mathbf{X}&=-\tilde{\mathbf{X}}^\dag\mathbf{L}_+^{\mathrm{UESC}},\quad \mathbf{L}_+^{\mathrm{UESC}}=\mathbf{F}^{LL}+\mathbf{F}^{LS}\mathbf{X},\label{XdecoupleUESC}
\end{align}
where $\mathbf{L}_+^{\mathrm{UESC}}$ is the UESC (unnormalized elimination of the small component) Hamiltonian\cite{NESC}
associated with the following eigenvalue problem,
\begin{align}
\mathbf{L}_+^{\mathrm{UESC}}\mathbf{C}^L_+&=\mathbf{M}^{LL}\mathbf{C}^L_+\mathbf{E}_+.\label{UESCeigen}
\end{align}
Further combined with Eq. \eqref{NESCeigen}, we have\cite{filatov2006comment}
\begin{align}
\mathbf{L}_+^{\mathrm{UESC}}=\mathbf{M}^{LL} \tilde{\mathbf{M}}^{-1}_+\mathbf{L}_+^{\mathrm{NESC}}.\label{NESC2UESC}
\end{align}
Eq. \eqref{XdecoupleUESC} can hence be rewritten as
\begin{align}
\mathbf{F}^{SL}+\mathbf{F}^{SS}\mathbf{X}&=-\tilde{\mathbf{X}}\mathbf{M}^{LL} \tilde{\mathbf{M}}^{-1}_+\mathbf{L}_+^{\mathrm{NESC}},\label{XdecoupleNESC}
\end{align}
which turns out to be more robust than Eq. \eqref{XdecoupleUESC} when solved iteratively\cite{X2C2007kutz}.
It deserves to be emphasized that both Eqs. \eqref{XdecoupleUESC} and \eqref{XdecoupleNESC},
proposed in Refs. \citenum{X2C2005} and \citenum{X2C2007kutz}, respectively, in different forms though,
are state-independent and hence fundamentally different from the state-dependent decoupling conditions proposed by Dyall\cite{NESC}.
The orthogonality condition $\mathbf{C}_+^\dag\mathbf{M}\mathbf{C}_-=\mathbf{0}$ leads to
\begin{equation}
\tilde{\mathbf{X}}=-(\mathbf{M}^{LL})^{-1}\mathbf{X}^{\dagger}\mathbf{M}^{SS}.\label{tildeX}
\end{equation}
It follows that $\tilde{\mathbf{X}}$ is determined directly by $\mathbf{X}$.
In other words, the two sets solutions, $\mathbf{X}$ and $\tilde{\mathbf{X}}$, of
the quadratic decoupling condition \eqref{XdecoupleUESC}/\eqref{XdecoupleNESC} are mutually related.
Without going into further details,
the renormalization matrices read\cite{X2C2009}
\begin{align}
\mathbf{R}_{+}&=[(\mathbf{M}^{LL})^{-1}\tilde{\mathbf{M}}_{+}]^{-\frac{1}{2}}=
(\mathbf{M}^{LL})^{-\frac{1}{2}}\mathbf{K}_+(\mathbf{M}^{LL})^{\frac{1}{2}},\nonumber\\
\mathbf{K}_+&=[(\mathbf{M}^{LL})^{-\frac{1}{2}}\tilde{\mathbf{M}}_{+}(\mathbf{M}^{LL})^{-\frac{1}{2}}]^{-\frac{1}{2}},\label{Rplus}\\
\mathbf{R}_{-}&=[(\mathbf{M}^{SS})^{-1}\tilde{\mathbf{M}}_{-}]^{-\frac{1}{2}}=
(\mathbf{M}^{SS})^{-\frac{1}{2}}\mathbf{K}_-(\mathbf{M}^{SS})^{\frac{1}{2}},\nonumber\\
\mathbf{K}_-&=[(\mathbf{M}^{SS})^{-\frac{1}{2}}\tilde{\mathbf{M}}_{-}(\mathbf{M}^{SS})^{-\frac{1}{2}}]^{-\frac{1}{2}},\label{Rminus}
\end{align}
where the relativistic metric for the NESs reads (cf. Eq. \eqref{RelMetric})
\begin{align}
\tilde{\mathbf{M}}_{-}&=\mathbf{M}^{SS}+\tilde{\mathbf{X}}^{\dagger}\mathbf{M}^{LL}\tilde{\mathbf{X}}.
\end{align}
It has been proven\cite{X2CSOC2} that the particular renormalization $\mathbf{R}_+$ \eqref{Rplus} renders
the resulting two-component spinors closest to the large components of the positive-energy M4Ss in the least-squares sense,
whereas the two-step\cite{BSS1,BSS2,Jensen2005} and multiple-step\cite{ReiherDKH-I,ReiherDKH-II,DKH-Peng2009}
transformations amount to adopting different renormalizations.
The $\mathbf{U}$-transformation \eqref{Utrans2} of Eq. \eqref{DEQMat} leads to
\begin{align}
(\mathbf{U}^\dag\mathbf{F}\mathbf{U})(\mathbf{U}^{-1}\mathbf{C})&=(\mathbf{U}^\dag\mathbf{M}\mathbf{U})(\mathbf{U}^{-1}\mathbf{C})\mathbf{E}\Leftrightarrow
\tilde{\mathbf{F}}\tilde{\mathbf{C}}=\mathbf{M}\tilde{\mathbf{C}} \mathbf{E},\label{hDtrans}
\end{align}
where
\begin{align}
\tilde{\mathbf{F}}&=\mathbf{U}^\dag\mathbf{F}\mathbf{U}=\begin{pmatrix} \mathbf{F}_{+}^{\mathrm{X2C}} & \mathbf{0} \\
\mathbf{0} & \mathbf{F}_{-}^{\mathrm{X2C}}\end{pmatrix},\quad\mathbf{U}^\dag\mathbf{M}\mathbf{U}=\begin{pmatrix}\mathbf{M}^{LL}&\mathbf{0}\\ \mathbf{0}&\mathbf{M}^{SS}\end{pmatrix}, \label{hDtrans2}\\
&\tilde{\mathbf{C}}=\mathbf{U}^{-1}\mathbf{C}=\mathbf{M}^{-1}\mathbf{U}^\dag \mathbf{M}\mathbf{C}
=\begin{pmatrix}\mathbf{R}_+^{-1}\mathbf{C}_+^L&\mathbf{0}\\
\mathbf{0}&\mathbf{R}_-^{-1}\mathbf{C}_-^S\end{pmatrix}
=\begin{pmatrix} \tilde{\mathbf{C}}_{+} & \mathbf{0} \\
\mathbf{0} & \tilde{\mathbf{C}}_{-}\end{pmatrix}.\label{CX}
\end{align}
The upper-left block of Eq. \eqref{hDtrans} defines the unique X2C equation\cite{X2C2005,X2C2009} for the PESs in the Schr\"odinger picture,
\begin{align}
\mathbf{F}_{+}^{\mathrm{X2C}}\tilde{\mathbf{C}}_{+}&=\mathbf{M}^{LL}\tilde{\mathbf{C}}_{+} \mathbf{E}_+,\label{FWeq}\\
\mathbf{F}_{+}^{\mathrm{X2C}}&=(\mathbf{U}^\dag\mathbf{F}\mathbf{U})^{LL}=\sum_{X,Y=L,S}(\mathbf{U}^\dag)^{LX}\mathbf{F}^{XY}\mathbf{U}^{YL}\nonumber\\
&=\mathbf{R}_{+}^{\dagger}\mathbf{L}^{\mathrm{X}}_{+}\mathbf{R}_{+},\quad
\mathrm{X}=\mathrm{NESC, SESC},\label{hplus}\\
\mathbf{L}^{\mathrm{SESC}}_{+}&=\frac{1}{2}[\tilde{\mathbf{M}}_{+}(\mathbf{M}^{LL})^{-1}\mathbf{L}^{\mathrm{UESC}}_{+}+\mathrm{c.c.}].\label{SESC}
\end{align}
Here, the SESC (symmetrized elimination of the small component)\cite{Q4CX2C,X2C2007kutz} Hamiltonian $\mathbf{L}^{\mathrm{SESC}}_{+}$
arises from the identity $\mathbf{L}^{\mathrm{SESC}}_{+}=\frac{1}{2}(\mathbf{L}^{\mathrm{NESC}}_{+}+\mathbf{L}^{\mathrm{NESC}}_{+})$
 and the relation \eqref{NESC2UESC}. The one-
 ($\mathbf{F}^{\mathrm{X2C,1e}}_{+}$) and two-electron ($\mathbf{F}^{\mathrm{X2C,2e}}_{+}$) terms of $\mathbf{F}_+^{\mathrm{X2C}}$ \eqref{hplus} read
\begin{align}
\mathbf{F}^{\mathrm{X2C,1e}}_{+}&=\mathbf{R}_+^\dag\mathbf{L}^{\mathrm{X,1e}}_+\mathbf{R}_+,\quad \mathrm{X=NESC, SESC},\\
\mathbf{L}^{\mathrm{SESC,1e}}_{+}&=\frac{1}{2}[\tilde{\mathbf{M}}_{+}(\mathbf{M}^{LL})^{-1}\mathbf{L}^{\mathrm{UESC,1e}}_{+}+\mathrm{c.c.}],
\quad \mathbf{L}^{\mathrm{UESC,1e}}_{+}=\mathbf{V}+\mathbf{TX},\label{SESC1e}\\
\mathbf{F}^{\mathrm{X2C,2e}}_{+}[\mathbf{D}]&=\mathbf{R}_+^\dag \mathbf{L}^{\mathrm{X,2e}}_+[\mathbf{D}]\mathbf{R}_+,\quad \mathrm{X=NESC, SESC},\label{hplus2e}\\
\mathbf{L}^{\mathrm{SESC,2e}}_+[\mathbf{D}]&=\frac{1}{2}[\tilde{\mathbf{M}}_{+}(\mathbf{M}^{LL})^{-1}
\mathbf{L}^{\mathrm{UESC,2e}}_{+}+\mathrm{c.c.}],\label{SESC2e}\\
\mathbf{L}^{\mathrm{UESC,2e}}_{+}&=\mathbf{G}^{LL}[\mathbf{D}]+\mathbf{G}^{LS}[\mathbf{D}]\mathbf{X}.\label{UESC2e}
\end{align}
At this stage, the two-electron term $\mathbf{F}^{\mathrm{X2C,2e}}_{+}$ \eqref{hplus2e} still involves explicitly the four-component
molecular density matrix $\mathbf{D}$ \eqref{AO4cDmat}, which can be eliminated as follows. In view of the relation
$\mathbf{C}=\mathbf{U}\tilde{\mathbf{C}}$ given by the first equality of Eq. \eqref{CX}, i.e.,
\begin{align}
\mathbf{C}^X_+=\mathbf{U}^{XL}\tilde{\mathbf{C}}_+,\quad X=L,S,\label{2C-C24C-C}
\end{align}
the components $\psi_i^X$ ($X=L, S$) of a positive-energy M4S $\psi_p$ can be expressed as
\begin{align}
\psi_p^X&=\sum_{\mu=1}^n \chi_{\mu}^X C_{+, \mu p}^X
= \sum_{\mu=1}^n \chi_{\mu}^X (\mathbf{U}^{XL}\tilde{\mathbf{C}}_+)_{\mu p}
= \sum_{\nu=1}^n \tilde{\chi}_{\nu}^X\tilde{\mathbf{C}}_{+,\nu p},\\
\tilde{\chi}_{\nu}^X&=\sum_{\mu=1}^{n}\chi_{\mu}^X U^{XL}_{\mu\nu},\quad \tilde{\chi}_{\nu}^{X\dag}=\sum_{\mu=1}^n (U^{XL\dag})_{\nu\mu}\chi_{\mu}^{X\dag}=\sum_{\mu=1}^n [(U^\dag)^{LX}]_{\nu\mu}\chi_{\mu}^{X\dag},\label{RenormBasis}
\end{align}
where $\{\tilde{\chi}^X_{\nu}\}_{\nu=1}^{n}$ ($X=L,S$) can be referred to as renormalized two-component spinor functions.
Likewise, the components $\mathbf{D}^{XY}$ ($X,Y=L,S$) of $\mathbf{D}$ \eqref{AO4cDmat} can be written as
\begin{align}
\mathbf{D}^{XY}&=\mathbf{U}^{XL}\tilde{\mathbf{D}}\mathbf{U}^{YL\dag}=\mathbf{U}^{XL}\tilde{\mathbf{D}}(\mathbf{U}^\dag)^{LY},\quad
\tilde{\mathbf{D}}=\tilde{\mathbf{C}}_+\mathbf{n}\tilde{\mathbf{C}}^\dag_+=\begin{pmatrix}\tilde{\mathbf{D}}^{\alpha\alpha}&\tilde{\mathbf{D}}^{\alpha\beta}\\
\tilde{\mathbf{D}}^{\beta\alpha}&\tilde{\mathbf{D}}^{\beta\beta}\end{pmatrix}.\label{DXY2D}
\end{align}
It then follows that $\mathbf{F}^{\mathrm{X2C,2e}}_{+}[\mathbf{D}]$ \eqref{hplus2e}
can be expressed in terms entirely of the renormalized quantities, viz.,
\begin{align}
\mathbf{F}^{\mathrm{X2C,2e}}_{+}[\mathbf{D}]&=\mathbf{R}_+^\dag\mathbf{L}^{\mathrm{X,2e}}_{+}[\tilde{\mathbf{D}}]\mathbf{R}_+,\quad \mathrm{X=NESC, SESC},\label{NESCSESC2e}\\
\mathbf{R}_+^\dag\mathbf{L}^{\mathrm{NESC,2e}}_{+}[\tilde{\mathbf{D}}]\mathbf{R}_+&=\tilde{G}^{LL}[\tilde{\mathbf{D}}]+\tilde{G}^{LS}[\tilde{\mathbf{D}}]
+\tilde{G}^{SL}[\tilde{\mathbf{D}}]+\tilde{G}^{SS}[\tilde{\mathbf{D}}],\label{FullX2C2e}\\
\mathbf{R}_+^\dag\mathbf{L}^{\mathrm{SESC,2e}}_{+}[\tilde{\mathbf{D}}]\mathbf{R}_+&=
\frac{1}{2}[(\bar{G}^{LL}[\tilde{\mathbf{D}}]+\bar{G}^{LS}[\tilde{\mathbf{D}}])+\mathrm{c.c.}],\label{FullSESC2e}
\end{align}
where
\begin{align}
\tilde{G}^{LL}_{\mu\nu}[\tilde{\mathbf{D}}]
&=(\tilde{\Omega}_{\mu\nu}^{LL}|g_{12}|\tr_2[\tilde{\mathbf{D}}_{\lambda\kappa}\tilde{\Omega}^{LL}_{\kappa\lambda}])
-(\tilde{\Omega}_{\mu\lambda}^{LL}|g_{12}|\tilde{\mathbf{D}}_{\lambda\kappa}\tilde{\Omega}^{LL}_{\kappa\nu})
+(\tilde{\Omega}_{\mu\nu}^{LL}|g_{12}|\tr_2[\tilde{\mathbf{D}}_{\lambda\kappa}\tilde{\Omega}^{SS}_{\kappa\lambda}])\nonumber\\
&-\mathrm{c_g}(\tilde{\boldsymbol{\Xi}}_{\mu\lambda}^{LS}|g_{12}\cdot|\tilde{\mathbf{D}}_{\lambda\kappa}\tilde{\boldsymbol{\Xi}}_{\kappa\nu}^{SL})
-\mathrm{c_b}(\tilde{\boldsymbol{\Xi}}_{\mu\lambda}^{LS}|\cdot\mathbf{b}_{12}\cdot|\tilde{\mathbf{D}}_{\lambda\kappa}\tilde{\boldsymbol{\Xi}}_{\kappa\nu}^{SL}),\label{4CGLL2C}\\
\tilde{G}^{LS}_{\mu\nu}[\tilde{\mathbf{D}}]
&=-(\tilde{\Omega}^{LL}_{\mu\lambda}|g_{12}|\tilde{\mathbf{D}}_{\lambda\kappa}\tilde{\Omega}^{SS}_{\kappa\nu})\nonumber\\
&+\mathrm{c_g}[(\tilde{\boldsymbol{\Xi}}_{\mu\nu}^{LS}|g_{12}\cdot|\tr_2[\tilde{\mathbf{D}}_{\lambda\kappa}\tilde{\boldsymbol{\Xi}}^{LS}_{\kappa\lambda}])
-(\tilde{\boldsymbol{\Xi}}^{LS}_{\mu\lambda}|g_{12}\cdot| \tilde{\mathbf{D}}_{\lambda\kappa}\tilde{\boldsymbol{\Xi}}^{LS}_{\kappa\nu})
+(\tilde{\boldsymbol{\Xi}}^{LS}_{\mu\nu}|g_{12}\cdot|\tr_2[\tilde{\mathbf{D}}_{\lambda\kappa}\tilde{\boldsymbol{\Xi}}^{SL}_{\kappa\lambda}])]\nonumber\\
&+\mathrm{c_b}[\tilde{\boldsymbol{\Xi}}_{\mu\nu}^{LS}|\cdot\mathbf{b}_{12}\cdot|\tr_2[\tilde{\mathbf{D}}_{\lambda\kappa}\tilde{\boldsymbol{\Xi}}^{LS}_{\kappa\lambda}])
-(\tilde{\boldsymbol{\Xi}}^{LS}_{\mu\lambda}|\cdot\mathbf{b}_{12}\cdot|\tilde{\mathbf{D}}_{\lambda\kappa}\tilde{\boldsymbol{\Xi}}^{LS}_{\kappa\nu})
+(\tilde{\boldsymbol{\Xi}}^{LS}_{\mu\nu}|\cdot\mathbf{b}_{12}\cdot|\tr_2[\tilde{\mathbf{D}}_{\lambda\kappa}\tilde{\boldsymbol{\Xi}}^{SL}_{\kappa\lambda}])],\label{4CGLS2C}\\
\tilde{G}^{SL}_{\mu\nu}[\tilde{\mathbf{D}}]
&=-(\tilde{\Omega}^{SS}_{\mu\lambda}|g_{12}|\tilde{\mathbf{D}}_{\lambda\kappa}\tilde{\Omega}^{LL}_{\kappa\nu})\nonumber\\
&+\mathrm{c_g}[\tilde{\boldsymbol{\Xi}}_{\mu\nu}^{SL}|g_{12}\cdot|\tr_2[\tilde{\mathbf{D}}_{\lambda\kappa}\tilde{\boldsymbol{\Xi}}^{SL}_{\kappa\lambda}])
-(\tilde{\boldsymbol{\Xi}}^{SL}_{\mu\lambda}|g_{12}\cdot| \tilde{\mathbf{D}}_{\lambda\kappa}\tilde{\boldsymbol{\Xi}}^{SL}_{\kappa\nu})
+(\tilde{\boldsymbol{\Xi}}^{SL}_{\mu\nu}|g_{12}\cdot|\tr_2[\tilde{\mathbf{D}}_{\lambda\kappa}\tilde{\boldsymbol{\Xi}}^{LS}_{\kappa\lambda}])]\nonumber\\
&+\mathrm{c_b}[(\tilde{\boldsymbol{\Xi}}_{\mu\nu}^{SL}|\cdot\mathbf{b}_{12}\cdot|\tr_2[\tilde{\mathbf{D}}_{\lambda\kappa}\tilde{\boldsymbol{\Xi}}^{SL}_{\kappa\lambda}])
-(\tilde{\boldsymbol{\Xi}}^{SL}_{\mu\lambda}|\cdot\mathbf{b}_{12}\cdot|\tilde{\mathbf{D}}_{\lambda\kappa}\tilde{\boldsymbol{\Xi}}^{SL}_{\kappa\nu})
+(\tilde{\boldsymbol{\Xi}}^{SL}_{\mu\nu}|\cdot\mathbf{b}_{12}\cdot|\tr_2[\tilde{\mathbf{D}}_{\lambda\kappa}\tilde{\boldsymbol{\Xi}}^{LS}_{\kappa\lambda}])]\nonumber\\
&=\tilde{G}^{LS*}_{\nu\mu},\label{4CGSL2C}\\
\tilde{G}^{SS}_{\mu\nu}[\tilde{\mathbf{D}}]
&=(\tilde{\Omega}_{\mu\nu}^{SS}|g_{12}|\tr_2[\tilde{\mathbf{D}}_{\lambda\kappa}\tilde{\Omega}^{SS}_{\kappa\lambda}])
-(\tilde{\Omega}_{\mu\lambda}^{SS}|g_{12}|\tilde{\mathbf{D}}_{\lambda\kappa}\tilde{\Omega}^{SS}_{\kappa\nu})
+(\tilde{\Omega}_{\mu\nu}^{SS}|g_{12}|\tr_2[\tilde{\mathbf{D}}_{\lambda\kappa}\tilde{\Omega}^{LL}_{\kappa\lambda}])\nonumber\\
&-\mathrm{c_g}(\tilde{\boldsymbol{\Xi}}_{\mu\lambda}^{SL}|g_{12}\cdot|\tilde{\mathbf{D}}_{\lambda\kappa}\tilde{\boldsymbol{\Xi}}_{\kappa\nu}^{LS})
-\mathrm{c_b}(\tilde{\boldsymbol{\Xi}}_{\mu\lambda}^{SL}|\cdot\mathbf{b}_{12}\cdot|\tilde{\mathbf{D}}_{\lambda\kappa}\tilde{\boldsymbol{\Xi}}_{\kappa\nu}^{LS}),\label{4CGSS2C}\\
\bar{G}^{LL}_{\mu\nu}[\tilde{\mathbf{D}}]
&=(\bar{\Omega}_{\mu\nu}^{LL}|g_{12}|\tr_2[\tilde{\mathbf{D}}_{\lambda\kappa}\tilde{\Omega}^{LL}_{\kappa\lambda}])
-(\bar{\Omega}_{\mu\lambda}^{LL}|g_{12}|\tilde{\mathbf{D}}_{\lambda\kappa}\tilde{\Omega}^{LL}_{\kappa\nu})
+(\bar{\Omega}_{\mu\nu}^{LL}|g_{12}|\tr_2[\tilde{\mathbf{D}}_{\lambda\kappa}\tilde{\Omega}^{SS}_{\kappa\lambda}])\nonumber\\
&-\mathrm{c_g}(\bar{\boldsymbol{\Xi}}_{\mu\lambda}^{LS}|g_{12}\cdot|\tilde{\mathbf{D}}_{\lambda\kappa}\tilde{\boldsymbol{\Xi}}_{\kappa\nu}^{SL})
-\mathrm{c_b}(\bar{\boldsymbol{\Xi}}_{\mu\lambda}^{LS}|\cdot\mathbf{b}_{12}\cdot|\tilde{\mathbf{D}}_{\lambda\kappa}\tilde{\boldsymbol{\Xi}}_{\kappa\nu}^{SL}),\label{4CGLL2CU}\\
\bar{G}^{LS}_{\mu\nu}[\tilde{\mathbf{D}}]
&=-(\bar{\Omega}^{LL}_{\mu\lambda}|g_{12}|\tilde{\mathbf{D}}_{\lambda\kappa}\tilde{\Omega}^{SS}_{\kappa\nu})\nonumber\\
&+\mathrm{c_g}[(\bar{\boldsymbol{\Xi}}_{\mu\nu}^{LS}|g_{12}\cdot|\tr_2[\tilde{\mathbf{D}}_{\lambda\kappa}\tilde{\boldsymbol{\Xi}}^{LS}_{\kappa\lambda}])
-(\bar{\boldsymbol{\Xi}}^{LS}_{\mu\lambda}|g_{12}\cdot| \tilde{\mathbf{D}}_{\lambda\kappa}\tilde{\boldsymbol{\Xi}}^{LS}_{\kappa\nu})
+(\bar{\boldsymbol{\Xi}}^{LS}_{\mu\nu}|g_{12}\cdot|\tr_2[\tilde{\mathbf{D}}_{\lambda\kappa}\tilde{\boldsymbol{\Xi}}^{SL}_{\kappa\lambda}])]\nonumber\\
&+\mathrm{c_b}[\bar{\boldsymbol{\Xi}}_{\mu\nu}^{LS}|\cdot\mathbf{b}_{12}\cdot|\tr_2[\tilde{\mathbf{D}}_{\lambda\kappa}\tilde{\boldsymbol{\Xi}}^{LS}_{\kappa\lambda}])
-(\bar{\boldsymbol{\Xi}}^{LS}_{\mu\lambda}|\cdot\mathbf{b}_{12}\cdot|\tilde{\mathbf{D}}_{\lambda\kappa}\tilde{\boldsymbol{\Xi}}^{LS}_{\kappa\nu})
+(\bar{\boldsymbol{\Xi}}^{LS}_{\mu\nu}|\cdot\mathbf{b}_{12}\cdot|\tr_2[\tilde{\mathbf{D}}_{\lambda\kappa}\tilde{\boldsymbol{\Xi}}^{SL}_{\kappa\lambda}])],\label{4CGLS2CU}\\
\tilde{\Omega}^{LL}_{\mu\lambda}&=\tilde{\chi}^{L\dag}_{\mu}\tilde{\chi}^L_{\nu},\quad \tilde{\Omega}^{SS}_{\mu\lambda}=\tilde{\chi}^{S\dag}_{\mu}\tilde{\chi}^S_{\nu},\\
\tilde{\boldsymbol{\Xi}}^{LS}_{\mu\lambda}&=\tilde{\chi}^{L\dag}_{\mu}\boldsymbol{\sigma}\tilde{\chi}^S_{\nu},\quad
\tilde{\boldsymbol{\Xi}}^{SL}_{\mu\lambda}=\tilde{\chi}^{S\dag}_{\mu}\boldsymbol{\sigma}\tilde{\chi}^L_{\nu},\\
\bar{\Omega}^{LL}_{\mu\lambda}&=\bar{\chi}^{L\dag}_{\mu}\tilde{\chi}^L_{\nu},\quad
\bar{\chi}^{L}_{\mu}=\sum_{\nu=1}^n \chi^L_{\nu}\bar{U}^{LL}_{\nu\mu},\quad \bar{\mathbf{U}}^{LL}= (\mathbf{M}^{LL})^{-1}\tilde{\mathbf{M}}_+\mathbf{R}_+,\\
\bar{\boldsymbol{\Xi}}^{LS}_{\mu\lambda}&=\bar{\chi}^{L\dag}_{\mu}\boldsymbol{\sigma}\tilde{\chi}^S_{\nu}.
\end{align}
Note that $\tilde{\mathbf{G}}^{XY}[\tilde{\mathbf{D}}]$ in Eqs. \eqref{4CGLL2C}-\eqref{4CGSS2C} are
completely parallel to $\mathbf{G}^{XY}[\mathbf{D}]$ in Eqs. \eqref{4CGLL}-\eqref{4CGSS},
in the sense that they can be obtained simply by replacing $\Omega^{VV}$, $\boldsymbol{\Xi}^{VW}$,
and $\mathbf{D}^{VW}$ ($V,W=L,S$) with $\tilde{\Omega}^{VV}$, $\tilde{\boldsymbol{\Xi}}^{VW}$,
and $\tilde{\mathbf{D}}$, respectively, in the latter. However, as far as implementation is concerned,
it is more advantageous to first construct $\mathbf{L}_+^{\mathrm{NESC/SESC,2e}}[\mathbf{D}]$ by transforming
$\tilde{\mathbf{D}}$ back to $\mathbf{D}^{XY}$ (cf. Eq. \eqref{DXY2D}), and then do
the matrix transformation \eqref{hplus2e}.

So far no approximation has been made, for what has been done is merely to
convert a single $4n$-by-$4n$ matrix DHF equation \eqref{DEQMat} for $2n$ PESs and $2n$ NESs to two mutually coupled
$2n$-by-$2n$ equations, Eqs. \eqref{FWeq} and \eqref{XdecoupleUESC}/\eqref{XdecoupleNESC}, for the $2n$ PESs alone.
Once these equations are solved via a dual-level iteration scheme\cite{X2C2005,X2C2007kutz}, the PESs of the parent matrix DHF equation \eqref{DEQMat}
can be reproduced up to machine accuracy, thereby justifying the name ``exact two-component''\cite{X2CName}. However,
this is neither computationally favorable (even for one-electron systems) nor necessary in practice. What really matters
is to find an accurate and easily accessible approximation to the decoupling matrix $\mathbf{X}$,
so as to solve the decoupling condition \eqref{XdecoupleUESC}/\eqref{XdecoupleNESC} indirectly.
Taking $\mathbf{X}$ as a matrix functional $\mathbf{X}[V_{\mathrm{eff}}]$ of the effective potential $V_{\mathrm{eff}}$
(which itself is a functional of the four-component molecular density matrix $\mathbf{D}$ \eqref{AO4cDmat}), and the interatomic
interaction strength as a formal expansion parameter, various approximations to  $\mathbf{X}$ can readily be
envisaged\cite{Q4CX2C}. The first approximation is obviously the AtomX \eqref{ATOMX}\cite{LCA4S}, where each
atomic $\mathbf{X}_A$ is derived from the eigenvectors of
the matrix DHF equation \eqref{DEQMat} for a spherically averaged and unpolarized atomic configuration.
Note that the renomalization $\mathbf{R}_+$ \eqref{Rplus} is still of full dimension.
This approximation for $\mathbf{L}^{\mathrm{NESC}}_+$ in $\mathbf{F}_+^{\mathrm{X2C}}$ stays obviously in the same spirit as P4C/Q4C\cite{Essential2020}, and
is hence unsurprisingly very accurate,
not only for ground state energies of molecular systems\cite{Q4C,Q4CX2C}, but also for electric\cite{X2CTDReO4,X2CTDYbO,X2CTDOsO4,X2CTDSOC,X2CEOMCC2017,RTDDFTrev} and magnetic\cite{X2CNMR2009,X2CNMR2012} response properties, analytic energy gradient and Hessian\cite{X2Cgrd2020}, as well as periodic systems\cite{X2C-PBC}.
There have been attempts\cite{DLU,Nakai2012a,Nakai2012b} to approximate
the renormalization matrix $\mathbf{R}_+$ \eqref{Rplus} also as the superposition of the atomic ones
(in conjunction with a `diagonal local X' obtained by solving the
one-electron Dirac equation that is block-diagonal
in atoms but including all nuclear attractions), so as
to make $\mathbf{U}$ \eqref{Utrans2} block diagonal in atoms. Since $\mathbf{R}_+$
is much less local than $\mathbf{X}$, such approximation does introduce discernible errors\cite{LiuGVVPT2},
which are only tolerable for large systems in view of the dramatic gain in computational efficiency (especially in
gradient and Hessian calculations\cite{X2Cgrd2020}).
An obvious improvement of the AtomX \eqref{ATOMX} is a diatomic approximation (DAX), i.e.,
\begin{align}
\mathbf{X}\approx \sum_F^\oplus\mathbf{X}_{F},\label{FragX}
\end{align}
where $\mathbf{X}_{F}$ is derived from the diagonalization of $\mathbf{F}^{\mathrm{Frag}}$ in Eq. \eqref{FragF} for every pair $F$ of atoms.
The common-atom blocks of fragmental matrices can simply be averaged in this case.
This option is not only necessary for $\mathbf{L}^{\mathrm{NESC}}_+$ in $\mathbf{F}_+^{\mathrm{X2C}}$ for situations
where two heavy atoms are located too closely (e.g., at a distance shorter than half of a regular bond),
such that the off-diagonal blocks of the molecular $\mathbf{X}$ become significant\cite{Essential2020},
but also a must\cite{Q4CX2C,X2CBook2017} for $\mathbf{F}_+^{\mathrm{X2C}}$ in conjunction with $\mathbf{L}^{\mathrm{SESC}}_+$
(which is less accurate\cite{NESC} than $\mathbf{L}^{\mathrm{NESC}}_+$ for an approximate $\mathbf{X}$).
More generally, the DAX can be extended to a fragmental approximation (FragX)\cite{Q4CX2C}, where a fragment
can be chosen to include one, two, or multiple atoms.
It is just that the chosen fragments should be kept fixed for all molecular geometries.
The limiting case, where the whole molecule is treated as a single fragment, has been called model-potential approximation (ModX)\cite{Q4CX2C}.
Here, the following four-component Fock matrix
\begin{align}
\mathbf{F}^{\mathrm{pmf}}=\mathbf{h}+\mathbf{G}[\mathbf{D}^0],\quad \mathbf{D}^0=\sum_{A}^\oplus \mathbf{D}_A \label{ModF}
\end{align}
is to be constructed and diagonalized. The term `model potential' (arising from the superposition of atomic densities)
was first introduced by van W\"ullen\cite{ZORA-MP1998} to fix the gauge problem
of the zeroth-order regular approximation\cite{CPD,ZORA1} in the context of DFT. Very recently, this ansatz
was also employed in the extended atomic mean-field (eamf) approach\cite{2ePCE2022}.
Since the ModX was already introduced\cite{Q4CX2C}
long before, and there is no difference between X2C-KS\cite{Q4CX2C} and X2C-HF\cite{2ePCE2022} in the
context of constructing the unitary transformation \eqref{Utrans2}, a new name is hardly justified. Nonetheless,
the term `pre-molecular mean-field' (pmf) seems to be a better characterization of Eq. \eqref{ModF}
than both `model potential' and `extended atomic mean-field',
for a pre-molecule is by definition the superposition of atoms.
Moreover, the abbreviation pmf is in closer analogy with mmf (molecular mean-field)\cite{mmf-X2C2009},
where it is the full, converged molecular four-component Fock matrix that is converted to $\mathbf{F}_+^{\mathrm{X2C}}$ \eqref{hplus}.
Therefore, the ModX will be renamed to pmfX from now on.
As an approximation of pmfX, the 1eX\cite{SaueX2C} obtained by diagonalizing only the first, one-electron term of Eq. \eqref{ModF}
is widely used. However, it is not accurate enough for magnetic properties\cite{X2CNMR2009,X2CNMR2012}, and cannot be applied to periodic systems.
As an alternative approximation of pmfX, Eq. \eqref{ModF} was replaced by
\begin{align}
\mathbf{F}^{\mathrm{amf}}=\mathbf{h}+\sum_A^\oplus \mathbf{G}_A[\mathbf{D}_A]\label{amfF}
\end{align}
in the so-called atomic mean-field (amf) approach\cite{2ePCE2022} for the correction of 2ePCEs.
However, $\mathbf{F}^{\mathrm{amf}}$ \eqref{amfF} is not a legitimate Hamiltonian,
for there is no physical justification for ignoring the two-electron but retaining the one-electron interatomic couplings.
In particular, the (scalar) electrostatic interaction between electrons is known to be long-ranged.
Numerical experimentations\cite{ChengNote} do reveal that $\mathbf{F}^{\mathrm{amf}}$ may
have spurious solutions, especially when uncontracted RKB basis sets are used. As such, this amfX
should be dumped definitely. In contrast, the Hamiltonians $\sum_A^{\oplus}\mathbf{F}_A$, $\mathbf{F}^{\mathrm{Frag}}$ \eqref{FragF},
$\mathbf{F}^{\mathrm{pmf}}$ \eqref{ModF}, and
$\mathbf{h}$ \eqref{h1e} employed for deriving the AtomX, FragX, pmfX, and 1eX, respectively,
are all well behaved.

Having discussed the various ways for approximating the decoupling matrix $\mathbf{X}$, possible approximations
to the elements of $\mathbf{F}_+^{\mathrm{X2C}}$ \eqref{hplus} should further be pursued.
The very first of these is the MDM approximation (an equivalent of MSCA\cite{Q4CX2C})
 underlying Q4C (see Sec. \ref{SecQ4C}). That is,
$\mathbf{G}^{\mathrm{NESC}}[\mathbf{D}]$ in $\mathbf{L}_+^{\mathrm{NESC,2e}}[\mathbf{D}]$ \eqref{NESC2eNR} is replaced
with $\mathbf{G}^{\mathrm{NESC}}[\mathbf{D}^0]$, so as to rewrite the NESC-based $\mathbf{F}_+^{\mathrm{X2C}}$ \eqref{hplus} as
\begin{align}
\tilde{\mathbf{F}}_+^{\mathrm{X2C}}&=\tilde{\mathbf{h}}_{\mathrm{eff}}^{\mathrm{NESC}}+\tilde{\mathbf{G}}^{\mathrm{NR}}[\tilde{\mathbf{D}}],\label{effX2C}\\
\tilde{\mathbf{h}}_{\mathrm{eff}}^{\mathrm{NESC}}&=\mathbf{R}_+^\dag\mathbf{L}_{+}^{\mathrm{NESC,1e}}\mathbf{R}_+ +\tilde{\mathbf{G}}^{\mathrm{NESC}}_{\mathrm{eff}},\quad
\tilde{\mathbf{G}}^{\mathrm{NESC}}_{\mathrm{eff}}=\mathbf{R}_+^\dag\mathbf{G}^{\mathrm{NESC}}[\mathbf{D}^0]\mathbf{R}_+,\label{effX2C1e}\\
\tilde{G}^{\mathrm{NR}}_{\mu\nu}[\tilde{\mathbf{D}}]&=(\tilde{\Omega}_{\mu\nu}^{LL}|g_{12}|\tr_2[\tilde{\mathbf{D}}_{\lambda\kappa}\tilde{\Omega}^{LL}_{\kappa\lambda}])
-(\tilde{\Omega}_{\mu\lambda}^{LL}|g_{12}|\tilde{\mathbf{D}}_{\lambda\kappa}\tilde{\Omega}^{LL}_{\kappa\nu})\label{GNRrenormalized}\\
&=(\mathbf{R}_+^\dag \mathbf{G}^{\mathrm{NR}}[\mathbf{D}^{LL}]\mathbf{R}_+)_{\mu\nu}.\label{X2CGNR}
\end{align}
As a matter of fact, $\tilde{\mathbf{F}}_+^{\mathrm{X2C}}$ \eqref{effX2C} can be obtained directly
 by the $\mathbf{U}$-transformation \eqref{Utrans2} of pmfDHF \eqref{pmf4C}. It
has the same structure as $\mathbf{F}_+^{\mathrm{Q4C}}$ \eqref{effQ4C}. However, there exists a subtle
but important difference in between:
unlike the renormalization $\mathbf{R}_+$ \eqref{Rplus} in $\tilde{\mathbf{G}}^{\mathrm{NR}}$ \eqref{X2CGNR},
the $\mathbf{a}$ matrix in $\bar{\mathbf{G}}^{\mathrm{NR}}$ \eqref{effQ4CGNR}/\eqref{barGNRpv} is merely collection of
the contraction coefficients of the A4Ss (cf. Eq. \eqref{LCA4S}). That is, $\bar{\mathbf{G}}^{\mathrm{NR}}$ \eqref{effQ4CGNR}/\eqref{barGNRpv}
is a true analog of the nonrelativistic Coulomb interaction, whereas $\tilde{\mathbf{G}}^{\mathrm{NR}}$ \eqref{X2CGNR} represents a
renormalization of such interaction, which has significant impact on the energy levels of heavy elements.
By virtue of the identity (cf. Eq \eqref{NESC2eNR}),
\begin{align}
\mathbf{G}^{\mathrm{NESC}}[\mathbf{D}^0]=\mathbf{L}_+^{\mathrm{NESC,2e}}[\mathbf{D}^0] -\mathbf{G}^{\mathrm{NR}}[\sum_A^\oplus\mathbf{D}^{LL}_A],\label{GNESCnew}
\end{align}
$\tilde{\mathbf{G}}^{\mathrm{NESC}}_{\mathrm{eff}}$ in Eq. \eqref{effX2C1e} can be written as
\begin{align}
\tilde{\mathbf{G}}^{\mathrm{NESC}}_{\mathrm{eff}}&=\mathbf{R}_+^\dag\mathbf{L}_+^{\mathrm{NESC,2e}}[\mathbf{D}^0]\mathbf{R}_+ -\mathbf{R}_+^\dag\mathbf{G}^{\mathrm{NR}}[\sum_A^\oplus\mathbf{D}^{LL}_A]\mathbf{R}_+\\
&=\bar{\mathbf{G}}^{\mathrm{NESC}}_{\mathrm{eff}}-\Delta \mathbf{G}^{\mathrm{NESC}},
\end{align}
where
\begin{align}
\bar{\mathbf{G}}^{\mathrm{NESC}}_{\mathrm{eff}}&=\mathbf{R}_+^\dag\mathbf{L}_+^{\mathrm{NESC,2e}}[\mathbf{D}^0]\mathbf{R}_+
-\mathbf{G}^{\mathrm{NR}}[\sum_A^\oplus\tilde{\mathbf{D}}_A]),\label{barGNESC}\\
\Delta\mathbf{G}^{\mathrm{NESC}}&=\mathbf{R}_+^\dag\mathbf{G}^{\mathrm{NR}}[\sum_A^\oplus\mathbf{D}^{LL}_A]\mathbf{R}_+
-\mathbf{G}^{\mathrm{NR}}[\sum_A^\oplus\tilde{\mathbf{D}}_A]\\
&=\mathbf{R}_+^\dag\mathbf{G}^{\mathrm{NR}}[\mathbf{D}^{LL}]\mathbf{R}_+ -\mathbf{G}^{\mathrm{NR}}[\tilde{\mathbf{D}}] \nonumber\\
&-\mathbf{R}_+^\dag(\mathbf{G}^{\mathrm{NR}}[\mathbf{D}^{LL}]-\mathbf{G}^{\mathrm{NR}}[\sum_A^\oplus\mathbf{D}^{LL}_A])\mathbf{R}_+
+(\mathbf{G}^{\mathrm{NR}}[\tilde{\mathbf{D}}]-\mathbf{G}^{\mathrm{NR}}[\sum_A^\oplus\tilde{\mathbf{D}}_A]) \\
&=\tilde{\mathbf{G}}^{\mathrm{NR}}[\tilde{\mathbf{D}}]-\mathbf{G}^{\mathrm{NR}}[\tilde{\mathbf{D}}]-\Delta\Delta\mathbf{G}^{\mathrm{NR}},\label{Extra2ePCE}\\
\Delta\Delta\mathbf{G}^{\mathrm{NR}}&=\mathbf{R}_+^\dag\mathbf{G}^{\mathrm{NR}}[\Delta\mathbf{D}^{LL}]\mathbf{R}_+
-\mathbf{G}^{\mathrm{NR}}[\Delta\tilde{\mathbf{D}}],\label{DD2ePCE}\\
\Delta{\mathbf{D}}^{LL}&=\mathbf{D}^{LL}-\sum_A^\oplus\mathbf{D}_A^{LL},\quad \Delta\tilde{\mathbf{D}}=\tilde{\mathbf{D}}-\sum_A^\oplus\tilde{\mathbf{D}}_A.\label{Diffden}
\end{align}
Since both $\Delta{\mathbf{D}}^{LL}$ and $\Delta\tilde{\mathbf{D}}$
characterize the deformation and polarization when going from free atoms to the molecule (due to bonding interactions between valence shells),
the two terms of Eq. \eqref{DD2ePCE} essentially cancel each other, thereby leading to
\begin{align}
\Delta\mathbf{G}^{\mathrm{NESC}}\approx \tilde{\mathbf{G}}^{\mathrm{NR}}[\tilde{\mathbf{D}}]-\mathbf{G}^{\mathrm{NR}}[\tilde{\mathbf{D}}].\label{Extra2ePCE-b}
\end{align}
Eq. \eqref{effX2C} can hence be reduced to
\begin{align}
\bar{\mathbf{F}}_+^{\mathrm{X2C}}&=\bar{\mathbf{h}}_{\mathrm{eff}}^{\mathrm{NESC}}+\mathbf{G}^{\mathrm{NR}}[\tilde{\mathbf{D}}],\label{pmfX2C}\\
\bar{\mathbf{h}}_{\mathrm{eff}}^{\mathrm{NESC}}&=\mathbf{R}_+^\dag\mathbf{L}_{+}^{\mathrm{NESC,1e}}\mathbf{R}_+ +\bar{\mathbf{G}}^{\mathrm{NESC}}_{\mathrm{eff}},\label{pmfX2C1e}
\end{align}
which amounts to shifting $\Delta\mathbf{G}^{\mathrm{NESC}}$ \eqref{Extra2ePCE-b} from $\tilde{\mathbf{G}}^{\mathrm{NR}}$ \eqref{GNRrenormalized}
to $\tilde{\mathbf{G}}^{\mathrm{NESC}}_{\mathrm{eff}}$ \eqref{effX2C1e}. That is, $\Delta\mathbf{G}^{\mathrm{NESC}}$ \eqref{Extra2ePCE-b} behaves
as an additional correction of the 2e-PCEs.
Eq. \eqref{pmfX2C}, in conjunction with the pmfX derived from Eq. \eqref{ModF},
has been dubbed eamfX2C\cite{2ePCE2022} (or preferably pmfX2C), where the expressions \eqref{pmfX2C} and \eqref{pmfX2C1e} arise naturally by
regarding Eq. \eqref{barGNESC} as the pmf approximation to the true term $\mathrm{R}_+^\dag\mathbf{L}_+^{\mathrm{NESC,2e}}[\mathbf{D}]\mathrm{R}_+
-\mathbf{G}^{\mathrm{NR}}[\tilde{\mathbf{D}}]$.
Note that any approximation to the decoupling matrix $\mathbf{X}$ can be used in $\bar{\mathbf{F}}_+^{\mathrm{X2C}}$ \eqref{pmfX2C}, although
the pmfX appears to be a natural choice, for $\bar{\mathbf{G}}^{\mathrm{NESC}}_{\mathrm{eff}}$ \eqref{barGNESC} can be obtained for free in this case.

Similarly, if the SESC-based $\mathbf{F}_+^{\mathrm{X2C}}$ \eqref{hplus} is to be adopted, we will have under the MDM approximation
\begin{align}
\tilde{\mathbf{F}}_+^{\mathrm{X2C}}&=\tilde{\mathbf{h}}_{\mathrm{eff}}^{\mathrm{SESC}}+\tilde{\mathbf{g}}^{\mathrm{NR}}[\tilde{\mathbf{D}}],\label{effSESC}\\
\tilde{\mathbf{h}}_{\mathrm{eff}}^{\mathrm{SESC}}&=\mathbf{R}_+^\dag\mathbf{L}_{+}^{\mathrm{SESC,1e}}\mathbf{R}_+
+\tilde{\mathbf{G}}^{\mathrm{SESC}}_{\mathrm{eff}},\quad \tilde{\mathbf{G}}^{\mathrm{SESC}}_{\mathrm{eff}}=\mathbf{R}_+^\dag\mathbf{G}^{\mathrm{SESC}}[\mathbf{D}^0]\mathbf{R}_+,\label{effSESC1e}\\
\tilde{g}^{\mathrm{NR}}_{\mu\nu}[\tilde{\mathbf{D}}]&=\frac{1}{2}\{[(\bar{\Omega}_{\mu\nu}^{LL}|g_{12}|\tr_2[\tilde{\mathbf{D}}_{\lambda\kappa}\tilde{\Omega}^{LL}_{\kappa\lambda}])
-(\bar{\Omega}_{\mu\lambda}^{LL}|g_{12}|\tilde{\mathbf{D}}_{\lambda\kappa}\tilde{\Omega}^{LL}_{\kappa\nu})]+\mathrm{c.c.}\}\\
&=\frac{1}{2}\{(\mathbf{R}_+^\dag \tilde{\mathrm{M}}_+(\mathbf{M}^{LL})^{-1}\mathbf{G}^{\mathrm{NR}}[\mathbf{D}^{LL}]\mathbf{R}_+)_{\mu\nu}+\mathrm{c.c.}\},\label{effSESC2e}
\end{align}
whereas $\mathbf{G}^{\mathrm{SESC}}_{\mathrm{eff}}[\mathbf{D}^0]$ collects all small-component-containing terms
of $\mathbf{L}_+^{\mathrm{SESC,2e}}[\mathbf{D}^0]$ \eqref{SESC2e}. Following the same procedure going from Eq. \eqref{GNESCnew} to Eq. \eqref{Extra2ePCE-b},
Eq. \eqref{effSESC} can be reduced to
\begin{align}
\bar{\mathbf{F}}_+^{\mathrm{X2C}}&=\bar{\mathbf{h}}_{\mathrm{eff}}^{\mathrm{SESC}}+\mathbf{G}^{\mathrm{NR}}[\tilde{\mathbf{D}}],\label{pmfSESC}\\
\bar{\mathbf{h}}_{\mathrm{eff}}^{\mathrm{SESC}}&=\mathbf{R}_+^\dag\mathbf{L}_{+}^{\mathrm{SESC,1e}}\mathbf{R}_+
+\bar{\mathbf{G}}^{\mathrm{SESC}}_{\mathrm{eff}},\label{pmfSESC1e}\\
\bar{\mathbf{G}}^{\mathrm{SESC}}_{\mathrm{eff}}&=\mathbf{R}_+^\dag\mathbf{L}_+^{\mathrm{SESC,2e}}[\mathbf{D}^0]\mathbf{R}_+
-\mathbf{G}^{\mathrm{NR}}[\sum_A^\oplus\tilde{\mathbf{D}}_A]),\label{barGSESC}
\end{align}
which is the SESC variant of pmfX2C.

It has been attempted to replace the eamf/pmf-2ePCE correction $\bar{\mathbf{G}}^{\mathrm{NESC}}_{\mathrm{eff}}$ \eqref{barGNESC} in Eq. \eqref{pmfX2C1e}  with
the superposition of the atomic ones, viz.,
\begin{align}
\bar{\mathbf{G}}^{\mathrm{NESC}}_{\mathrm{eff}}=\sum_A^\oplus \bar{\mathbf{G}}^{\mathrm{NESC}}_{\mathrm{eff},A},
\quad \bar{\mathbf{G}}^{\mathrm{NESC}}_{\mathrm{eff},A}= \mathrm{R}_{+,A}^\dag\mathbf{L}_{+,A}^{\mathrm{NESC,2e}}[\mathbf{D}_A]\mathrm{R}_{+,A}-\mathbf{G}^{\mathrm{NR}}_A[\tilde{\mathbf{D}}_A],
\end{align}
so as to reduce eamfX2C/pmfX2C to amfX2C\cite{2ePCE2022}.
However, even when combined with a proper approximation (e.g., 1eX) to the decoupling matrix $\mathbf{X}$
(in lieu of the original amfX\cite{2ePCE2022} derived from the ill-behaved Hamiltonian \eqref{amfF}),
this amfX2C has sizeable errors (up to 5 $m\mathrm{E}_h$) for the energy levels of frontier spinors of polar systems (e.g., HI)
with extended basis sets\cite{ChengNote}, a point that was not observed in Ref. \citenum{2ePCE2022} due to the use of low-quality basis sets.
The reason should be ascribed to the long-range nature of electrostatic electron-electron interactions.
In contrast, when $\bar{\mathbf{G}}^{\mathrm{NESC}}_{\mathrm{eff},A}$ is composed only of one-center
two-electron spin-orbit (2eSO) integrals (contracted with the four-component atomic density matrix $\mathbf{D}_A$),
the resulting SOX2CAMF approach\cite{X2CAMF2018} turns out to be very successful.
It appears that two-electron scalar relativistic (2eSC) PCEs are either ignored completely or accounted for by the pmf scheme.
For the same reason, the scalar Gaunt/Breit integrals should also be removed from SOX2CAMF that incorporates the full atomic mean-field Gaunt/Breit
integrals\cite{X2CAMF2022}.

It has been a common practice to take the spin-free part of $\bar{\mathbf{F}}_+^{\mathrm{X2C}}$ \eqref{pmfX2C}, in conjunction with $\bar{\mathbf{G}}^{\mathrm{NESC}}_{\mathrm{eff}}=0$,
as the Hamiltonian in scalar relativistic mean-field calculations.
Since the spin-free part of $\bar{\mathbf{G}}^{\mathrm{NESC}}_{\mathrm{eff}}$ \eqref{barGNESC}
can readily be obtained, it is strongly recommended to correct the 2eSC-PCEs even in scalar relativistic calculations, so as to achieve a
balanced description of core and valence electrons.

At this stage, some general remarks should be made.
\begin{enumerate}[(1)]
\item Although precisely the same in structure, Q4C\cite{Q4CX2C,LiuMP,X2CBook2017} is much simpler than eamfX2C/pmfX2C\cite{2ePCE2022}
(defined by Eqs. \eqref{ModF}, \eqref{barGNESC}, and \eqref{pmfX2C}),
especially for response properties (due to the presence of pmfX and renormalization $\mathbf{R}_+$ in eamfX2C; see Ref. \citenum{X2Cgrd2020}).
Moreover, unlike pmfX2C, Q4C does not suffer from 2cPCEs even in the fluctuation potential governing electron correlation.
It is hence unclear why eamX2C, instead of Q4C, was characterized
as a `fundamental milestone toward a universal and reliable relativistic
two-component quantum-chemical approach'\cite{2ePCE2022}.

\item Since $\tilde{\mathbf{G}}^{\mathrm{NR}}[\tilde{\mathbf{D}}]$ has little computational overhead compared to
$\mathbf{G}^{\mathrm{NR}}[\tilde{\mathbf{D}}]$, $\tilde{\mathbf{F}}_+^{\mathrm{X2C}}$ \eqref{effX2C}/\eqref{effSESC}, in conjunction with
any acceptable approximation to the decoupling matrix $\mathbf{X}$, is also a valuable variant of X2C. Conceptually,
it is even preferred over $\bar{\mathbf{F}}_+^{\mathrm{X2C}}$ \eqref{pmfX2C}/\eqref{pmfSESC}, for it is
the only formulation that is consistent with the renormalization procedure
when going from pmfDHF \eqref{pmf4C} to pmfX2C. In particular, the difference between $\tilde{\mathbf{F}}_+^{\mathrm{X2C}}$
and $\bar{\mathbf{F}}_+^{\mathrm{X2C}}$, i.e., $\Delta\Delta\mathbf{G}^{\mathrm{NR}}$ \eqref{DD2ePCE},
has discernible effects on the innermost shells of heavy elements and is hence relevant for properties (e.g., contact densities and shifts\cite{2ePCE2022})
that are very sensitive to such electronic shells.

\item SESC is much simpler than NESC. As can be seen from Eq. \eqref{FullSESC2e},
SESC does not require the very expensive two-electron term $\tilde{\mathbf{G}}^{SS}$ \eqref{4CGSS2C} that
enters Eq. \eqref{FullX2C2e} for NESC.
Even the one-electron term $\mathbf{L}_+^{\mathrm{SESC,1e}}$ \eqref{SESC1e} of SESC is much simpler than that [Eq. \eqref{NESC1e}] of NESC.
In particular, for a pure density functional, $\mathbf{R}_+^\dag \mathbf{L}_+^{\mathrm{SESC,1e}}\mathbf{R}_+$ would be the full SESC-KS Hamiltonian, which
does not have an explicit spin-orbit operator (since $\mathbf{L}_+^{\mathrm{UESC,1e}}=\mathbf{T}+\mathbf{V}_{\mathrm{KS}}[\rho]\mathbf{X}$ in $\mathbf{L}_+^{\mathrm{SESC,1e}}$), but does not miss any spin-orbit couplings described by the given functional\cite{Q4CX2C}. However, when
the pmfX is employed, the simplicity of SESC in the form of Eq. \eqref{effSESC}/\eqref{pmfSESC} over NESC \eqref{effX2C}/\eqref{pmfX2C} is lost, for the expensive
$\mathbf{G}^{\mathrm{NESC}}[\sum_A^\oplus \mathbf{D}_A]$ required for assembling $\tilde{\mathbf{G}}^{\mathrm{NESC}}_{\mathrm{eff}}$/$\bar{\mathbf{G}}^{\mathrm{NESC}}_{\mathrm{eff}}$ is already available in this case. Nonetheless, SESC is still much simpler than NESC if the MDM approximation is not employed.
\item The 1CSC approximation of the ERIs [cf. Eqs. \eqref{SSERI} to \eqref{LSLS1c}] can also be
applied to  $\mathbf{F}^{\mathrm{Frag}}$ \eqref{FragF},
$\mathbf{F}_+^{\mathrm{Q4C}}$ \eqref{effQ4C}/\eqref{Q4Ccross}, $\mathbf{F}_+^{\mathrm{X2C}}$ \eqref{hplus}, $\mathbf{F}^{\mathrm{pmf}}$ \eqref{ModF}, $\tilde{\mathbf{F}}_+^{\mathrm{X2C}}$ \eqref{effX2C}/\eqref{effSESC}, and $\bar{\mathbf{F}}_+^{\mathrm{X2C}}$ \eqref{pmfX2C}/\eqref{pmfSESC}, so as to render them share exactly the same ERIs as the parent four-component Fock matrix $\mathbf{F}$ \eqref{DEQMat}.

\item Although only defined algebraically, it is still possible to separate the various forms of the X2C Hamiltonian into spin-free
and spin-dependent terms. As a matter of fact, taking the spin-free part of X2C (sf-X2C) as the zeroth order,
a family of spin-dependent operators of finite orders in spin-orbit (SO) interaction can be obtained by means of matrix perturbation theory\cite{X2CSOC1,X2CSOC2}
or derivative technique\cite{ChengSOCop2014,ChengSOCop2020} (see Ref. \citenum{X2CSOCBook2017} for a comprehensive review). Among these, the Douglas-Kroll-Hess type of SO operators (so-DKHn)\cite{X2CSOC1,X2CSOC2} are particularly simple and variationally stable (so-DKH1 is even computationally the same as the Breit-Pauli spin-orbit
operator).

\item As a conceptual point, it should be pointed that, while the decoupling matrix $\mathbf{X}$ can
in principle be obtained within the two-component
framework\cite{X2C2005,X2C2007kutz}, it is the borrowing of easily accessible four-component information that renders X2C really effective.
The same applies also to P4C and Q4C.
\item If wanted, the unoccupied NESs can be obtained by a single diagonalization of Eq. \eqref{NESeigen} upon convergence of the PESs,
\begin{align}
\mathbf{F}_{-}^{\mathrm{X2C}}\tilde{\mathbf{C}}_{-}&=\mathbf{M}^{SS}\tilde{\mathbf{C}}_{-} \mathbf{E}_-,\label{NESeigen}\\
\mathbf{F}_{-}^{\mathrm{X2C}}&=(\mathbf{U}^\dag\mathbf{F}\mathbf{U})^{SS}=\sum_{X,Y=L,S}(\mathbf{U}^\dag)^{SX}\mathbf{F}^{XY}\mathbf{U}^{YS}\nonumber\\
&=\mathbf{R}_{-}^{\dagger}\mathbf{L}^{\mathrm{X}}_{-}\mathbf{R}_{-},\quad\mathrm{X}=\mathrm{NESC, SESC},\label{hminus}\\
\mathbf{L}^{\mathrm{NESC}}_{-}&=\mathbf{F}^{SS}+\mathbf{F}^{SL}\tilde{\mathbf{X}}+
\tilde{\mathbf{X}}^{\dagger}\mathbf{F}^{LS}+\tilde{\mathbf{X}}^{\dagger}\mathbf{F}^{LL}\tilde{\mathbf{X}},\label{NESCminus}\\
\mathbf{L}^{\mathrm{SESC}}_{-}&=\frac{1}{2}[\tilde{\mathbf{M}}_{-}(\mathbf{M}^{SS})^{-1}\mathbf{L}^{\mathrm{UESC}}_{-}+\mathrm{c.c.}],\\
\mathbf{L}^{\mathrm{UESC}}_{-}&=\mathbf{F}^{SS}+\mathbf{F}^{SL}\tilde{\mathbf{X}},\label{SESCminus}\\
\mathbf{F}^{XY}&=\mathbf{h}^{XY}+\mathbf{G}^{XY}[\mathbf{D}].
\end{align}
This is a cheap step, for the relevant two-electron ERIs $G^{XY}_{\mu\nu}[\mathbf{D}]$ are already available.
Here, the four-component density matrix $\mathbf{D}$ is back-transformed from
the converged two-component density matrix $\tilde{\mathbf{D}}$ (cf. Eq. \eqref{DXY2D}).
Note that the same equation \eqref{NESeigen} can also be employed to obtain the NESs of Q4C.
It is just that the four-component density matrix is $\bar{\mathbf{D}}$, back-transformed from the Q4C density matrix
$\bar{\mathbf{P}}$ (cf. Eq. \eqref{P4Cden}).
Alternatively, the four-component DHF equation \eqref{DEQMat} can be assembled upon convergence of the X2C/Q4C calculation
and is then diagonalized to obtain the PESs and NESs simultaneously.

\item The quaternion form of X2C can be obtained by the $\mbox{}^{\mathrm{Q}}\mathbf{U}$ transformation \eqref{QUmatC}
of $\mathbf{F}^{\mathrm{X2C}}_+$/$\bar{\mathbf{F}}^{\mathrm{X2C}}_+$. The X4C variant of KROHF
can then readily be formulated (see Appendix  \ref{AppendixA}).

\item Since the leading quantum electrodynamics effect (i.e., vacuum polarization and electron self-energy) can be described by
an effective, HF-like one-body potential \cite{eQED,PhysRep} that can be added to
the one-body Dirac operator \eqref{FockOperator}, it is readily accounted for in DHF, Q4C or X2C calculations.
The situation is particularly simple when the eQED potential is fitted into a model spectral form\cite{ShabaevModelSE,ShabaevModelSEcode2018}.
For a deep understanding of the fundamentals of QED, see comprehensive reviews\cite{X2C2016,eQEDBook2017,Essential2020,CommentQED}.
\end{enumerate}

\section{Many-electron Relativistic Hamiltonians}\label{SecMB}
Having determined the PESs (and NESs) by DHF, Q4C or X2C, a generic normal-ordered many-electron Hamiltonian
can be written down
\begin{eqnarray}
H_n=H-\langle 0|H|0\rangle=F^{\mathrm{X}}_{pq}\{a^p_q\}_n+\frac{1}{2}g_{pq}^{rs}\{a^{pq}_{rs}\}_n,\quad \mathrm{X=DHF, Q4C, X2C, eQED},\label{MB-Hamiltonian}
\end{eqnarray}
for post-HF calculations. Here, $|0\rangle$ is a reference state, with respect to which the normal ordering
of the one-body ($a^p_q=a_p^\dag a_q$) and two-body ($a^{pq}_{rs}=a_p^\dag a_q^\dag a_s a_r$) excitation operators has been made.
Like the 4C case, the two-body operator of Q4C incorporates automatically the full Coulomb-Gaunt/Breit interaction \eqref{V12full}.
Since the large and small components of the spinor basis functions in Q4C share
the same expansion coefficients, the AO ERIs can be transformed
to the MO representation as a whole, instead of component-wise as in the 4C case.
When the AO ERIs refer to those of the renormalized basis functions (cf. Eq. \eqref{RenormBasis}),
the two-body operator of X2C will also incorporate the full Coulomb-Gaunt/Breit interaction, which is equivalent to
working with the M4Ss recovered from the X2C spinors (cf. Eq. \eqref{2C-C24C-C}). As such,
the index transformation of the two-body operator of X2C is as expensive as that of 4C and more expensive than that of Q4C.
The computation of X2C is simplified greatly by using only the untransformed two-body operator (i.e., $\{g_{pq}^{rs}\}$
are just the ERIs of the Coulomb interaction over the X2C spinors). However, X2C will then suffer from 2ePCEs in
electron-electron interaction, which are particularly prominent for deep-core properties
of heavy elements\cite{2ePCE2022,Many-body2ePCE}. Not only so, the lack of genuine 2e-SO interaction
will fail to describe SO splittings between states differing by two electrons. The simplest remedy\cite{OneCenterSOint2000,AccuracyOfmfSO} of this failure is
to include the one-center 2e-SO AO ERIs (resulting from spin separation of
the transformed X2C two-body operator\cite{X2CSOC1,X2CSOC2}) and transform them only to the active orbitals spanning the target states.
Beyond these no-pair relativistic Hamiltonians is the effective QED Hamiltonian\cite{eQED,PhysRep}, where the normal ordering is taken with respective
to the filled Dirac sea but meanwhile incorporating charge conjugation symmetry\cite{Essential2020}. This
is the only correct and complete QED Hamiltonian\cite{CommentQED} in the same form of Eq. \eqref{MB-Hamiltonian}. It is just
that the one-body operator further includes the effective potential describing vacuum polarization and electron self-energy
and that the orbital indices refer to both PESs and NESs. Unlike no-pair correlation,
the full QED correlation energy is independent of the ways of generating
the orbitals (like nonrelativistic full configuration interaction)\cite{PhysRep}, for the \emph{filled} NESs are also correlated therein.
Note that the frequency-dependent Breit interaction must be employed for this purpose.
For possible means of treating relativity, correlation, and QED effects simultaneously, see Ref. \citenum{LiuWires}.

The above spinor-based Hamiltonians are imperative for core properties of heavy atoms or valence properties involving $np$ ($n\ge 5$) orbitals.
However, for chemical systems with moderate spin-orbit couplings, it is more appealing to invoke a two-step Hamiltonian
that treats scalar relativity and spin-orbit coupling separately. The
sf-X2C+so-DKHn variant\cite{X2CSOC1,X2CSOC2} is arguably the best one for this purpose. The great advantage here lies in that
real-valued orbitals can be used, so as to facilitate the treatment of electron correlation, on top of which
spin-orbit couplings can further be added in one way or another. Methodologies and applications along this line
have recently been summarized\cite{SOiCI,SOiCISCF,ChengSOCpt2023} and are hence not repeated here.

%
%

\section{Conclusions and outlook}\label{SecConclusion}
It has been shown that the DHF, Q4C, and X2C equations can be recast into the same mean-field form by
making use of the MDM approximation for the small-component charge/current density functions.
They also share exactly the same relativistic integrals that can be simplified by using the 1CSC approximation.
As such, it is a matter of taste which variant is to be adopted for real-life applications. Nevertheless,
Q4C is manifestly most efficient due to the use of most compact bases and all-together integral transformation for subsequent
treatment of electron correlation. In particular, it has no 2ePCEs even in electron-election interaction
and is simpler than X2C for response properties. QED effects, including not only the one-body vacuum polarization and electron self-energy but also
the two-body correlation of NESs, can readily be accounted for in precision spectroscopic calculations.

\begin{acknowledgments}
This work was supported by the National Natural Science Foundation of China (Grant Nos. 22373057, 21833001, and 21973054) and
Mount Tai Scholar Climbing Project of Shandong Province.
\end{acknowledgments}

\section*{Data Availability Statement}
The data that support the findings of this study are available within the article.

\section*{Conflicts of interest}
There are no conflicts to declare.


\appendix
\section{Quaternion DHF, KUHF, and KROHF}\label{AppendixA}
Although the first (third) and second (fourth) columns of the RKB basis $|\xi_{\mu}\rangle$ \eqref{RKBbasis} are manifestly time-reversal related (cf. Eq. \eqref{Tdef}), it turns out that the overall time-reversal structure of the 4-by-4 matrix \eqref{RKBbasis} can be made more transparent only by the following unitary transformation\cite{SaueMolPhys1997},
\begin{align}
\mathbf{Q}&=\begin{pmatrix}1&0&0&0 \\ 0&0&1&0 \\ 0&1&0&0\\ 0&0&0&1\end{pmatrix}=\mathbf{Q}^{-1}=\mathbf{Q}^\dag,\label{Qdef}\\
\mbox{}^{\mathrm{Q}}\xi_{\mu}&=\mathbf{Q}\xi_{\mu}\mathbf{Q}^\dag
=\begin{pmatrix}\xi^{\alpha}_{\mu}&\bar{\xi}^{\beta}_{\mu}\\
\bar{\xi}^{\alpha}_{\mu}&\xi^{\beta}_{\mu}\end{pmatrix}=\begin{pmatrix}\xi^{\alpha}_{\mu}&\bar{\xi}^{\beta}_{\mu}\\
-\bar{\xi}^{\beta *}_{\mu}&\xi^{\alpha *}_{\mu}\end{pmatrix},\quad \mu\in[1,n],\label{TRS-RKB}\\
\xi^{\alpha}_{\mu}&=\begin{pmatrix}g_{\mu}&0\\ 0&-\frac{\ii}{2c} g^z_{\mu}\end{pmatrix},\quad
\bar{\xi}^{\beta}_{\mu}=\begin{pmatrix}0&0\\ 0&-\frac{\ii}{2c}(g^x_{\mu}-\ii g^y_{\mu})\end{pmatrix}\label{xialpha}.
\end{align}
$\mbox{}^{\mathrm{Q}}\xi_{\mu}$ \eqref{TRS-RKB} is clearly time-reversal symmetric (cf. Eq. \eqref{Ohmat})
and can hence be block-diagonalized by the quaternion unitary transformation $\mbox{}^{\mathrm{Q}}\mathbf{U}$ \eqref{QUmatC}, viz.,
\begin{align}
\mbox{}^{\mathrm{Q}}\mathbf{U} \mbox{}^{\mathrm{Q}}\xi_{\mu}\mbox{}^{\mathrm{Q}}\mathbf{U}^\dag&=\begin{pmatrix}\mbox{}^{\mathrm{q}}\xi_{\mu}&0\\
0&\mbox{}^{\mathrm{q}}\xi_{\mu}\end{pmatrix},\\
\mbox{}^{\mathrm{q}}\xi_{\mu}&=\xi^{\alpha}_{\mu}+\bar{\xi}^{\beta}_{\mu}\check{j}
=\mbox{}^0X_{\mu}+\mathbf{X}_{\mu},\quad \mathbf{X}_{\mu}=\sum_{i=1}^3\mbox{}^iX_{\mu}e_i,\label{qRKB}\\
\mbox{}^0X_{\mu}&=\begin{pmatrix}g_{\mu}&0\\ 0&0\end{pmatrix},\quad \mbox{}^iX_{\mu}=-\frac{1}{2c}\begin{pmatrix}0&0\\ 0&g^i_{\mu}\end{pmatrix},
\end{align}
where the quaternion units $(e_1, e_2, e_3)$  refer to the ordering $(z,y,x)$ instead of the usual ordering $(x,y,z)$ ($=-(z,y,x)$),
see Eq. \eqref{QuaternionDef}. It is said that the 4-by-4 matrix $\mbox{}^{\mathrm{Q}}\xi_{\mu}$ \eqref{TRS-RKB}
is algebraically isomorphic to the 2-by-2 (real) quaternion matrix $\mbox{}^{\mathrm{q}}\xi_{\mu}$ \eqref{qRKB} [cf. Eq. \eqref{RealQuaternionB}].

The $\mathbf{Q}$-transformation \eqref{Qdef} of an M4S $\psi_p$ amounts to swapping its second and third components,
\begin{align}
\mbox{}^{\mathrm{Q}}\psi_p&=\mathbf{Q}\psi_p=\begin{pmatrix}\mbox{}^{\mathrm{Q}}\psi_p^{\alpha}\\ \mbox{}^{\mathrm{Q}}\psi_p^{\beta}\end{pmatrix}
=\mbox{}^{\mathrm{Q}}\xi_{\mu}\mbox{}^{\mathrm{Q}}C_{\mu p},
\quad \mbox{}^{\mathrm{Q}}\psi_p^{\sigma}=\begin{pmatrix}\mbox{}^{\mathrm{Q}}\psi_p^{\sigma L}\\
\mbox{}^{\mathrm{Q}}\psi_p^{\sigma S}\end{pmatrix},\\
\mbox{}^{\mathrm{Q}}C_{\mu p}&=\mathbf{Q} C_{\mu p}=\begin{pmatrix}\mbox{}^{\mathrm{Q}}C^{\alpha}_{\mu p}\\ \mbox{}^{\mathrm{Q}}C^{\beta}_{\mu p}\end{pmatrix},\quad
\mbox{}^{\mathrm{Q}}C_{\mu p}^{\sigma}=\begin{pmatrix}\mbox{}^{\mathrm{Q}}C^{\sigma L}_{\mu p}\\ \mbox{}^{\mathrm{Q}}C^{\sigma S}_{\mu p}\end{pmatrix},
\end{align}
such that the resulting 4-spinor $\mbox{}^{\mathrm{Q}}\psi_p$ is grouped according to spin labels $(\alpha,\beta)$ instead of large and small components $(L,S)$.
The Dirac operator \eqref{hop} is transformed accordingly,
\begin{align}
\mbox{}^{\mathrm{Q}}h&=\mathbf{Q}h\mathbf{Q}^\dag=\begin{pmatrix}\mbox{}^{\mathrm{Q}}h^{\alpha\alpha}&\mbox{}^{\mathrm{Q}}h^{\alpha\beta}\\ \mbox{}^{\mathrm{Q}}h^{\beta\alpha}&\mbox{}^{\mathrm{Q}}h^{\beta\beta}\end{pmatrix}
=\begin{pmatrix}\mbox{}^{\mathrm{Q}}h^{\alpha\alpha}&\mbox{}^{\mathrm{Q}}h^{\alpha\beta}\\ -(\mbox{}^{\mathrm{Q}}h^{\alpha\beta})^*&(\mbox{}^{\mathrm{Q}}h^{\alpha\alpha})^*\end{pmatrix},\label{QhQop}\\
\mbox{}^{\mathrm{Q}}h^{\alpha\alpha}&=\begin{pmatrix}\mbox{}^{\mathrm{Q}}h^{\alpha\alpha, LL}&\mbox{}^{\mathrm{Q}}h^{\alpha\alpha, LS}\\
\mbox{}^{\mathrm{Q}}h^{\alpha\alpha, SL}&\mbox{}^{\mathrm{Q}}h^{\alpha\alpha, SS}\end{pmatrix}
=\begin{pmatrix}V&cp_z \\ cp_z& V-2c^2\end{pmatrix}=(\mbox{}^{\mathrm{Q}}h^{\beta\beta})^*,\\
\mbox{}^{\mathrm{Q}}h^{\alpha\beta}&=\begin{pmatrix}\mbox{}^{\mathrm{Q}}h^{\alpha\beta, LL}&\mbox{}^{\mathrm{Q}}h^{\alpha\beta, LS}\\
\mbox{}^{\mathrm{Q}}h^{\alpha\beta, SL}&\mbox{}^{\mathrm{Q}}h^{\alpha\beta, SS}\end{pmatrix}
=\begin{pmatrix}0&c(p_x-\ii p_y)\\ c(p_x-\ii p_y)&0\end{pmatrix}=-(\mbox{}^{\mathrm{Q}}h^{\beta\alpha})^*,
%
\end{align}
which can also be block-diagonalized by the quaternion unitary transformation \eqref{QUmatC},
\begin{align}
\mbox{}^{\mathrm{Q}}\mathbf{U} \mbox{}^{\mathrm{Q}}h\mbox{}^{\mathrm{Q}}\mathbf{U}^\dag&=\begin{pmatrix}\mbox{}^{\mathrm{q}}h&0\\
0&\mbox{}^{\mathrm{q}}h\end{pmatrix},
\end{align}
where
\begin{align}
\mbox{}^{\mathrm{q}}h&=\mbox{}^{\mathrm{Q}}h^{\alpha\alpha}+\mbox{}^{\mathrm{Q}}h^{\alpha\beta}\check{j}
=\sum_{i=0}^3\mbox{}^ih e_i,\label{qDop}\\
\mbox{}^0h_D&=\begin{pmatrix}V& 0 \\ 0 & V-2c^2\end{pmatrix},\quad
\mbox{}^ih=\begin{pmatrix}0&-c\partial_i\\-c\partial_i&0\end{pmatrix}.
\end{align}
Precisely in the same way, the one-body matrix elements $h_{\mu\nu}$ can be transformed to quaternion form,
\begin{align}
\mbox{}^{\mathrm{Q}}h_{\mu\nu}&=\mathbf{Q}h_{\mu\nu}\mathbf{Q}^\dag=\begin{pmatrix}\mbox{}^{\mathrm{Q}}h^{\alpha\alpha}_{\mu\nu}&\mbox{}^{\mathrm{Q}}h^{\alpha\beta}_{\mu\nu}\\
\mbox{}^{\mathrm{Q}}h^{\beta\alpha}_{\mu\nu}&\mbox{}^{\mathrm{Q}}h^{\beta\beta}_{\mu\nu}\end{pmatrix}
=\begin{pmatrix}\mbox{}^{\mathrm{Q}}h^{\alpha\alpha}_{\mu\nu}&\mbox{}^{\mathrm{Q}}h^{\alpha\beta}_{\mu\nu}\\
-\mbox{}^{\mathrm{Q}}h^{\alpha \beta *}_{\mu\nu}&\mbox{}^{\mathrm{Q}}h^{\alpha\alpha *}_{\mu\nu}\end{pmatrix},\\
\mbox{}^{\mathrm{Q}}\mathbf{U}\mbox{}^{\mathrm{Q}}h_{\mu\nu}\mbox{}^{\mathrm{Q}}\mathbf{U}^\dag&=\begin{pmatrix}\mbox{}^{\mathrm{q}}h_{\mu\nu}&0\\
0&\mbox{}^{\mathrm{q}}h_{\mu\nu}\end{pmatrix},\\
\mbox{}^{\mathrm{q}}h_{\mu\nu}&=
\langle\mbox{}^{\mathrm{q}}\xi_{\mu}|\mbox{}^{\mathrm{q}}h|\mbox{}^{\mathrm{q}}\xi_{\nu}\rangle=
\mbox{}^{\mathrm{Q}}h^{\alpha\alpha}_{\mu\nu}+\mbox{}^{\mathrm{Q}}h^{\alpha\beta}_{\mu\nu}\check{j}=\sum_{i=0}^3\mbox{}^ih_{\mu\nu}e_i,\\
\mbox{}^{\mathrm{Q}}h^{\alpha\alpha}_{\mu\nu}&=\begin{pmatrix}V_{\mu\nu}&T_{\mu\nu}\\
T_{\mu\nu}&\frac{1}{4c^2}\langle g_{\mu}^j|V_{N}|g_{\nu}^j\rangle-T_{\mu\nu}\end{pmatrix}
+\frac{1}{4c^2}\begin{pmatrix}0&0\\
0&\ii \epsilon_{zjk}\langle g_{\mu}^j|V|g_{\nu}^k\rangle\end{pmatrix},\quad j,k\in x, y, z,\\
\mbox{}^{\mathrm{Q}}h^{\alpha\beta}_{\mu\nu}&=\frac{1}{4c^2}\begin{pmatrix}0&0\\
0&(\epsilon_{yjk}+\ii \epsilon_{xjk})\langle g_{\mu}^j|V|g_{\nu}^k\rangle\end{pmatrix},\quad j,k\in x, y, z.
\end{align}
To obtain the quaternion form $\mbox{}^{\mathrm{q}}G_{\mu\nu}[\mbox{}^{\mathrm{q}}\mathbf{D}]$ of
the two-electron terms $G_{\mu\nu}[\mathbf{D}]$ \eqref{Fmat0},
we first calculate the quaternion form $\mbox{}^{\mathrm{q}}\Omega_{\mu\nu}$ of the overlap charge distribution function $\Omega_{\mu\nu}$
\eqref{4COmegadef}:
\begin{align}
&\mbox{}^{\mathrm{Q}}\Omega_{\mu\nu}=\mathbf{Q}\Omega_{\mu\nu}\mathbf{Q}^\dag
=\begin{pmatrix}\mbox{}^{\mathrm{Q}}\Omega_{\mu\nu}^{\alpha\alpha}&\mbox{}^{\mathrm{Q}}\Omega_{\mu\nu}^{\alpha\beta}\\
\mbox{}^{\mathrm{Q}}\Omega_{\mu\nu}^{\beta\alpha}&\mbox{}^{\mathrm{Q}}\Omega_{\mu\nu}^{\beta\beta}\end{pmatrix}
=\begin{pmatrix}\mbox{}^{\mathrm{Q}}\Omega_{\mu\nu}^{\alpha\alpha}&\mbox{}^{\mathrm{Q}}\Omega_{\mu\nu}^{\alpha\beta}\\
-(\mbox{}^{\mathrm{Q}}\Omega_{\mu\nu}^{\alpha\beta})^*&(\mbox{}^{\mathrm{Q}}\Omega_{\mu\nu}^{\alpha\alpha})^*\end{pmatrix},\\
&\mbox{}^{\mathrm{Q}}\mathbf{U} \mbox{}^{\mathrm{Q}}\Omega_{\mu\nu}\mbox{}^{\mathrm{Q}}\mathbf{U}^\dag =\begin{pmatrix}\mbox{}^{\mathrm{q}}\Omega_{\mu\nu}&0\\
0&\mbox{}^{\mathrm{q}}\Omega_{\mu\nu}\end{pmatrix},
\end{align}
where
\begin{align}
\mbox{}^{\mathrm{q}}\Omega_{\mu\nu}&
=\mbox{}^{\mathrm{Q}}\Omega_{\mu\nu}^{\alpha\alpha}+\mbox{}^{\mathrm{Q}}\Omega_{\mu\nu}^{\alpha\beta}\check{j}=(\mbox{}^{\mathrm{q}}\xi_{\mu})^\dag \mbox{}^{\mathrm{q}}\xi_{\nu}=\sum_{i=0}^3\mbox{}^i\Omega_{\mu\nu}e_i,\label{qOmega}\\
\mbox{}^{\mathrm{Q}}\Omega_{\mu\nu}^{\alpha\alpha}&=\begin{pmatrix}g_{\mu}g_{\nu}&0\\ 0& \frac{1}{4c^2}g^j_\mu g^j_{\nu}\end{pmatrix}
+\frac{1}{4c^2}\begin{pmatrix}0&0\\ 0& \ii \epsilon_{zjk}g^j_\mu g^k_{\nu}\end{pmatrix},\quad j,k\in x, y, z,\\
\mbox{}^{\mathrm{Q}}\Omega_{\mu\nu}^{\alpha\beta}&
=\frac{1}{4c^2}\begin{pmatrix}0&0\\ 0& (\epsilon_{yjk}+\ii \epsilon_{xjk})g^j_\mu g^k_{\nu}\end{pmatrix},\quad j,k\in x, y, z.
\end{align}
Likewise, the complex quaternion form $\mbox{}^{\mathrm{q}}\Xi^{i}_{\mu\nu}$ (cf. Eq. \eqref{ComplexQuaternion}) of the time-reversal antisymmetric
overlap current distribution vectors $\Xi^i_{\mu\nu}$ \eqref{4CGammadef} can be obtained
by block-diagonalizing $\mbox{}^{\mathrm{Q}}\Xi^i_{\mu\nu}$,
\begin{align}
&\mbox{}^{\mathrm{Q}}\Xi^i_{\mu\nu}=\mathbf{Q}\Xi^i_{\mu\nu}\mathbf{Q}^\dag=h \begin{pmatrix}\mbox{}^{\mathrm{Q}}\Xi^{i,\alpha\alpha}_{\mu\nu}&\mbox{}^{\mathrm{Q}}\Xi^{i,\alpha\beta}_{\mu\nu}\\
\mbox{}^{\mathrm{Q}}\Xi^{i,\beta\alpha}_{\mu\nu}&\mbox{}^{\mathrm{Q}}\Xi^{i,\beta\beta}_{\mu\nu}\end{pmatrix}
=h \begin{pmatrix}\mbox{}^{\mathrm{Q}}\Xi^{i,\alpha\alpha}_{\mu\nu}&\mbox{}^{\mathrm{Q}}\Xi^{i,\alpha\beta}_{\mu\nu}\\
-(\mbox{}^{\mathrm{Q}}\Xi^{i,\alpha\beta}_{\mu\nu})^*&(\mbox{}^{\mathrm{Q}}\Xi^{i,\alpha\alpha}_{\mu\nu})^*\end{pmatrix},\label{hnotation}\\
&\mbox{}^{\mathrm{Q}}\mathbf{U}\mbox{}^{\mathrm{Q}}\Xi^i_{\mu\nu} \mbox{}^{\mathrm{Q}}\mathbf{U}^\dag=\begin{pmatrix}\mbox{}^{\mathrm{q}}\Xi^{i}_{\mu\nu}&0\\
0&\mbox{}^{\mathrm{q}}\Xi^{i}_{\mu\nu}\end{pmatrix},
\end{align}
where
\begin{align}
\mbox{}^{\mathrm{q}}\Xi^{i}_{\mu\nu}&=
h(\mbox{}^{\mathrm{Q}}\Xi^{i,\alpha\alpha}_{\mu\nu}+\mbox{}^{\mathrm{Q}}\Xi^{i,\alpha\beta}_{\mu\nu}\check{j}),\quad i\in x, y, z,\label{qiVecaa}\\
\mbox{}^{\mathrm{Q}}\Xi^{x,\alpha\alpha}_{\mu\nu}&=\frac{1}{2c}\begin{pmatrix}0&-g_{\mu}g_{\nu}^x-\ii g_{\mu}g_{\nu}^y \\ g_{\mu}^x g_{\nu}-\ii g_{\mu}^yg_{\nu}&0\end{pmatrix}, \\
\mbox{}^{\mathrm{Q}}\Xi^{x,\alpha\beta}_{\mu\nu}&=\frac{1}{2c}\begin{pmatrix}0&g_{\mu}g_{\nu}^z \\ g_{\mu}^z g_{\nu}&0\end{pmatrix},\\
\mbox{}^{\mathrm{Q}}\Xi^{y,\alpha\alpha}_{\mu\nu}&=\frac{1}{2c}\begin{pmatrix}0&-g_{\mu}g_{\nu}^y+\ii g_{\mu}g_{\nu}^x \\ g_{\mu}^y g_{\nu}+\ii g_{\mu}^xg_{\nu}&0\end{pmatrix}, \\
\mbox{}^{\mathrm{Q}}\Xi^{y,\alpha\beta}_{\mu\nu}&=\frac{1}{2c}\begin{pmatrix}0&-\ii g_{\mu}g_{\nu}^z \\ -\ii g_{\mu}^z g_{\nu}&0\end{pmatrix},\\
\mbox{}^{\mathrm{Q}}\Xi^{z,\alpha\alpha}_{\mu\nu}&=\frac{1}{2c}\begin{pmatrix}0&-g_{\mu}g_{\nu}^z \\ g_{\mu}^z g_{\nu}&0\end{pmatrix}, \\
\mbox{}^{\mathrm{Q}}\Xi^{z,\alpha\beta}_{\mu\nu}&=\frac{1}{2c}\begin{pmatrix}0&-g_{\mu}g_{\nu}^x+\ii g_{\mu}g_{\nu}^y \\ -g_{\mu}^x g_{\nu}+\ii g_{\mu}^yg_{\nu}&0\end{pmatrix}.
\end{align}
Note that the symbol $h$ in Eqs. \eqref{hnotation} and \eqref{qiVecaa}
is an alternative notation for the square root of minus one, to emphasize that it commutes with the quaternion units $e_i$ in the context of
biquaternion algebra. For a double check, we recalculate directly $\mbox{}^{\mathrm{q}}\Xi^{i}_{\mu\nu}$ \eqref{qiVecaa}, by noting first that
\begin{align}
Q\widetilde{\alpha}_iQ^\dag&=\widetilde{\sigma}_i\otimes\sigma_x\simeq \mbox{}^{\mathrm{q}}\alpha_i=-h \sigma_xe_i,\quad he_i=e_ih, \quad i\in 1,2,3,
\end{align}
where $(\widetilde{\alpha}_1,\widetilde{\alpha}_2,\widetilde{\alpha}_3)=(\alpha_z,\alpha_y,\alpha_x)$.
We then have (cf. \eqref{qrqr123})
\begin{align}
\mbox{}^{\mathrm{q}}\xi^{\alpha\dag}_{\mu}\mbox{}^{\mathrm{q}}\alpha_i&=-h (\mbox{}^0X_{\mu}-\mathbf{X}_{\mu})\sigma_x e_i\\
&=-h(\mbox{}^iX_{\mu}\sigma_x+\mbox{}^0X_{\mu}\sigma_x e_i-e_j\epsilon_{jki}\mbox{}^kX_{\mu}\sigma_x),\quad i,j,k\in 1,2,3.
\end{align}
Further considering the fact that among the 16 products $\mbox{}^iX_{\mu}\sigma_x\mbox{}^jX_{\nu}$ ($i,j\in [0,3]$),
only $\mbox{}^iX_{\mu}\sigma_x\mbox{}^0X_{\nu}$ and $\mbox{}^0X_{\mu}\sigma_x\mbox{}^iX_{\nu}$ ($i\in[1,3]$) are nonvanishing,
we readily obtain
\begin{align}
\mbox{}^{\mathrm{q}}\Xi^{i,\alpha\alpha}_{\mu\nu}&=\mbox{}^{\mathrm{q}}\xi^{\alpha\dag}_{\mu}\mbox{}^{\mathrm{q}}\alpha_i\mbox{}^{\mathrm{q}}\xi^{\alpha}_{\nu}
=\sum_{j=0}^3\mbox{}^j\Xi^i_{\mu\nu}e_j,\label{qXiAll}\\
\mbox{}^0\Xi^{i,\alpha\alpha}_{\mu\nu}&=-h (\mbox{}^iX_{\mu}\sigma_x\mbox{}^0X_{\nu}-\mbox{}^0X_{\mu}\sigma_x\mbox{}^iX_{\nu}),\quad i\in 1,2,3\\
&=\frac{h}{2c}\begin{pmatrix}0&-g_{\mu} g^i_{\nu}\\ g^i_{\mu}g_{\nu}&0\end{pmatrix},\label{0Xi-i}\\
\mbox{}^j\Xi^{i,\alpha\alpha}_{\mu\nu}&=h \epsilon_{jki}(\mbox{}^kX_{\mu}\sigma_x\mbox{}^0X_{\nu}+\mbox{}^0X_{\mu}\sigma_x\mbox{}^kX_{\nu}),\quad i,j,k\in 1,2,3\\
&=\frac{h}{2c}\epsilon_{jki}\begin{pmatrix}0&-g_{\mu} g^k_{\nu}\\ - g^k_{\mu} g_{\nu}&0\end{pmatrix},\quad j\ne i.\label{jXi-i}
\end{align}
It can readily be verified that Eq. \eqref{qXiAll} agrees with Eq. \eqref{qiVecaa}.

In contrast, the density matrix $\mathbf{D}$ \eqref{AO4cDmat} is generally neither time-reversal symmetric nor antisymmetric, but
can be decomposed into a sum of time-reversal symmetric ($\tilde{\mathbf{D}}$)
and antisymmetric ($\ii \bar{\mathbf{D}}$) components, viz.,
\begin{align}
\mbox{}^{\mathrm{Q}}\mathbf{D}&=\mathbf{Q}\mathbf{D}\mathbf{Q}^\dag=\mbox{}^{\mathrm{Q}}\mathbf{C}\mathbf{n}\mbox{}^{\mathrm{Q}}\mathbf{C}^\dag
=\begin{pmatrix}\mbox{}^{\mathrm{Q}}\mathbf{D}^{\alpha\alpha}&\mbox{}^{\mathrm{Q}}\mathbf{D}^{\alpha\beta}\\
\mbox{}^{\mathrm{Q}}\mathbf{D}^{\beta\alpha}&\mbox{}^{\mathrm{Q}}\mathbf{D}^{\beta\beta}\end{pmatrix}
=\mbox{}^{\mathrm{Q}}\tilde{\mathbf{D}}+\ii \mbox{}^{\mathrm{Q}}\bar{\mathbf{D}},\\
\mbox{}^{\mathrm{Q}}\tilde{\mathbf{D}}&=\begin{pmatrix}\mbox{}^{\mathrm{Q}}\tilde{\mathbf{D}}^{\alpha\alpha}&\mbox{}^{\mathrm{Q}}\tilde{\mathbf{D}}^{\alpha\beta}\\
\mbox{}^{\mathrm{Q}}\tilde{\mathbf{D}}^{\beta\alpha}&\mbox{}^{\mathrm{Q}}\tilde{\mathbf{D}}^{\beta\beta}\end{pmatrix}
=\begin{pmatrix}\mbox{}^{\mathrm{Q}}\tilde{\mathbf{D}}^{\alpha\alpha}&\mbox{}^{\mathrm{Q}}\tilde{\mathbf{D}}^{\alpha\beta}\\
-\mbox{}^{\mathrm{Q}}\tilde{\mathbf{D}}^{\alpha\beta *}&\mbox{}^{\mathrm{Q}}\tilde{\mathbf{D}}^{\alpha\alpha *}\end{pmatrix},\\
\mbox{}^{\mathrm{Q}}\bar{\mathbf{D}}&=\begin{pmatrix}\mbox{}^{\mathrm{Q}}\bar{\mathbf{D}}^{\alpha\alpha}&\mbox{}^{\mathrm{Q}}\bar{\mathbf{D}}^{\alpha\beta}\\
\mbox{}^{\mathrm{Q}}\bar{\mathbf{D}}^{\beta\alpha}&\mbox{}^{\mathrm{Q}}\bar{\mathbf{D}}^{\beta\beta}\end{pmatrix}
=\begin{pmatrix}\mbox{}^{\mathrm{Q}}\bar{\mathbf{D}}^{\alpha\alpha}&\mbox{}^{\mathrm{Q}}\bar{\mathbf{D}}^{\alpha\beta}\\
-\mbox{}^{\mathrm{Q}}\bar{\mathbf{D}}^{\alpha\beta *}&\mbox{}^{\mathrm{Q}}\bar{\mathbf{D}}^{\alpha\alpha *}\end{pmatrix},
\end{align}
such that
\begin{align}
\mbox{}^{\mathrm{Q}}\mathbf{U}\mbox{}^{\mathrm{Q}}\mathbf{D}\mbox{}^{\mathrm{Q}}\mathbf{U}^\dag=
\begin{pmatrix}\mbox{}^{\mathrm{q}}\mathbf{D}&\mathbf{0}\\
\mathbf{0}&\mbox{}^{\mathrm{q}}\mathbf{D}\end{pmatrix}=
\begin{pmatrix}\mbox{}^{\mathrm{q}}\tilde{\mathbf{D}}+h \mbox{}^{\mathrm{q}}\bar{\mathbf{D}}&\mathbf{0}\\
\mathbf{0}&\mbox{}^{\mathrm{q}}\tilde{\mathbf{D}}+h\mbox{}^{\mathrm{q}}\bar{\mathbf{D}}\end{pmatrix}.\label{qDENmat}
\end{align}
Specific expressions for $\mbox{}^{\mathrm{q}}\tilde{\mathbf{D}}$ and $\mbox{}^{\mathrm{q}}\bar{\mathbf{D}}$
can be read out from Eq. \eqref{SymmAsymm} with $\mbox{}^{\mathrm{Q}}\mathbf{D}$ for $\mathbf{M}$.

In terms of the above quantities, the $\mbox{}^{\mathrm{Q}}\mathbf{U}$ transformation \eqref{QUmatC} of
the $\mathbf{Q}$-transformed DHF equation \eqref{GHF},
\begin{align}
\mbox{}^{\mathrm{Q}}\mathbf{F}\mbox{}^{\mathrm{Q}}\mathbf{C}_p&=\mbox{}^{\mathrm{Q}}\mathbf{M}\mbox{}^{\mathrm{Q}}\mathbf{C}_p\epsilon_p\Leftrightarrow\\
\begin{pmatrix}\mbox{}^{\mathrm{Q}}\mathbf{F}^{\alpha\alpha}&\mbox{}^{\mathrm{Q}}\mathbf{F}^{\alpha\beta}\\ \mbox{}^{\mathrm{Q}}\mathbf{F}^{\beta\alpha}&\mbox{}^{\mathrm{Q}}\mathbf{F}^{\beta\beta}\end{pmatrix}
\begin{pmatrix}\mbox{}^{\mathrm{Q}}\mathbf{C}^{\alpha}_p\\
\mbox{}^{\mathrm{Q}}\mathbf{C}^{\beta}_p\end{pmatrix}&=\begin{pmatrix}\mbox{}^{\mathrm{Q}}\mathbf{M}^{\alpha\alpha}&\mathbf{0}\\ \mathbf{0}&\mbox{}^{\mathrm{Q}}\mathbf{M}^{\beta\beta}\end{pmatrix}
\begin{pmatrix}\mbox{}^{\mathrm{Q}}\mathbf{C}^{\alpha}_p\\
\mbox{}^{\mathrm{Q}}\mathbf{C}^{\beta}_p\end{pmatrix}\epsilon_p,\quad \epsilon_p\in\mathbb{R},\label{QDEQMat}
\end{align}
gives rise to
\begin{align}
\mbox{}^{\mathrm{q}}\mathbf{F}\mbox{}^{\mathrm{q}}\mathbf{C}_p=
\mbox{}^{\mathrm{q}}\mathbf{M}\mbox{}^{\mathrm{q}}\mathbf{C}_p\epsilon_p,\quad \mbox{}^{\mathrm{q}}\mathbf{C}_p\in\mathbb{Q}^{2n\times 1},\quad \epsilon_p\in\mathbb{R},\label{qGHF}
\end{align}
where $\mbox{}^{\mathrm{q}}F_{\mu\nu}$ has both time-reversal symmetric ($\mbox{}^{\mathrm{q}}\tilde{F}_{\mu\nu}$) and antisymmetric ($\mbox{}^{\mathrm{q}}\bar{F}_{\mu\nu}$)
components, viz.,
 \begin{align}
\mbox{}^{\mathrm{q}}F_{\mu\nu}&
=\mbox{}^{\mathrm{q}}\tilde{F}_{\mu\nu} + \mbox{}^{\mathrm{q}}\bar{F}_{\mu\nu},\label{qF+-}\\
\mbox{}^{\mathrm{q}}\tilde{F}_{\mu\nu}&=\mbox{}^{\mathrm{q}}h_{\mu\nu}+
[(\mbox{}^{\mathrm{q}}\Omega_{\mu\nu}|g_{12}|\tr[\mbox{}^{\mathrm{q}}\tilde{\mathbf{D}}_{\lambda\kappa}\mbox{}^{\mathrm{q}}\Omega_{\kappa\lambda}])
-(\mbox{}^{\mathrm{q}}\Omega_{\mu\lambda}|g_{12}|\mbox{}^{\mathrm{q}}\tilde{\mathbf{D}}_{\lambda\kappa}\mbox{}^{\mathrm{q}}\Omega_{\kappa\nu})]\nonumber\\
&
-\mathrm{c_g}(\mbox{}^{\mathrm{q}}\Xi^{i}_{\mu\lambda}|g_{12}|\mbox{}^{\mathrm{q}}\tilde{\mathbf{D}}_{\lambda\kappa}\mbox{}^{\mathrm{q}}\Xi^{i}_{\kappa\nu})
-\mathrm{c_b}(\mbox{}^{\mathrm{q}}\Xi^{i}_{\mu\lambda}|b_{12}^{ij}|\mbox{}^{\mathrm{q}}\tilde{\mathbf{D}}_{\lambda\kappa}\mbox{}^{\mathrm{q}}\Xi^{j}_{\kappa\nu})],\label{qF+}\\
\mbox{}^{\mathrm{q}}\bar{F}_{\mu\nu}&=
\mathrm{c_g}(\mbox{}^{\mathrm{q}}\Xi^{i}_{\mu\nu}|g_{12}|\tr[h\mbox{}^{\mathrm{q}}\bar{\mathbf{D}}_{\lambda\kappa}\mbox{}^{\mathrm{q}}\Xi^{i}_{\kappa\lambda}])
+\mathrm{c_b}(\mbox{}^{\mathrm{q}}\Xi^{i}_{\mu\nu}|b_{12}^{ij}|\tr[h\mbox{}^{\mathrm{q}}\bar{\mathbf{D}}_{\lambda\kappa}\mbox{}^{\mathrm{q}}\Xi^{j}_{\kappa\lambda}])\nonumber\\
&-(\mbox{}^{\mathrm{q}}\Omega_{\mu\lambda}|g_{12}|h\mbox{}^{\mathrm{q}}\bar{\mathbf{D}}_{\lambda\kappa}\mbox{}^{\mathrm{q}}\Omega_{\kappa\nu})
-\mathrm{c_g}(\mbox{}^{\mathrm{q}}\Xi^{i}_{\mu\lambda}|g_{12}|h\mbox{}^{\mathrm{q}}\bar{\mathbf{D}}_{\lambda\kappa}\mbox{}^{\mathrm{q}}\Xi^{i}_{\kappa\nu})\nonumber\\
&
-\mathrm{c_b}(\mbox{}^{\mathrm{q}}\Xi^{i}_{\mu\lambda}|b_{12}^{ij}|h\mbox{}^{\mathrm{q}}\bar{\mathbf{D}}_{\lambda\kappa}\mbox{}^{\mathrm{q}}\Xi^{j}_{\kappa\nu}),\label{qF-}\\
\mbox{}^{\mathrm{q}}M_{\mu\nu}&=\mbox{}^{\mathrm{Q}}M^{\alpha\alpha}_{\mu\nu}=\mbox{}^{\mathrm{Q}}M^{\beta\beta}_{\mu\nu}
=\begin{pmatrix}S_{\mu\nu}&0\\0&\frac{1}{2c^2}T_{\mu\nu}\end{pmatrix}.
\end{align}
Use of the fact that
the time-reversal antisymmetric ($h\mbox{}^{\mathrm{q}}\bar{\mathbf{D}}$) and
symmetric ($\mbox{}^{\mathrm{q}}\tilde{\mathbf{D}}$) parts of
the density matrix $\mbox{}^{\mathrm{q}}\mathbf{D}$ \eqref{qDENmat} do not contribute to the
number density \eqref{4Cdensity} and current density \eqref{4Ccurrent}, respectively, has been made
to derive the Fock matrix elements.
Since the quaternion density matrix element $\mbox{}^{\mathrm{q}}\tilde{\mathbf{D}}_{\lambda\kappa}$ ($h\mbox{}^{\mathrm{q}}\bar{\mathbf{D}}_{\lambda\kappa}$) is generally nonzero,
each of the five nonzero scalar elements in
$\mbox{}^{\mathrm{q}}\Omega_{\mu\nu}$ \eqref{qOmega} (3 in $\mbox{}^{\mathrm{Q}}\Omega^{\alpha\alpha}_{\mu\nu}$ and 2 in $\mbox{}^{\mathrm{Q}}\Omega^{\alpha\beta}_{\mu\nu}$) can `interact' with each of the five nonzero scalar elements in $\mbox{}^{\mathrm{q}}\Omega_{\kappa\lambda}$, 
implying that there will be in total 25 real-valued scalar integrals to evaluate and process just for a single term of $\mbox{}^{\mathrm{q}}G_{\mu\nu}[\mbox{}^{\mathrm{q}}\mathbf{D}]$
under the Coulomb interaction\cite{ReSpect-JCP2020}.
On the other hand, each $\mbox{}^{\mathrm{q}}\Xi^{i}_{\mu\nu}$ \eqref{qiVecaa} has 18 nonzero scalar elements
(6 in each of the three Cartesian components), there will be 324 real-valued scalar integrals
to evaluate and process just for a single term of $\mbox{}^{\mathrm{q}}G_{\mu\nu}[\mbox{}^{\mathrm{q}}\mathbf{D}]$ under the Gaunt or Breit interaction.
Use of such observations has been made to achieve very efficient implementations of the Gaunt\cite{LiXSGauntInt} and Breit\cite{LiXSBreitInt} integrals.

If wanted, the eigenvector $\mbox{}^{\mathrm{Q}}\mathbf{C}_p$ of Eq. \eqref{QDEQMat} can be mapped out from $\mbox{}^{\mathrm{q}}\mathbf{C}_p=\sum_{i=0}^3\mbox{}^{i}\mathbf{C}_p$
(cf. Eq. \eqref{X2Y}),
\begin{align}
\mbox{}^{\mathrm{Q}}\mathbf{C}_p&=\psi(\mbox{}^{\mathrm{q}}\mathbf{C}_p)\begin{pmatrix}\mathbf{I}_{2n}\\ \mathbf{0}\end{pmatrix}=
\begin{pmatrix}\mbox{}^{0}\mathbf{C}_p+\ii \mbox{}^{1}\mathbf{C}_p\\
-\mbox{}^{2}\mathbf{C}_p+\ii\mbox{}^{3}\mathbf{C}_p\end{pmatrix}=\begin{pmatrix}\mbox{}^{\mathrm{Q}}\mathbf{C}^{\alpha}_p\\
\mbox{}^{\mathrm{Q}}\mathbf{C}^{\beta}_p\end{pmatrix}.\label{X2YC}
\end{align}
The quaternion density matrix $\mbox{}^{\mathrm{q}}\mathbf{D}$ can be calculated as
\begin{align}
\mbox{}^{\mathrm{q}}\mathbf{D}&=\mbox{}^{\mathrm{q}}\mathbf{C}\mathbf{n}\mbox{}^{\mathrm{q}}\mathbf{C}^\dag=\sum_{i=0}^3\mbox{}^{i}\mathbf{D}e_i \in\mathbb{Q}^{2n\times 2n},
\end{align}
which is isomorphic to (cf. Eq. \eqref{qM2QM})
\begin{align}
\mbox{}^{\mathrm{Q}}\mathbf{D}=\psi(\mbox{}^{\mathrm{q}}\mathbf{D})=\mbox{}^{\mathrm{Q}}\mathbf{C}\mathbf{n}\mbox{}^{\mathrm{Q}}\mathbf{C}^\dag
\in\mathbb{C}^{4n\times 4n}.
\end{align}

For a closed-shell system in the absence of external magnetic fields, the time-reversal antisymmetric component $\mbox{}^{\mathrm{q}}\bar{\mathbf{F}}$ vanishes,
such that $\mbox{}^{\mathrm{q}}\mathbf{F}=\mbox{}^{\mathrm{q}}\tilde{\mathbf{F}}\in \mathbb{Q}^{2n\times 2n}$ can be diagonalized very efficiently\cite{RoschQuaternion1983,Qmatdiag1984a,Qmatdiag1984b,SaueQuaternion1999,shiozaki2017quaternion},
faster than the diagonalization of $\mbox{}^{\mathrm{Q}}\mathbf{F}=\mathbf{Q}\mathbf{F}\mathbf{Q}^\dag=\mbox{}^{\mathrm{Q}}\tilde{\mathbf{F}}+
\mbox{}^{\mathrm{Q}}\bar{\mathbf{F}}\in\mathbb{C}^{4n\times 4n}$ by up to a factor of two\cite{shiozaki2017quaternion}.
Yet, in the presence of $\mbox{}^{\mathrm{q}}\bar{F}$, no gain in efficiency can be achieved in the matrix diagonalization. Therefore, it is more 
appealing\cite{ReSpect-JCP2020} to first construct $\mbox{}^{\mathrm{q}}\tilde{\mathbf{F}}$ and $\mbox{}^{\mathrm{q}}\bar{\mathbf{F}}$ and then map them to
$\mbox{}^{\mathrm{Q}}\tilde{\mathbf{F}}$ and $\mbox{}^{\mathrm{Q}}\bar{\mathbf{F}}$
(cf. Eqs. \eqref{SymmAsymm-A} and \eqref{SymmAsymm-B}), respectively.

Apart from the above generalized quaternion DHF equation \eqref{qGHF}, a Kramers-unrestriced quaternion DHF (KUHF) scheme can also be formulated for
open-shell systems (see Ref. \citenum{Nakai-KRUHF} for the two-component counterpart), by drawing analogy with the spin-free UHF approach. The working equations read
\begin{align}
\mbox{}^{\mathrm{q}}\mathbf{F}^{\mathrm{x}}\mbox{}^{\mathrm{q}}\mathbf{C}^{\mathrm{x}}&=\mbox{}^{\mathrm{q}}\mathbf{M}
\mbox{}^{\mathrm{q}}\mathbf{C}^{\mathrm{x}}\epsilon^{\mathrm{x}}_p,\quad \mathrm{x}=\mathrm{u},\mathrm{d}, \label{KUHFup}\\
\mbox{}^{\mathrm{q}}F^{\mathrm{x}}_{\mu\nu}&
=\mbox{}^{\mathrm{q}}\tilde{F}_{\mu\nu} + \mbox{}^{\mathrm{q}}\bar{F}_{\mu\nu},\label{qF+-UHF}\\
\mbox{}^{\mathrm{q}}\tilde{F}_{\mu\nu}&=\mbox{}^{\mathrm{q}}h_{\mu\nu}+
[(\mbox{}^{\mathrm{q}}\Omega_{\mu\nu}|g_{12}|\tr[\mbox{}^{\mathrm{q}}\tilde{\mathbf{D}}_{\lambda\kappa}\mbox{}^{\mathrm{q}}\Omega_{\kappa\lambda}])
-(\mbox{}^{\mathrm{q}}\Omega_{\mu\lambda}|g_{12}|\mbox{}^{\mathrm{q}}\tilde{\mathbf{D}}^{\mathrm{x}}_{\lambda\kappa}\mbox{}^{\mathrm{q}}\Omega_{\kappa\nu})]\nonumber\\
&
-\mathrm{c_g}(\mbox{}^{\mathrm{q}}\Xi^{i}_{\mu\lambda}|g_{12}|\mbox{}^{\mathrm{q}}\tilde{\mathbf{D}}^{\mathrm{x}}_{\lambda\kappa}\mbox{}^{\mathrm{q}}\Xi^{i}_{\kappa\nu})
-\mathrm{c_b}(\mbox{}^{\mathrm{q}}\Xi^{i}_{\mu\lambda}|b_{12}^{ij}|\mbox{}^{\mathrm{q}}\tilde{\mathbf{D}}^{\mathrm{x}}_{\lambda\kappa}\mbox{}^{\mathrm{q}}\Xi^{j}_{\kappa\nu})],\label{qF+x}\\
\mbox{}^{\mathrm{q}}\bar{F}_{\mu\nu}&=
\mathrm{c_g}(\mbox{}^{\mathrm{q}}\Xi^{i}_{\mu\nu}|g_{12}|\tr[h\mbox{}^{\mathrm{q}}\bar{\mathbf{D}}_{\lambda\kappa}\mbox{}^{\mathrm{q}}\Xi^{i}_{\kappa\lambda}])
+\mathrm{c_b}(\mbox{}^{\mathrm{q}}\Xi^{i}_{\mu\nu}|b_{12}^{ij}|\tr[h\mbox{}^{\mathrm{q}}\bar{\mathbf{D}}_{\lambda\kappa}\mbox{}^{\mathrm{q}}\Xi^{j}_{\kappa\lambda}])\nonumber\\
&-(\mbox{}^{\mathrm{q}}\Omega_{\mu\lambda}|g_{12}|h\mbox{}^{\mathrm{q}}\bar{\mathbf{D}}^{\mathrm{x}}_{\lambda\kappa}\mbox{}^{\mathrm{q}}\Omega_{\kappa\nu})
-\mathrm{c_g}(\mbox{}^{\mathrm{q}}\Xi^{i}_{\mu\lambda}|g_{12}|h\mbox{}^{\mathrm{q}}\bar{\mathbf{D}}^{\mathrm{x}}_{\lambda\kappa}\mbox{}^{\mathrm{q}}\Xi^{i}_{\kappa\nu})\nonumber\\
&
-\mathrm{c_b}(\mbox{}^{\mathrm{q}}\Xi^{i}_{\mu\lambda}|b_{12}^{ij}|h\mbox{}^{\mathrm{q}}\bar{\mathbf{D}}^{\mathrm{x}}_{\lambda\kappa}\mbox{}^{\mathrm{q}}\Xi^{j}_{\kappa\nu}),\label{qF-x}
\end{align}
where the density matrices for the `up' ($\mbox{}^{\mathrm{q}}\psi_p=\mbox{}^{\mathrm{q}}\xi_{\mu}\mbox{}^{\mathrm{q}}C^{\mathrm{u}}_{\mu p}$)
and `down' ($\mbox{}^{\mathrm{q}}\bar{\psi}_p= \mbox{}^{\mathrm{q}}\xi_{\mu}\mbox{}^{\mathrm{q}}C^{\mathrm{d}}_{\mu p}$) quaternion Kramers partners
are defined as
\begin{align}
\mbox{}^{\mathrm{q}}\mathbf{D}^{\mathrm{u}}&=\mbox{}^{\mathrm{q}}\mathbf{C}^{\mathrm{u}}\mathbf{n}^{\mathrm{u}}(\mbox{}^{\mathrm{q}}\mathbf{C}^{\mathrm{u}})^\dag=
\mbox{}^{\mathrm{q}}\tilde{\mathbf{D}}^{\mathrm{u}}+h\mbox{}^{\mathrm{q}}\bar{\mathbf{D}}^{\mathrm{u}},\\
\mbox{}^{\mathrm{q}}\mathbf{D}^{\mathrm{d}}&=\mbox{}^{\mathrm{q}}\mathbf{C}^{\mathrm{d}}\mathbf{n}^{\mathrm{d}}(\mbox{}^{\mathrm{q}}\mathbf{C}^{\mathrm{d}})^\dag=
\mbox{}^{\mathrm{q}}\tilde{\mathbf{D}}^{\mathrm{d}}+h\mbox{}^{\mathrm{q}}\bar{\mathbf{D}}^{\mathrm{d}},\\
\mbox{}^{\mathrm{q}}\tilde{\mathbf{D}}&=\mbox{}^{\mathrm{q}}\tilde{\mathbf{D}}^{\mathrm{u}}+\mbox{}^{\mathrm{q}}\tilde{\mathbf{D}}^{\mathrm{d}},\quad
\mbox{}^{\mathrm{q}}\bar{\mathbf{D}}=\mbox{}^{\mathrm{q}}\bar{\mathbf{D}}^{\mathrm{u}}+\mbox{}^{\mathrm{q}}\bar{\mathbf{D}}^{\mathrm{d}}.
\end{align}
Compared with Eq. \eqref{qF+-}, the contributions of opposite Kramers partners
to the exchange type of interactions have been neglected in $\mbox{}^{\mathrm{q}}\mathbf{F}^{\mathrm{x}}$ \eqref{qF+-UHF}, e.g., both $n_j^{\mathrm{u}}(\mbox{}^{\mathrm{q}}\xi_{\mu}\mbox{}^{\mathrm{q}}\psi_j|V(1,2)|\mbox{}^{\mathrm{q}}\psi_j\mbox{}^{\mathrm{q}}\xi_{\nu})
=(\mbox{}^{\mathrm{q}}\Omega_{\mu\lambda}|V(1,2)|\mbox{}^{\mathrm{q}}\mathbf{D}^{\mathrm{u}}_{\lambda\kappa}\mbox{}^{\mathrm{q}}\Omega_{\kappa\nu})$
and $n_j^{\mathrm{d}}(\mbox{}^{\mathrm{q}}\xi_{\mu}\mbox{}^{\mathrm{q}}\bar{\psi}_j|V(1,2)|\mbox{}^{\mathrm{q}}\bar{\psi}_j\mbox{}^{\mathrm{q}}\xi_{\nu})
=(\mbox{}^{\mathrm{q}}\Omega_{\mu\lambda}|V(1,2)|\mbox{}^{\mathrm{q}}\mathbf{D}^{\mathrm{d}}_{\lambda\kappa}\mbox{}^{\mathrm{q}}\Omega_{\kappa\nu})$
appear in $\mbox{}^{\mathrm{q}}\mathbf{F}$ \eqref{qF+-}, but only the former/latter appears in $\mbox{}^{\mathrm{q}}\mathbf{F}^{\mathrm{u/d}}$ \eqref{qF+-UHF}.
This is necessary to make Eq. \eqref{KUHFup} as the stationarity condition of the UHF energy functional,
\begin{align}
E_{\mathrm{KUHF}}&=\frac{1}{2}\tr\left[\mbox{}^{\mathrm{q}}\mathbf{D}^{\mathrm{u}}(\mbox{}^{\mathrm{q}}\mathbf{h}+\mbox{}^{\mathrm{q}}\mathbf{F}^{\mathrm{u}})+
\mbox{}^{\mathrm{q}}\mathbf{D}^{\mathrm{d}}(\mbox{}^{\mathrm{q}}\mathbf{h}+\mbox{}^{\mathrm{q}}\mathbf{F}^{\mathrm{d}})\right].\label{E-KUHF}
\end{align}

The major problem here lies in that there is no unique means to separate the Kramers partners into `up' and `down' subsets. One way\cite{Nakai-KRUHF} is to assign
the occupation numbers $\mathbf{n}^{\mathrm{u}}$ and $\mathbf{n}^{\mathrm{d}}$ according to the $\alpha$ and $\beta$ components of
$\psi(\mbox{}^{\mathrm{q}}\mathbf{C}^{\mathrm{u}}_p)(\mathbf{I}_{2n},\mathbf{0})^T$ and $\psi(\mbox{}^{\mathrm{q}}\mathbf{C}^{\mathrm{d}}_p)(\mathbf{I}_{2n},\mathbf{0})^T$,
respectively (cf. Eq. \eqref{X2YC}).
Alternatively, one can take the A4Ss as the
basis (where the A4Ss with positive and negative $m_j$ values are considered as `up' and `down', respectively),
so as to facilitate the assignment of `up' and `down' molecular Kramers partners.
Note in passing that such a KU scheme has been employed\cite{MomentPolarization-a,MomentPolarization-b,NCOLL} long ago
in relativistic Kohn-Sham (KS) theory for open-shell systems, under the name of `moment polarization'.

One can further introduce a relativistic Kramers restricted open-shell DHF scheme (KROHF), again following
the spin-free ROHF theory\cite{RoothaanROHF,CanonicalROHF} (see Ref. \citenum{Nakai-KROHF} for the two-component counterpart),
\begin{align}
\mbox{}^{\mathrm{q}}\mathbf{F}^{\mathrm{KR}}\mbox{}^{\mathrm{q}}\mathbf{C}^{\mathrm{KR}}&=\mbox{}^{\mathrm{q}}\mathbf{M}
\mbox{}^{\mathrm{q}}\mathbf{C}^{\mathrm{KR}}\epsilon^{\mathrm{KR}}_p,\label{KRODHF}\\
\mbox{}^{\mathrm{q}}\mathbf{F}^{\mathrm{KR}}&=\begin{pmatrix}\mbox{}^{\mathrm{q}}\mathbf{R}_{\mathrm{CC}}
&\mbox{}^{\mathrm{q}}\mathbf{R}_{\mathrm{CO}}&\mbox{}^{\mathrm{q}}\mathbf{R}_{\mathrm{CV}}\\
\mbox{}^{\mathrm{q}}\mathbf{R}_{\mathrm{OC}}&\mbox{}^{\mathrm{q}}\mathbf{R}_{\mathrm{OO}}&\mbox{}^{\mathrm{q}}\mathbf{R}_{\mathrm{OV}}\\
\mbox{}^{\mathrm{q}}\mathbf{R}_{\mathrm{VC}}&\mbox{}^{\mathrm{q}}\mathbf{R}_{\mathrm{VO}}&\mbox{}^{\mathrm{q}}\mathbf{R}_{\mathrm{VV}}\end{pmatrix}
=(\mbox{}^{\mathrm{q}}\mathbf{F}^{\mathrm{KR}})^\dag,\\
\mbox{}^{\mathrm{q}}\mathbf{R}_{\mathrm{CO}}&=\mbox{}^{\mathrm{q}}\mathbf{F}^{\mathrm{d}},\quad
\mbox{}^{\mathrm{q}}\mathbf{R}_{\mathrm{CV}}=\frac{1}{2}(\mbox{}^{\mathrm{q}}\mathbf{F}^{\mathrm{u}}+\mbox{}^{\mathrm{q}}\mathbf{F}^{\mathrm{d}}),\quad
\mbox{}^{\mathrm{q}}\mathbf{R}_{\mathrm{OV}}=\mbox{}^{\mathrm{q}}\mathbf{F}^{\mathrm{u}},\label{Offdiag}\\
\mbox{}^{\mathrm{q}}\mathbf{R}_{\mathrm{CC}}&=\mbox{}^{\mathrm{q}}\mathbf{F}^{\mathrm{d}},\quad
\mbox{}^{\mathrm{q}}\mathbf{R}_{\mathrm{OO}}=\mbox{}^{\mathrm{q}}\mathbf{F}^{\mathrm{u}},\quad
\mbox{}^{\mathrm{q}}\mathbf{R}_{\mathrm{VV}}=\mbox{}^{\mathrm{q}}\mathbf{F}^{\mathrm{u}},\label{DiagR}
\end{align}
where subscripts C, O, and V denote closed-, open-, and vacant-shells, respectively.
Up convergence, the off-diagonal blocks \eqref{Offdiag} vanish, whereas the canonical form\cite{CanonicalROHF} for the diagonal blocks \eqref{DiagR}
has been chosen to satisfy Koopman's theorem\cite{koopmans1934} for electron ionization and attachment.
Unlike KUHF, time-reversal symmetry is fully incorporated in KROHF. Because of this,
the closed shells do not contribute to the direct Gaunt/gauge term, resulting in a significant reduction of the computational cost.
Such KROHF usually works only for `high-spin' open-shell systems, for which the energy functional can be expressed the same as Eq. \eqref{E-KUHF}.

It should be clear that both KUHF \eqref{KUHFup} and KROHF \eqref{KRODHF} are not rigorous, for they are not derived
from the true relativistic energy functionals, but are just induced from the corresponding nonrelativistic counterparts.
One major difference between nonrelativistic and relativistic ROHF lies in that in the former,
all $\alpha$ spin orbitals are automatically orthogonal to all $\beta$ spin orbitals,
but in the latter, a spinor $\psi_p$ is only orthogonal to its own time-reversed partner $\bar{\psi}_p$ but generally nonorthogonal
to other time-reversed spinors $\{\bar{\psi}_q|q\ne p\}$.
The latter fact renders the rigorous formulation of KROHF overly complicated. A much simpler yet rigorous
formulation of relativistic open-shell mean-field theory is the average-of-configuration (AOC) approach\cite{ThyssenAOC},
where the energy is averaged equally over all the determinants $\{D_k\}_{k=1}^K$ that can be generated by distributing $n_e(S)$ electrons in $n_o(S)$ active
spinors of shell $S$ (i.e., $E_{\mathrm{av}}=\sum_{k=1}^K\langle D_k|H|D_k\rangle/K$), precisely in the same way as the nonrelativistic counterpart\cite{McWeenyAOC}.
It is then relatively straightforward to derive the stationarity conditions\cite{ThyssenAOC}, which give rise to optimized and
fully symmetry adapted spinors for subsequent correlated calculations.

The iVI approach\cite{iVI,iVI-TDDFT} can also be modified to obtain directly the occupied states
of quaternion equations \eqref{qGHF}, \eqref{KUHFup}, and \eqref{KRODHF}, so as to achieve
a speedup factor of $2n/N_o$ as compared with the full quaternion matrix diagonalization.

\section{time reversal and quaternion algebra}\label{AppendixB}
The \emph{untiunitary} time-reversal operator $\mathcal{T}$ is defined as
\begin{equation}\label{Tdef}
\mathcal{T}=\begin{cases}
K_0& \text{if one-component},\quad \mathcal{T}^{-1}=\mathcal{T}^\dag=\mathcal{T} \cr
 -\ii \sigma_y K_0& \text{if two-component}, \quad \mathcal{T}^{-1}=\mathcal{T}^\dag=-\mathcal{T}  \cr
-\ii [I_2\otimes \sigma_y]K_0&\text{if four-component},\quad \mathcal{T}^{-1}=\mathcal{T}^\dag=-\mathcal{T},
\end{cases}
\end{equation}
where $K_0$ represents complex conjugation. For a time-reversal symmetric ($t=1$) or antisymmetric ($t=-1$), Hermitian
($h=1$) or anti-Hermitian ($h=-1$), two- or four-component operator $O$ (i.e.,
$\mathcal{T}O\mathcal{T}^{-1}=\mathcal{T}^{-1}O\mathcal{T}=t O$ and $O^\dag=hO$),
its matrix elements in the basis of Kramers pairs $|\phi\rangle=\{|\mu\rangle\}_{\mu=1}^n\cup\{|\bar{\mu}\rangle=\mathcal{T}|\mu\rangle\}_{\mu=1}^{n}$ read
\begin{align}
\begin{pmatrix}O_{\mu\nu}&O_{\mu\bar{\nu}}\\ O_{\bar{\mu}\nu}&O_{\bar{\mu}\bar{\nu}}\end{pmatrix}
&=\begin{pmatrix}A_{\mu\nu}&B_{\mu\nu}\\ -t B^*_{\mu\nu}&t A^*_{\mu\nu}\end{pmatrix},\quad \mathbf{A}=h\mathbf{A}^\dag, \quad \mathbf{B}=-th\mathbf{B}^T,\label{Ohmat}\\
O_{\bar{\mu}\bar{\nu}}&=\langle \mathcal{T}\mu|O|\mathcal{T}\nu\rangle=\langle\mathcal{T}\mu|\mathcal{T}\mathcal{T}^{-1}O\mathcal{T}\nu\rangle
=t\langle\mathcal{T}\mu|\mathcal{T}(O\nu)\rangle\nonumber\\
&=t \langle \mu|O|\nu\rangle^*=t A_{\mu\nu}^*= th A_{\nu\mu},\\
O_{\bar{\mu}\nu}&=\langle\mathcal{T}\mu|O| \nu\rangle=\langle\mathcal{T}\mu|\mathcal{T}\mathcal{T}^{-1}O\mathcal{T}\mathcal{T}^{-1}\nu\rangle=t\langle \mathcal{T}\mu|\mathcal{T}(O\mathcal{T}^{-1}\nu)\rangle\nonumber\\
&=t \langle \mu|O|\mathcal{T}^{-1}\nu\rangle^*=-t\langle \mu|O|\bar{\nu}\rangle^*=-t B_{\mu\nu}^*,\\
B_{\mu\nu}&=\langle \mu|O|\bar{\nu}\rangle=\langle\mathcal{T}\mathcal{T}^{-1}\mu|\mathcal{T}\mathcal{T}^{-1}O\mathcal{T}\nu\rangle\nonumber\\
&=t\langle\mathcal{T}^{-1}\mu|O|\nu\rangle^*=-th\langle\nu|O|\bar{\mu}\rangle=-th(B^T)_{\mu\nu}.
\end{align}
It can readily be verify that the complex conjugate of $\mathbf{O}$ reads
\begin{align}
\mathbf{O}^*&=t\mathbf{O}_T=t\mathbf{U}_T^\dag\mathbf{O}\mathbf{U}_T,\quad \mathbf{U}_T=\begin{pmatrix}0&-1\\ 1&0\end{pmatrix}\otimes \mathbf{I}_n,\label{TtransM}
\end{align}
where  $\mathbf{O}_T$ is the matrix representation of $O$ in the time-reversed basis $\mathcal{T}|\phi\rangle=|\phi\rangle \mathbf{U}_T$.

If $\mathbf{O}$ is time-reversal symmetric, we have
\begin{align}
\mathbf{O}&=
\begin{pmatrix}\mathbf{A}&\mathbf{B}\\-\mathbf{B}^*&\mathbf{A}^*\end{pmatrix}
=\begin{pmatrix}\Re\mathbf{A}+\ii \Im \mathbf{A}&\Re \mathbf{B}+\ii \Im \mathbf{B}\\
-\Re \mathbf{B}+\ii \Im\mathbf{B} &\Re\mathbf{A}-\ii \Im \mathbf{A}\end{pmatrix}\label{Omat}\\
&=\sigma_0\otimes\Re\mathbf{A}+[\ii \sigma_z]\otimes\Im \mathbf{A}
+[\ii \sigma_y]\otimes\Re\mathbf{B}+[\ii \sigma_x]\otimes\Im\mathbf{B},\label{TRSquaternionstructure}
\end{align}
which is algebraically isomorphic to the following (real) quaternion matrix,
\begin{align}
\mbox{}^{\mathrm{q}}\mathbf{O}&= \Re\mathbf{A}+\check{i}\Im\mathbf{A}+\check{j}\Re\mathbf{B}+\check{k}\Im\mathbf{B}\label{RealQuaternion}\\
&\equiv \mathbf{A}+\mathbf{B}\check{j},\label{RealQuaternionB}
\end{align}
through the bijective map $\psi: \mathbb{Q}\rightarrow \mathbb{C}^{2\times 2}$ satisfying
\begin{align}
\psi(e_0=1)&=\mathbf{I}_2=\sigma_0,\quad
\psi(e_1=\check{i})=\ii \sigma_z=\ii\widetilde{\sigma}_1,\quad \psi(e_2=\check{j})=\ii \sigma_y=\ii\widetilde{\sigma}_2, \nonumber\\
\psi(e_3=\check{k})&= \ii \sigma_x=\ii\widetilde{\sigma}_3.\label{QuaternionDef}
\end{align}
Note in passing that the complex matrix $\mathbf{A}$ (and similarly $\mathbf{B}$) in Eq. \eqref{RealQuaternionB}
is a short-hand notation for $\Re\mathbf{A}+\check{i}\Im\mathbf{A}$, which is legitimate for
$\check{i}$ and $\ii$ behave the same under complex conjugation ($*$), transposition ($T$), as well as Hermitian conjugation ($\dagger$).
Moreover, the quaternion units have the following properties,
\begin{align}
 e_i^\dag &=-e_i,\quad e_i^*=(-1)^i e_i,\quad i\in 1,2,3,\label{QuaternionIdentityC}\\
 \check{i}\mathbf{A}&=\mathbf{A}\check{i},\quad \check{j}\mathbf{A}=\mathbf{A}^*\check{j},\quad \check{k}\mathbf{A}=\mathbf{A}^*\check{k},
 \quad (\mathbf{A}e_i)^\dag=-e_i\mathbf{A}^\dag. \label{QuaternionIdentityD}
\end{align}
Since
\begin{align}
\sigma_i\sigma_j=\delta_{ij}+\ii \epsilon_{ijk} \sigma_k,\quad i,j,k\in x, y, z,
\end{align}
and hence
\begin{align}
\widetilde{\sigma}_i\widetilde{\sigma}_j=\delta_{ij}-\ii \epsilon_{ijk} \widetilde{\sigma}_k,\quad i,j,k\in 1,2,3,
\end{align}
we have
\begin{align}
e_ie_j&=-\widetilde{\sigma}_i\widetilde{\sigma}_j\nonumber\\
&=-\delta_{ij}+\epsilon_{ijk}e_k,\quad i,j,k\in 1,2,3,\label{QuaternionIdentitya}\\
&=-\delta_{ij}-\epsilon_{ijk}e_k,\quad i,j,k\in x,y,z,\label{QuaternionIdentityb}
\end{align}
where $\epsilon$ is the Levi-Civita symbol (which satisfies $\epsilon_{ijk}\epsilon_{imn}=\delta_{jm}\delta_{kn}-\delta_{jn}\delta_{km}$
and $\epsilon_{ijk}\epsilon_{ijm}=2\delta_{km}$).
The change of sign when going from Eq. \eqref{QuaternionIdentitya} to Eq. \eqref{QuaternionIdentityb} stems from the fact that
the quaternion units $(e_1, e_2, e_3)$  refer to the ordering $(z,y,x)$ instead of the usual ordering $(x,y,z)$ ($=-(z,y,x)$).

In contrast, if $\mathbf{O}$ is time-reversal antisymmetric, we will have
\begin{align}
%
\mathbf{O}&=\begin{pmatrix}\mathbf{A}&\mathbf{B}\\\mathbf{B}^*&-\mathbf{A}^*\end{pmatrix}
=\ii\begin{pmatrix}-\ii\mathbf{A}&-\ii\mathbf{B}\\-\ii\mathbf{B}^*&\ii\mathbf{A}^*\end{pmatrix} \label{T-TRAMat}\\
&=\ii\left[\sigma_0\otimes\Im\mathbf{A}+[\ii \sigma_z]\otimes(-\Re \mathbf{A})
+[\ii \sigma_y]\otimes\Im\mathbf{B}+[\ii \sigma_x]\otimes(-\Re\mathbf{B})\right],\label{TRAquaternionstructure}
\end{align}
which is isomorphic to the complex quaternion,
\begin{align}
\mbox{}^{\mathrm{q}}\mathbf{O}= h\mbox{}^{\mathrm{q}}\bar{\mathbf{O}}, \quad \mbox{}^{\mathrm{q}}\bar{\mathbf{O}}=\Im\mathbf{A} +\check{i}(- \Re\mathbf{A})+ \check{j}\Im\mathbf{B} +\check{k}(-\Re\mathbf{B})\equiv -\ii\mathbf{A}-\ii\mathbf{B}\check{j},\quad he_i=e_ih,\label{ComplexQuaternion}
\end{align}
where $h$ is an alternative notation of the square root of minus one and commutes with the quaternion units $e_i$.

Whether time-reversal symmetric or antisymmetric, $\mathbf{O}$ \eqref{Ohmat} can be diagonalized as
\begin{align}
\mathbf{O}\mathbf{z}_p&=\begin{pmatrix}\mathbf{A}&\mathbf{B}\\-t\mathbf{B}^*&t\mathbf{A}^*\end{pmatrix}\begin{pmatrix}\mathbf{x}_p\\ \mathbf{y}_p\end{pmatrix}
=\begin{pmatrix}\mathbf{x}_p\\ \mathbf{y}_p\end{pmatrix}\epsilon_p,\quad \mathbf{z}_p^\dag \mathbf{z}_q=\delta_{pq},\quad \epsilon_p\in \mathbb{R}\label{Aeq}
\end{align}
as long as it is Hermitian.
In view of  Eq. \eqref{TtransM},
complex conjugation of Eq. \eqref{Aeq} gives rise to
\begin{align}
\mathbf{O}\bar{\mathbf{z}}_p&=\bar{\mathbf{z}}_p(t\epsilon_p), \quad \bar{\mathbf{z}}_p^\dag \bar{\mathbf{z}}_q=\delta_{pq},\label{Aeq2}
\end{align}
where
\begin{align}
\bar{\mathbf{z}}_p=\mathbf{U}_T\mathbf{z}_p^*=\begin{pmatrix}-\mathbf{y}_p^*\\ \mathbf{x}_p^*\end{pmatrix}\label{ReversedVec}
\end{align}
is the time-reversed solution that is orthogonal to all time-forward solutions $\{\mathbf{z}_q\}$, viz. $\bar{\mathbf{z}}_p^\dag\mathbf{z}_q=0$.
Eq. \eqref{Aeq2} implies that $\bar{\mathbf{z}}_p$ is associated with the same and opposite eigenvalue
as $\mathbf{z}_p$ in the time-reversal symmetric and antisymmetric cases, respectively.
Eqs. \eqref{Aeq} and \eqref{Aeq2} can be combined together,
\begin{align}
\mathbf{O}\mathbf{Z}_p=
\begin{pmatrix}\mathbf{A}&\mathbf{B}\\\mp \mathbf{B}^*&\pm \mathbf{A}^*\end{pmatrix}\begin{pmatrix}\mathbf{x}_p&-\mathbf{y}^*_p\\
\mathbf{y}_p&\mathbf{x}_p^*\end{pmatrix}
=\begin{pmatrix}\mathbf{x}_p&-\mathbf{y}^*_p\\
\mathbf{y}_p&\mathbf{x}^*_p\end{pmatrix}\boldsymbol{\epsilon}^{\pm}_p,\quad \boldsymbol{\epsilon}^{\pm}_p=\begin{pmatrix}\epsilon_p&0\\ 0&\pm \epsilon_p\end{pmatrix},\label{KReq}
\end{align}
which can be block-diagonalized by the quaternion unitary transformation\cite{RoschQuaternion1983,SaueQuaternion1999},
\begin{align}
\mbox{}^{\mathrm{Q}}\mathbf{U}&=\frac{1}{\sqrt{2}}\begin{pmatrix}\mathbf{I}_n&-\check{j}\mathbf{I}_n\\ -\check{j}\mathbf{I}_n&\mathbf{I}_n\end{pmatrix}.\label{QUmatorg}
\end{align}
Specifically, for the time-reversal symmetric case, we have
\begin{align}
\begin{pmatrix}\mbox{}^{\mathrm{q}}\mathbf{O}&\mathbf{0}\\
\mathbf{0}&-\check{k}\mbox{}^{\mathrm{q}}\mathbf{O}\check{k}\end{pmatrix}
\begin{pmatrix}\mbox{}^{\mathrm{q}}\mathbf{z}_p&\mathbf{0}\\
\mathbf{0}&-\check{k}\mbox{}^{\mathrm{q}}\mathbf{z}_p\check{k}\end{pmatrix}&=\begin{pmatrix}\mbox{}^{\mathrm{q}}\mathbf{z}_p&\mathbf{0}\\
\mathbf{0}&-\check{k}\mbox{}^{\mathrm{q}}\mathbf{z}_p\check{k}\end{pmatrix}\begin{pmatrix}\epsilon_p&0\\ 0&\epsilon_p\end{pmatrix},
%
%
\end{align}
where
\begin{align}
\mbox{}^{\mathrm{q}}\mathbf{O}&=\mathbf{A}+\mathbf{B}\check{j},\quad
\mbox{}^{\mathrm{q}}\mathbf{z}_p=\mathbf{x}_p-\mathbf{y}^*_p\check{j}.\label{TRSeig}
\end{align}
For the time-reversal antisymmetric case, we first rewrite $\mathbf{O}$ as $\ii\bar{\mathbf{O}}$ (cf. Eq. \eqref{T-TRAMat})
and $\boldsymbol{\epsilon}^-_p$ as $\ii\bar{\boldsymbol{\epsilon}}^-_p$, with $\bar{\mathbf{O}}$ and $\bar{\boldsymbol{\epsilon}}^-_p$ being time-reversal symmetric.
Eq. \eqref{KReq} can then be transformed to
\begin{align}
\begin{pmatrix}\mbox{}^{\mathrm{q}}\bar{\mathbf{O}}&\mathbf{0}\\
\mathbf{0}&-\check{k}\mbox{}^{\mathrm{q}}\bar{\mathbf{O}}\check{k}\end{pmatrix}
\begin{pmatrix}\mbox{}^{\mathrm{q}}\mathbf{z}_p&\mathbf{0}\\
\mathbf{0}&-\check{k}\mbox{}^{\mathrm{q}}\mathbf{z}_p\check{k}\end{pmatrix}&=\begin{pmatrix}\mbox{}^{\mathrm{q}}\mathbf{z}_p&\mathbf{0}\\
\mathbf{0}&-\check{k}\mbox{}^{\mathrm{q}}\mathbf{z}_p\check{k}\end{pmatrix}\begin{pmatrix}-\epsilon_p\check{i}&0\\ 0&
-\check{k}(-\epsilon_p\check{i})\check{k}\end{pmatrix},\label{AntiBlock}
\end{align}
with $\mbox{}^{\mathrm{q}}\bar{\mathbf{O}}$ given in Eq. \eqref{ComplexQuaternion} and $\mbox{}^{\mathrm{q}}\mathbf{z}_p$
in Eq. \eqref{TRSeig}.
The eigenvalue problem \eqref{KReq} can hence be reduced to a quaternion eigenvalue equation of half the dimension,
\begin{align}
\mbox{}^{\mathrm{q}}\mathbf{O}\mbox{}^{\mathrm{q}}\mathbf{z}_p=\mbox{}^{\mathrm{q}}\mathbf{z}_p\epsilon_p,
\quad \mbox{}^{\mathrm{q}}\mathbf{z}_p^\dag \mbox{}^{\mathrm{q}}\mathbf{z}_q=\delta_{pq}\label{QuaterEIG-S}
\end{align}
for the time-reversal symmetric case, and
\begin{align}
\mbox{}^{\mathrm{q}}\bar{\mathbf{O}}\mbox{}^{\mathrm{q}}\mathbf{z}_p=\mbox{}^{\mathrm{q}}\mathbf{z}_p (-\epsilon_p\check{i}),
\quad \mbox{}^{\mathrm{q}}\mathbf{z}_p^\dag \mbox{}^{\mathrm{q}}\mathbf{z}_q=\delta_{pq}\label{QuaterEIG-A}
\end{align}
for the time-reversal antisymmetric case. The price to pay for the latter case lies in that
the eigenvalues themselves become quaternions.
Given the quaternion solution $\mbox{}^{\mathrm{q}}\mathbf{z}_p$, the eigenvector $\mathbf{z}_p$ of $\mathbf{O}$ can be obtained as
\begin{align}
\mathbf{z}_p&=\begin{pmatrix}\mathbf{x}_p\\\mathbf{y}_p\end{pmatrix}=\begin{pmatrix}\mbox{}^{0}\mathbf{z}_p+\ii \mbox{}^{1}\mathbf{z}_p\\
-\mbox{}^{2}\mathbf{z}_p+\ii \mbox{}^{3}\mathbf{z}_p\end{pmatrix}.\label{Zivec}
\end{align}
If wanted, the time-reversed eigenvector $\bar{\mathbf{z}}_p$ can be generated according to Eq. \eqref{ReversedVec}.
That $\bar{\mathbf{z}}_p$ is associated with the same or opposite eigenvalue as $\mathbf{z}_p$ is dictated by
the real or imaginary nature of $\mbox{}^{\mathrm{q}}\mathbf{O}$.

It is obvious that $\mbox{}^{\mathrm{Q}}\mathbf{U}$ \eqref{QUmatorg} can be modified to
\begin{align}
\mbox{}^{\mathrm{Q}}\mathbf{U}=\frac{1}{\sqrt{2}}\begin{pmatrix}\mathbf{I}_n&\mathbf{0}\\\mathbf{0}&\check{k}\mathbf{I}_n\end{pmatrix}
\begin{pmatrix}\mathbf{I}_n&-\check{j}\mathbf{I}_n\\ -\check{j}\mathbf{I}_n&\mathbf{I}_n\end{pmatrix}
=\frac{1}{\sqrt{2}}\begin{pmatrix}\mathbf{I}_n&-\check{j}\mathbf{I}_n\\\check{i}\mathbf{I}_n&\check{k}\mathbf{I}_n\end{pmatrix},\label{QUmat}
\end{align}
so as to make the diagonal blocks of $\mbox{}^{\mathrm{Q}}\mathbf{U}\mathbf{O}\mbox{}^{\mathrm{Q}}\mathbf{U}^\dag$ identical
for both the time-reversal symmetric and antisymmetric cases. Still, however,
the notion that the complex matrices $\mathbf{A}$ and $\mathbf{B}$ in $\mathbf{O}$ \eqref{Ohmat} should be viewed
as quaternions (i.e., $\mathbf{A}=\Re\mathbf{A}+\check{i}\Im\mathbf{A}$) is uneasy. For this reason,
the more consistent notion of biquaternions (complexified quaternion) should be adopted, where the quaternion units $e_i$ \eqref{QuaternionDef}
commute with their complex coefficients. In particular, a new basis,
\begin{align}
e_{11}&=\frac{1}{2}(1-\ii e_1),\quad e_{21}=\frac{1}{2}(-e_2-\ii e_3),\quad e_{12}=\frac{1}{2}(e_2-\ii e_3),\quad e_{22}=\frac{1}{2}(1+\ii e_1),
\end{align}
can be introduced\cite{Biquaternions} to simply the calculation,
for they satisfy the following simple relations,
\begin{align}
e_{pq}e_{rs}=\delta_{qr}e_{ps},\quad e_{pq}^\dag=e_{qp},\quad e_{11}+e_{22}=1.
\end{align}
It is then very easy to verify that any matrix $\mathbf{M}\in\mathbb{C}^{2n\times 2n}$ can be block-diagonalized by the quaternion unitary transformation,
\begin{align}
\mbox{}^{\mathrm{Q}}\mathbf{U}=\begin{pmatrix}e_{11}\mathbf{I}_n&e_{21}\mathbf{I}_n\\ e_{12}\mathbf{I}_n&e_{22}\mathbf{I}_n\end{pmatrix},\label{QUmatC}
\end{align}
that is,
\begin{align}
\mbox{}^{\mathrm{Q}}\mathbf{U}\mathbf{M}\mbox{}^{\mathrm{Q}}\mathbf{U}^\dag&=\begin{pmatrix}\mbox{}^{\mathrm{q}}\mathbf{M}&\mathbf{0}\\
\mathbf{0}&\mbox{}^{\mathrm{q}}\mathbf{M}\end{pmatrix},\quad \mbox{}^{\mathrm{q}}\mathbf{M}\in\mathbb{Q}^{n\times n},\\
\mbox{}^{\mathrm{q}}\mathbf{M}&=\sum_{p,q=1}^2\mathbf{M}_{pq}e_{pq}=\sum_{p,q=1}^2e_{pq}\mathbf{M}_{pq}\\
&=\sum_{i=0}^3 \mbox{}^i\mathbf{M} e_i=\sum_{i=0}^3 e_i\mbox{}^i\mathbf{M},\label{Mcomponents}\\
\mbox{}^{\mathrm{0}}\mathbf{M}&=\frac{\mathbf{M}_{11}+\mathbf{M}_{22}}{2},\quad \mbox{}^{1}\mathbf{M}=\frac{\mathbf{M}_{11}-\mathbf{M}_{22}}{2\ii},\nonumber\\
\mbox{}^{2}\mathbf{M}&=\frac{\mathbf{M}_{12}-\mathbf{M}_{21}}{2},\quad \mbox{}^{3}\mathbf{M}=\frac{\mathbf{M}_{12}+\mathbf{M}_{21}}{2\ii},\\
\mbox{}^{\mathrm{q}}\mathbf{M}^\dag&=\mbox{}^{0}\mathbf{M}^\dag -\sum_{i=1}^3\mbox{}^{i}\mathbf{M}^\dag e_i.
\end{align}
Literally, for any matrix $\mathbf{M}\in\mathbb{C}^{2n\times 2n}$ there exists a unique biquaternion matrix $\mbox{}^{\mathrm{q}}\mathbf{M}\in\mathbb{Q}^{n\times n}$
or vice versa, through the bijective map:
\begin{align}
\psi(\mbox{}^{\mathrm{q}}\mathbf{M})&=\mathbf{M}=\begin{pmatrix}\mbox{}^{0}\mathbf{M}+\ii \mbox{}^{1}\mathbf{M}&\mbox{}^{2}\mathbf{M}+\ii \mbox{}^{3}\mathbf{M}\\-\mbox{}^{2}\mathbf{M}+\ii \mbox{}^{3}\mathbf{M}&\mbox{}^{0}\mathbf{M}-\ii \mbox{}^{1}\mathbf{M}\end{pmatrix},
\quad \mbox{}^{i}\mathbf{M}\in \mathbb{C}^{n\times n},\quad \mbox{}^{\mathrm{q}}\mathbf{M}\in\mathbb{Q}^{n\times n}.\label{qM2QM}
\end{align}
Note in passing that $\mathbf{M}\in\mathbb{C}^{2n\times 2n}$ can further be separated into time-reversal symmetric ($\tilde{\mathbf{M}}$)
and antisymmetric ($h\bar{\mathbf{M}}$) components,
\begin{align}
\mathbf{M}&=\tilde{\mathbf{M}}+h\bar{\mathbf{M}}\Leftrightarrow \mbox{}^{\mathrm{q}}\mathbf{M}=\mbox{}^{\mathrm{q}}\tilde{\mathbf{M}}+h\mbox{}^{\mathrm{q}}\bar{\mathbf{M}},\label{SymmAsymm}\\
\tilde{\mathbf{M}}&=\psi(\mbox{}^{\mathrm{q}}\tilde{\mathbf{M}})=\begin{pmatrix}\mbox{}^{0}\tilde{\mathbf{M}}+\ii\mbox{}^{1}\tilde{\mathbf{M}} &\mbox{}^{2}\tilde{\mathbf{M}}+\ii \mbox{}^{3}\tilde{\mathbf{M}}\\ -\mbox{}^{2}\tilde{\mathbf{M}}+\ii\mbox{}^{3}\tilde{\mathbf{M}}&\mbox{}^{0}\tilde{\mathbf{M}}-\ii\mbox{}^{1}\tilde{\mathbf{M}}\end{pmatrix},\quad \mbox{}^{\mathrm{q}}\tilde{\mathbf{M}}=\sum_{i=0}^3 \mbox{}^{i}\tilde{\mathbf{M}} e_i,\label{SymmAsymm-A}\\
\bar{\mathbf{M}}&=\psi(\mbox{}^{\mathrm{q}}\bar{\mathbf{M}})=\begin{pmatrix}\mbox{}^{0}\bar{\mathbf{M}}+\ii\mbox{}^{1}\bar{\mathbf{M}} &\mbox{}^{2}\bar{\mathbf{M}}+\ii \mbox{}^{3}\bar{\mathbf{M}}\\ -\mbox{}^{2}\bar{\mathbf{M}}+\ii\mbox{}^{3}\bar{\mathbf{M}}&\mbox{}^{0}\bar{\mathbf{M}}-\ii\mbox{}^{1}\bar{\mathbf{M}}\end{pmatrix},
\quad \mbox{}^{\mathrm{q}}\bar{\mathbf{M}}=\sum_{i=0}^3 \mbox{}^{i}\bar{\mathbf{M}} e_i,\label{SymmAsymm-B}
\end{align}
where
\begin{align}
\mbox{}^0\tilde{\mathbf{M}}&=\frac{1}{2}\Re(\mathbf{M}_{11}+\mathbf{M}_{22}),\quad \mbox{}^0\bar{\mathbf{M}}=\frac{1}{2}\Im(\mathbf{M}_{11}+\mathbf{M}_{22}),\\
\mbox{}^1\tilde{\mathbf{M}}&=\frac{1}{2}\Im(\mathbf{M}_{11}-\mathbf{M}_{22}),\quad \mbox{}^1\bar{\mathbf{M}}=-\frac{1}{2}\Re(\mathbf{M}_{11}-\mathbf{M}_{22}),\\
\mbox{}^2\tilde{\mathbf{M}}&=\frac{1}{2}\Re(\mathbf{M}_{12}-\mathbf{M}_{21}),\quad \mbox{}^2\bar{\mathbf{M}}=\frac{1}{2}\Im(\mathbf{M}_{12}-\mathbf{M}_{21}),\\
\mbox{}^3\tilde{\mathbf{M}}&=\frac{1}{2}\Im(\mathbf{M}_{12}+\mathbf{M}_{21}),\quad
\mbox{}^3\bar{\mathbf{M}}=-\frac{1}{2}\Re(\mathbf{M}_{12}+\mathbf{M}_{21}).
\end{align}
$\mbox{}^{\mathrm{q}}\tilde{\mathbf{M}}$ ($h\mbox{}^{\mathrm{q}}\bar{\mathbf{M}}$) is usually called real (imaginary) part of
biquaternion $\mbox{}^{\mathrm{q}}\mathbf{M}$.
If $\mathbf{M}$ is Hermitian (i.e., $\tilde{\mathbf{M}}=\tilde{\mathbf{M}}^\dag$
and $\bar{\mathbf{M}}=-\bar{\mathbf{M}}^\dag$),
$\mbox{}^0\tilde{\mathbf{M}}$ and $\mbox{}^k\bar{\mathbf{M}}$ ($k\in 1,2,3$) would be symmetric, whereas
$\mbox{}^k\tilde{\mathbf{M}}$ ($k\in 1,2,3$) and $\mbox{}^0\bar{\mathbf{M}}$ antisymmetric.
As an example, applying the $\mbox{}^{\mathrm{Q}}\mathbf{U}$  transformation \eqref{QUmatC} to
the time-reversal antisymmetric case of Eq. \eqref{KReq} leads directly to
\begin{align}
\begin{pmatrix}\mbox{}^{\mathrm{q}}\mathbf{O}&\mathbf{0}\\
\mathbf{0}&\mbox{}^{\mathrm{q}}\mathbf{O}\end{pmatrix}\begin{pmatrix}\mbox{}^{\mathrm{q}}\mathbf{z}_p&
\mathbf{0}\\ \mathbf{0}&\mbox{}^{\mathrm{q}}\mathbf{z}_p\end{pmatrix}&=
\begin{pmatrix}\mbox{}^{\mathrm{q}}\mathbf{z}_p&
\mathbf{0}\\ \mathbf{0}&\mbox{}^{\mathrm{q}}\mathbf{z}_p\end{pmatrix}
\begin{pmatrix}-\ii\epsilon_p e_1 &0\\ 0&-\ii\epsilon_p e_1\end{pmatrix},\label{AsymmEQ}
\end{align}
with $\mbox{}^{\mathrm{q}}\mathbf{O}$ given in Eq. \eqref{ComplexQuaternion}.

Given a right eigenpair of
\begin{align}
\mbox{}^{\mathrm{q}}\mathbf{M}\mbox{}^{\mathrm{q}}\mathbf{X}_p=\mbox{}^{\mathrm{q}}\mathbf{X}_p\lambda_p,\quad \mbox{}^{\mathrm{q}}\mathbf{X}_p=\sum_{i=0}^3
\mbox{}^{i}\mathbf{X}_pe_i\in\mathbb{Q}^{n\times 1},\quad \lambda_p\in\mathbb{C},\label{QuaterEQ}
\end{align}
the corresponding eigenvector $\mathbf{Y}_p$ of
\begin{align}
\psi(\mbox{}^{\mathrm{q}}\mathbf{M})\mathbf{Y}_p=\mathbf{Y}_p\lambda_p,\quad \mathbf{Y}_p\in\mathbb{C}^{2n\times 1},\quad \lambda_p\in\mathbb{C}\label{ComplexEQ}
\end{align}
is simply the complex adjoint vector $\vec{\mbox{}^{\mathrm{q}}\mathbf{X}}_p$ of $\mbox{}^{\mathrm{q}}\mathbf{X}_p$,
\begin{align}
\mathbf{Y}_p&=\vec{\mbox{}^{\mathrm{q}}\mathbf{X}}_p=\begin{pmatrix}\mbox{}^{0}\mathbf{X}_p+\ii \mbox{}^{1}\mathbf{X}_p\\
-\mbox{}^{2}\mathbf{X}_p+\ii\mbox{}^{3}\mathbf{X}_p\end{pmatrix}
=\psi(\mbox{}^{\mathrm{q}}\mathbf{X}_p)\begin{pmatrix}\mathbf{I}_n\\ \mathbf{0}\end{pmatrix},\label{X2Y}
\end{align}
which is in line with Eq. \eqref{Zivec}. This arises from the complex adjoint vectors of
the left and right hand sides of Eq. \eqref{ComplexEQ},
\begin{align}
\psi(\mbox{}^{\mathrm{q}}\mathbf{M}\mbox{}^{\mathrm{q}}\mathbf{X}_p)\begin{pmatrix}\mathbf{I}_n\\ \mathbf{0}\end{pmatrix}=
\psi(\mbox{}^{\mathrm{q}}\mathbf{M})\psi(\mbox{}^{\mathrm{q}}\mathbf{X})\begin{pmatrix}\mathbf{I}_n\\ \mathbf{0}\end{pmatrix}=
\psi(\mbox{}^{\mathrm{q}}\mathbf{X})\psi(\lambda_p)\begin{pmatrix}\mathbf{I}_n\\ \mathbf{0}\end{pmatrix}=
\psi(\mbox{}^{\mathrm{q}}\mathbf{X})\begin{pmatrix}\mathbf{I}_n\\ \mathbf{0}\end{pmatrix}\lambda_p,\quad \lambda_p\in\mathbb{C}.\label{X2Ychain}
\end{align}
Conversely, given a complex eigenpair of Eq. \eqref{ComplexEQ}, the right eigenvector of Eq. \eqref{QuaterEQ}
can be obtained as\cite{Biquaternions}
\begin{align}
\mbox{}^{\mathrm{q}}\mathbf{X}_p&=\mathbf{E}_{2n}\mathbf{Y}_p,\quad \mathbf{E}_{2n}=\begin{pmatrix}e_{11}\mathbf{I}_n&e_{21}\mathbf{I}_{n}\end{pmatrix},\label{Y2X}
\end{align}
which stems from the following identities,
\begin{align}
\mbox{}^{\mathrm{q}}\mathbf{M}&=\mathbf{E}_{2n}\psi(\mbox{}^{\mathrm{q}}\mathbf{M})\mathbf{E}_{2n}^\dag,
\quad\psi(\mbox{}^{\mathrm{q}}\mathbf{M})\mathbf{E}_{2n}^\dag\mathbf{E}_{2n}=\mathbf{E}_{2n}^\dag\mathbf{E}_{2n}\psi(\mbox{}^{\mathrm{q}}\mathbf{M}),\quad
\mathbf{E}_{2n}\mathbf{E}_{2n}^\dag=\mathbf{I}_n,\\
\mbox{}^{\mathrm{q}}\mathbf{M}\mbox{}^{\mathrm{q}}\mathbf{X}&=(\mathbf{E}_{2n}\psi(\mbox{}^{\mathrm{q}}\mathbf{M})\mathbf{E}_{2n}^\dag)(\mathbf{E}_{2n}\mathbf{Y}_p)
=\mathbf{E}_{2n}\psi(\mbox{}^{\mathrm{q}}\mathbf{M})\mathbf{Y}_p=\mathbf{E}_{2n}\mathbf{Y}_p\lambda_p=\mbox{}^{\mathrm{q}}\mathbf{X}_p\lambda_p.
\end{align}
However, care should be taken when the right eigenvalue $\lambda_p$ of $\mbox{}^{\mathrm{q}}\mathbf{M}$ is not a complex number but instead a quaternion,
$\mbox{}^{\mathrm{q}}\lambda_p=\sum_{i=0}^3\mbox{}^{i}\lambda_p e_i$ with $\sum_{i=1}^3 (\mbox{}^{i}\lambda_p)^2=\tau^2_p$ (cf. Eq. \eqref{AsymmEQ}).
In this case, the last equality of Eq. \eqref{X2Ychain} does not hold. Yet,
in view of the characteristic polynomial of $\psi(\mbox{}^{\mathrm{q}}\lambda_p)$,
\begin{align}
\vline\gamma\mathbf{I}_2-\psi(\mbox{}^{\mathrm{q}}\lambda_p)\vline&=
\vline \begin{array}{cc}\gamma-(\mbox{}^{0}\lambda_p+\ii \mbox{}^{1}\lambda_p)&- \mbox{}^{2}\lambda_p-\ii \mbox{}^{3}\lambda_p\\
\mbox{}^{2}\lambda_p-\ii \mbox{}^{3}\lambda_p&\gamma-(\mbox{}^{0}\lambda_p-\ii \mbox{}^{1}\lambda_p)\end{array}\vline\nonumber\\
&=(\gamma-\mbox{}^{0}\lambda_p)^2
+\tau^2_p,\quad \tau^2_p=(\mbox{}^{1}\lambda_p)^2+(\mbox{}^{2}\lambda_p)^2+(\mbox{}^{3}\lambda_p)^2,
\end{align}
a similarity transformation $\mathbf{P}_p$ can be performed if $\tau^2_p\ne 0$, such that
\begin{align}
\psi(\lambda_p)&=\begin{pmatrix}\mbox{}^{0}\lambda_p+\ii \mbox{}^{1}\lambda_p & \mbox{}^{2}\lambda_p+\ii \mbox{}^{3}\lambda_p\\
-\mbox{}^{2}\lambda_p+\ii \mbox{}^{3}\lambda_p&\mbox{}^{0}\lambda_p-\ii \mbox{}^{1}\lambda_p\end{pmatrix}=\mathbf{P}_p\begin{pmatrix}\mbox{}^{0}\lambda_p+\ii\tau_p&0\\
0&\mbox{}^{0}\lambda_p-\ii\tau_p\end{pmatrix}\mathbf{P}_p^{-1}.
\end{align}
and hence
\begin{align}
\mbox{}^{\mathrm{q}}\lambda_p&=\mbox{}^{\mathrm{q}}P_p[\mbox{}^{0}\lambda_p+\tau_p e_1]\mbox{}^{\mathrm{q}}P_p^{-1},\quad \mbox{}^{\mathrm{q}}P_p=\mathbf{E}_{2}\mathbf{P}_p\mathbf{E}_{2}^\dag,
\quad \mbox{}^{\mathrm{q}}P_p^{-1}=\mathbf{E}_{2}\mathbf{P}_p^{-1}\mathbf{E}_{2}^\dag,
\end{align}
in view of the correspondence
\begin{align}
\psi(\mbox{}^{\mathrm{q}}\mathbf{A})=\psi(\mbox{}^{\mathrm{q}}\mathbf{B})&\Leftrightarrow\mbox{}^{\mathrm{q}}\mathbf{A}=\mbox{}^{\mathrm{q}}\mathbf{B}.
\end{align}
Eq. \eqref{QuaterEQ} can therefore be reexpressed as
\begin{align}
\mbox{}^{\mathrm{q}}\mathbf{M}(\mbox{}^{\mathrm{q}}\mathbf{X}_p\mbox{}^{\mathrm{q}}P_p)&=(\mbox{}^{\mathrm{q}}\mathbf{X}_p\mbox{}^{\mathrm{q}}P_p)
[\mbox{}^{0}\lambda_p+\tau_pe_1],
\end{align}
from which we obtain
\begin{align}
\psi(\mbox{}^{\mathrm{q}}\mathbf{M})\psi(\mbox{}^{\mathrm{q}}\mathbf{X}_p\mbox{}^{\mathrm{q}}P_p)
&=\psi(\mbox{}^{\mathrm{q}}\mathbf{X}_p\mbox{}^{\mathrm{q}}P_p)\begin{pmatrix}\mbox{}^{0}\lambda_p+\ii\tau_p&0\\
0&\mbox{}^{0}\lambda_p-\ii\tau_p\end{pmatrix}.
\end{align}
It follows that the two columns of $\psi(\mbox{}^{\mathrm{q}}\mathbf{X}_p\mbox{}^{\mathrm{q}}P_p)$ are two eigenvectors of $\psi(\mbox{}^{\mathrm{q}}\mathbf{M})$
associated with eigenvalues $\mbox{}^{0}\lambda_p\pm\ii\tau_p$.
Applying such general manipulation to the special case of Eq. \eqref{AsymmEQ},
it is easily found that $\mbox{}^{0}\lambda_p=0$, $\tau_p=-\ii\epsilon_p$, $\mathbf{P}_p=\mathbf{I}_2$, and $\mbox{}^{\mathrm{q}}P_p=1$,
such that the two columns of
$\psi(\mbox{}^{\mathrm{q}}\mathbf{z}_p)$, i.e., $\mathbf{z}_p$ \eqref{Zivec} and $\bar{\mathbf{z}}_p$ \eqref{ReversedVec},
are two eigenvectors of $\mathbf{O}$ \eqref{T-TRAMat} associated with eigenvalues $\pm\epsilon_p$.

On the other hand, if $\tau^2_p= 0$, $\psi(\mbox{}^{\mathrm{q}}\lambda_p)$ would not be diagonalizable but which can be made similar to the Jordan canonical form,
\begin{align}
\psi(\mbox{}^{\mathrm{q}}\lambda_p)&=\mathbf{Q}_p\begin{pmatrix}\mbox{}^{0}\lambda_p&1\\
0&\mbox{}^{0}\lambda_p\end{pmatrix}\mathbf{Q}_p^{-1},
\end{align}
which leads to
\begin{align}
\mbox{}^{\mathrm{q}}\lambda_p&=\mbox{}^{\mathrm{q}}Q_p[\mbox{}^{0}\lambda_p+\frac{1}{2}e_2-\frac{1}{2}\ii e_3]\mbox{}^{\mathrm{q}}Q_p^{-1},\quad
\mbox{}^{\mathrm{q}}Q_p=\mathbf{E}_{2}\mathbf{Q}_p\mathbf{E}_{2}^\dag,\\
\mbox{}^{\mathrm{q}}\mathbf{M}(\mbox{}^{\mathrm{q}}\mathbf{X}_p\mbox{}^{\mathrm{q}}Q_p)&=(\mbox{}^{\mathrm{q}}\mathbf{X}_p\mbox{}^{\mathrm{q}}Q_p)
(\mbox{}^{0}\lambda_p+\frac{1}{2}e_2-\frac{1}{2}\ii e_3),
\end{align}
and hence
\begin{align}
\psi(\mbox{}^{\mathrm{q}}\mathbf{M})\psi(\mbox{}^{\mathrm{q}}\mathbf{X}_p \mbox{}^{\mathrm{q}}Q_p)&=\psi(\mbox{}^{\mathrm{q}}\mathbf{X}_p \mbox{}^{\mathrm{q}}Q_p)\psi(\mbox{}^{0}\lambda_p+\frac{1}{2}e_2-\frac{1}{2}\ii e_3)=
\psi(\mbox{}^{\mathrm{q}}\mathbf{X}_p \mbox{}^{\mathrm{q}}Q_p)\begin{pmatrix}\mbox{}^{0}\lambda_p&1\\ 0& \mbox{}^{0}\lambda_p\end{pmatrix}.
\end{align}
It follows that the first column of $\psi(\mbox{}^{\mathrm{q}}\mathbf{X}_p \mbox{}^{\mathrm{q}}Q_p)$ is an eigenvector of
$\psi(\mbox{}^{\mathrm{q}}\mathbf{M})$ associated with eigenvalue $\mbox{}^{0}\lambda_p$.

As a final note, the product of two real quaternion matrices, $\mbox{}^{\mathrm{q}}\mathbf{A}=\mathbf{A}+\mathbf{B}\check{j}=\sum_{i=0}^3 \mbox{}^i\mathbf{A}e_i$
and $\mbox{}^{\mathrm{q}}\mathbf{C}=\mathbf{C}+\mathbf{D}\check{j}=\sum_{i=0}^3 \mbox{}^i\mathbf{C}e_i$, can be expressed
in terms of their components,
\begin{align}
\mbox{}^{\mathrm{q}}\mathbf{A}\mbox{}^{\mathrm{q}}\mathbf{C}&=(\mathbf{A}\mathbf{C}-\mathbf{B}\mathbf{D}^*)+(\mathbf{A}\mathbf{D}+\mathbf{B}\mathbf{C}^*)\check{j}\\
&=(\mbox{}^0\mathbf{B}\mbox{}^0\mathbf{C}-\mbox{}^i\mathbf{B}\mbox{}^i\mathbf{C})+e_i (\mbox{}^0\mathbf{B}\mbox{}^i\mathbf{C}+\mbox{}^i\mathbf{B}\mbox{}^0\mathbf{C}
+\epsilon_{ijk}\mbox{}^j\mathbf{B}\mbox{}^k\mathbf{C}),\quad i,j,k\in 1,2,3,\label{qrqr123}\\
&=(\mbox{}^0\mathbf{B}\mbox{}^0\mathbf{C}-\mbox{}^i\mathbf{B}\mbox{}^i\mathbf{C})+e_i (\mbox{}^0\mathbf{B}\mbox{}^i\mathbf{C}+\mbox{}^i\mathbf{B}\mbox{}^0\mathbf{C}
-\epsilon_{ijk}\mbox{}^j\mathbf{B}\mbox{}^k\mathbf{C}),\quad i,j,k\in x, y, z.\label{qrqrxyz}
%
%
%
\end{align}

\clearpage
\newpage

\bibliographystyle{apsrev4-2}
\bibliography{BDFlib}

\end{document}